\documentclass{aa}  
\usepackage{graphicx}
\usepackage{txfonts}
\usepackage{lscape}

\newcommand{\cmc}{\mbox{$\mbox{cm}^{-3}$}}
\newcommand{\kms}{\mbox{km$\,$s$^{-1}$}}
\newcommand{\cmg}{\mbox{$\mbox{cm}^{2} \, \mbox{g}^{-1}$}} 
\newcommand{\lsun}{\mbox{$L_\odot$}}
\newcommand{\msun}{\mbox{$M_\odot$}}
\newcommand{\mjb}{\mbox{$\mbox{mJy} \, \mbox{beam}^{-1}$}}
\newcommand{\htwo}{\mbox{$_{\mbox{\tiny H$_2$}}$}}
\newcommand{\K}{\mbox{K}}
\newcommand{\hii}{H\mbox{\sc ~ii} }       
\newcommand{\lbol}{\mbox{$L_{\mbox{\tiny bol}}$}}

\newcommand{\speak}{\mbox{$S_{\mbox{\tiny 1.2~mm}}^{\mbox{\tiny ~peak}}$}}
\newcommand{\sint}{\mbox{$S_{\mbox{\tiny 1.2~mm}}^{\mbox{\tiny ~int}}$}}

\newcommand{\tdust}{\mbox{$T_{\mbox{\tiny dust}}$}}
\newcommand{\kmm}{\mbox{$\kappa_{\mbox{\tiny 1.2~mm}}$}}
\newcommand{\Bmm}{\mbox{$B_{\mbox{\tiny 1.2~mm}}$(\tdust)}}
\newcommand{\msmm}{\mbox{$M_{\mbox{\tiny 1.2~mm}}$}}
\newcommand{\mvir}{\mbox{$M_{\mbox{\tiny vir}}$}}

\begin{document}
   \title{The earliest phases of high-mass star formation:\\
a 3 square degree millimeter continuum mapping of Cygnus~X\thanks{Tables~\ref{t:densecores}--\ref{t:clumps}, Figs.~\ref{f:msxnorth_app}--\ref{f:sio_app}, and Appendices~A and B are only available in electronic form at http://www.aanda.org. Fits images associated with Figs.~\ref{f:mambo}a--c and catalogs built from Tables~\ref{t:densecores}--\ref{t:clumps} can be queried from the CDS via anonymous ftp to cdsarc.u-strasbg.fr (130.79.128.5) or via http://cdsweb.u-strasbg.fr/cgi-bin/qcat?J/A+A/.}}

   \subtitle{}

   \author{F. Motte\inst{1,2}
          \and S. Bontemps\inst{3}
          \and P. Schilke\inst{4}
          \and N. Schneider\inst{1,5}
          \and K. M. Menten\inst{4}
          \and D. Brogui\`ere\inst{6}}

   \offprints{F. Motte}

   \institute{Laboratoire AIM, CEA/DSM - CNRS - Universit\'e Paris Diderot, DAPNIA/Service d'Astrophysique, B\^at. 709, CEA-Saclay, F-91191 Gif-sur-Yvette C\'edex, France\\
              \email{motte@cea.fr}
         \and California Institute of Technology, Downs Laboratory of Physics, Mail Stop 320-47, 1200 E California Blvd, Pasadena, CA 91125, USA
         \and OASU/LAB-UMR~5804, CNRS - Universit\'e Bordeaux 1, 2 rue de l'Observatoire, BP 89, 33270 Floirac, France
         \and Max-Planck-Institut f\"ur Radioastronomie, Auf dem H\"ugel 69, 53121 Bonn, Germany
         \and I. Physik. Institut, Universit\"at K\"oln, 50937 K\"oln, Germany
         \and IRAM, 300 rue de la Piscine, F-38406 St. Martin d'H\`eres, France}

   \date{Received 10 May 2007/ Accepted 23 July 2007}

\abstract
{}
{Our current knowledge of high-mass star formation is mainly based on follow-up studies of bright sources found by \emph{IRAS}, and is thus biased against its earliest phases, inconspicuous at infrared wavelengths. We therefore started searching, in an unbiased way and in the closest high-mass star-forming complexes, for the high-mass analogs of low-mass pre-stellar cores and class~0 protostars.}
{We have made an extensive 1.2~mm continuum mosaicing study of the \object{Cygnus~X} molecular cloud complex using the MAMBO cameras at the IRAM 30~m telescope. The $\sim 3 ^{\circ^2}$ imaged areas cover all the high-column density ($A_\mathrm{V} \ge 15$~mag) clouds of this nearby ($\sim 1.7$~kpc) cloud complex actively forming OB stars. We then compared our millimeter maps with mid-infrared images, and have made SiO(2-1) follow-up observations of the best candidate progenitors of high-mass stars.}
{Our complete study of Cygnus~X with $\sim 0.09$~pc resolution provides, for the first time, an unbiased census of massive young stellar objects. We discover 129 massive dense cores ({\it FWHM} size $\sim 0.1$~pc, $\msmm = 4-950~\msun$, volume-averaged density $\sim 10^5~\cmc$), among which $\sim 42$ are probable precursors of high-mass stars. A large fraction of the Cygnus~X dense cores ($2/3$ of the sample) remain undetected by the  \emph{MSX} satellite, regardless of the mass range considered. Among the most massive ($>40~\msun$) cores, infrared-quiet objects are driving powerful outflows traced by SiO emission. Our study qualifies 17 cores as good candidates for hosting massive infrared-quiet protostars, while up to 25 cores potentially host high-luminosity infrared protostars. We fail to discover in the high-mass analogs of pre-stellar dense cores ($\sim 0.1$~pc, $> 10^4~\cmc$) in Cygnus~X, but find several massive starless clumps ($\sim 0.8$~pc, $7 \times 10^3~\cmc$) that might be gravitationally bound.}
{Since our sample is derived from a single molecular complex and covers every embedded phase of  high-mass star formation, it gives the first statistical estimates of their lifetime. In contrast to what is found for low-mass class~0 and class~I phases, the infrared-quiet protostellar phase of high-mass stars may last as long as their better-known high-luminosity infrared phase. The statistical lifetimes of high-mass protostars and pre-stellar cores ($\sim 3 \times 10^4$~yr and $< 10^3$~yr) in Cygnus~X are one and two order(s) of magnitude smaller, respectively, than what is found in nearby, low-mass star-forming regions. We therefore propose that high-mass pre-stellar and protostellar cores are in a highly dynamic state, as expected in a molecular cloud where turbulent processes dominate.}

\titlerunning{High-mass star formation in the Cygnus~X complex}
\maketitle
%
\section{Introduction}
High-mass (OB, $>8~\msun$) stars, though few in number, play a major role in the energy budget of galaxies.  Our current understanding of their formation, however, remains very schematic, especially concerning the earliest phases of the process.  High-mass stars are known to form in dense cores within molecular cloud complexes, by accretion (\cite{YS02}; \cite{KKMK07}) and/or coalescence (e.g. \cite{bonn01}).  The copious UV flux emitted by a newly-formed central star heats and ionizes its parental molecular cloud, leading to the formation and development of a hot core (e.g. \cite{HvD97}) and afterwards an \hii region (see a review by \cite{chur99}).

In the past few years, the progenitors of ultracompact \hii (UCH\mbox{\sc \,ii}) regions (sources  that have not yet begun to ionize their environment in a detectable way) have been searched among \emph{IRAS} point sources.  The criteria generally used aim at selecting high-luminosity ($> 10^3~\lsun$) stellar embryos embedded in a massive envelope and associated with hot gas but no UC\hii region (see \cite{kurtz00}). In fact, most surveys focussed on red \emph{IRAS} sources with colors satisfying the \cite{WC89} criteria (originally set for selecting UC\hii regions) but not detected in radio centimeter surveys.  The presence of an envelope around the  \emph{IRAS} sources is confirmed by their association with high-density gas, and the existence of hot gas by the detection of a hot core and water or methanol masers (e.g. \cite{BNM96}; \cite{plum97}).  In this way, studies by e.g., \cite{mol00} and  \cite{srid02} have identified massive young stellar objects harboring high-luminosity infrared protostars (see also \cite{muel02}; \cite{faun04}).

Such surveys are biased against younger and, thus probably colder, massive young stellar objects which could be inconspicuous in the mid-infrared bands of \emph{IRAS} and the \emph{Midcourse Space Experiment} (\emph{MSX}). These high-mass analogs of low--mass class~0 protostars and pre-stellar cores (cf. \cite{AWB00}) should be best detected via far-infrared to (sub)millimeter dust continuum and high-density molecular line tracers. Serendipitous discoveries of a few infrared-quiet massive young stellar objects have been made in submillimeter continuum maps (e.g. \cite{mol98a}; \cite{MSL03}; \cite{SS04}; \cite{hill05}) and as absorption regions in mid-infrared images (often called ``infrared dark clouds", e.g. \cite{car00}; \cite{thom05}; \cite{rath06}). These studies suggest that, in the high-mass star formation process also, infrared-quiet phases must exist and be associated with significant populations of young protostars and/or pre-stellar cores. The lack of statistics, homogeneity, and angular resolution of studies such as the above, however, prevents the determination of the basic characteristics  of these infrared-quiet objects.

To make significant progress, one needs to search in a systematic way for the earliest phases of high-mass star formation in nearby complexes. One of the most appropriate methods is to survey entire molecular cloud complexes in which high-mass stars are forming, using tracers of high-density clouds and protostellar activity signatures.  We therefore started a multitracer study of Cygnus~X, which is one of the richest molecular and \hii complexes located at less than 3~kpc from the Sun. The molecular cloud complex is massive ($4\times 10^6~\msun$) and extends over $\sim 100$~pc in diameter (cf. \cite{LT92}). Found to be part of the local spiral arm, this complex has been  located at 1.7~kpc from the Sun by the recent $^{13}$CO study of \cite{schn06}.  Cygnus~X has a rich collection of \hii regions (see centimeter free-free emission maps by \cite{WHL91}, and infrared images from \emph{MSX} presented in Figs.~7--9 of Schneider et al. 2006), indicating recent high-mass star formation. It is associated with several OB associations (see \cite{uyan01} and references therein), amongst them one of the largest in our Galaxy (Cyg~OB2, cf. \cite{knod00}). It also contains several well-known massive young stellar objects like DR21, DR21(OH), AFGL~2591, S106-IR, and W75N (e.g. \cite{SED98}; \cite{vdT99}; \cite{schn02}; \cite{SKT04}).  Small parts of the Cygnus~X cloud complex have already been observed in (sub)millimeter continuum emission (e.g. \cite{CGC93}; \cite{rich93}; \cite{VF06}; \cite{davi07}).

The present paper tackles the following key questions: do high-mass pre-stellar cores exist? What are the lifetimes of  high-mass pre-stellar cores, infrared-quiet protostars, and high-luminosity infrared protostars? What are the main physical processes leading from pre-stellar objects to high-mass protostars? We report an extensive 1.2~mm continuum study of the Cygnus~X complex complemented by SiO(2-1) follow-up observations of the best candidate progenitors of high-mass stars. From the MAMBO-2 imaging ($\sim 3 ^{\circ^2}$) of the entire complex presented in Sect.~\ref{s:obs}, we make a complete census of the compact cloud fragments and larger-scale structures (Sect.~\ref{s:structure}). We determine the main characteristics of these new millimeter sources in Sect.~\ref{s:structure}, and search for signposts of protostellar activity including SiO emission (a tracer of outflow activity) in Sect.~\ref{s:nature}. Section~\ref{s:discussion} identifies 129 massive dense cores, among which 42 are probable precursors of high-mass stars. This unbiased sample of young embedded massive stars reveals 17 dense cores that are excellent candidates for harboring high-mass protostars in their infrared-quiet phase. The present paper also gives the first statistical results on the lifetime of high-mass protostars and pre-stellar cores. Finally, Sect.~\ref{s:summary} summarizes our conclusions, and  Appendices~A and B describe our data reduction and source extraction techniques.

\section{Observations and data reduction}\label{s:obs}

\begin{figure}
\centering
\includegraphics[width=9cm]{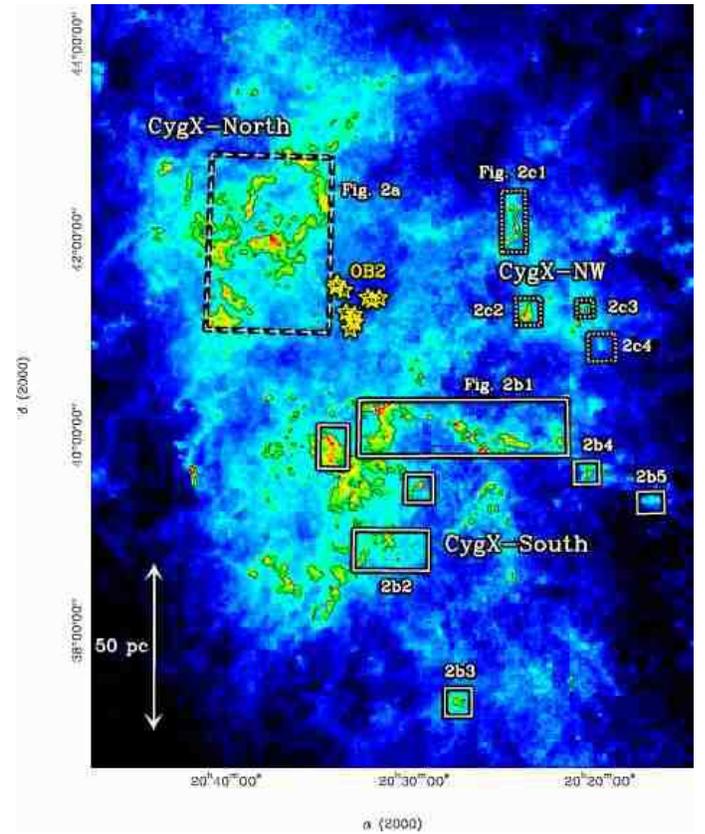}
\caption[]{Extinction map of the Cygnus~X complex derived by Bontemps et al. (in prep.) from the stellar reddening of background stars in JHK, using the 2MASS database and a pixel size of $1.3\arcmin$. The high-extinction regions are yellow and outlined by the 15~mag contour level. The three molecular cloud ensembles identified by \cite{schn06} are labeled and the nearby Cygnus~OB2 association is indicated. The fields mapped with MAMBO or MAMBO-2 are schematically outlined and add up to $\sim 3^{\circ^2}$.}
\label{f:extinct}
\end{figure}

\subsection{Imaging the entire molecular complex}

The Cygnus~X molecular cloud complex covers up to $\sim 30 ^{\circ^2}$ (see Fig.~1 of  Schneider et al. 2006) but the sites of high-mass star formation are found in its highest-density parts.  We therefore used an infrared extinction map produced from 2MASS\footnote{The Two Micron All Sky Survey was performed by the University of Massachussets and IPAC/Caltech and funded by NASA and NSF.} data (cf. Fig.~\ref{f:extinct} and \cite{Bontemps06})  to select the high-column density ($A_\mathrm{V} \ge 15$~mag) clouds of Cygnus~X to be mapped in millimeter continuum emission. Figure~\ref{f:extinct} and \cite{schn06} show that the Cygnus~X molecular complex contains several ensembles of dense molecular clouds  which extend over $\sim 50$ parsecs.  In the following, we use the name ``CygX-North" for the north-eastern group of clouds which harbors the well-known sources DR21 and W75N, ``CygX-South" for the southern region containing AFGL~2591 and S106-IR, and ``CygX-NW" for the north-western filamentary clouds. The mappings necessary to cover the high-column density clouds of Cygnus~X are outlined in Fig.~\ref{f:extinct} and cover a total area of $\sim 3^{\circ^2}$.

\subsection{Dust continuum observations at 1.2~mm}

The Cygnus~X molecular complex  was imaged at 1.2~mm with the MPIfR bolometer arrays installed at the IRAM\footnote{IRAM is supported by INSU/CNRS (France), MPG (Germany) and IGN (Spain).}  30~m telescope at Pico Veleta (Spain).  The passband of the MPIfR bolometers has an equivalent width of $\approx 70$~GHz and is centered at $\nu_{\rm eff} \approx 240$~GHz (\cite{krey98}).  The MAMBO (37 channels) camera was used in February-April 1999, January-March 2000, and December 2000-May 2001 for a total integration time of $\sim 30$~hours. The MAMBO-2 (117 channels) camera was used for $\sim 33$~hours as part of the observing pools organized by IRAM between November 2002 and March 2003.  A total of 142 large on-the-fly maps were taken using a fast-mapping technique similar to that described by \cite{TS98}.  These maps consist of a series of rows scanned in azimuth with a velocity of $8\arcsec$/sec and spaced by $22\arcsec$ (respectively $46\arcsec$) steps in elevation when using MAMBO or MAMBO-2.  The typical azimuthal size of individual maps is $\sim 7.5\arcmin$ with MAMBO versus $\sim 11\arcmin$ with MAMBO-2.  The signal of each bolometer is modulated by the secondary mirror which is wobbling with a frequency of  2~Hz  and a throw in azimuth of $70\arcsec$.  The resulting dual-beam maps were reduced with the IRAM software for bolometer-array data (NIC; cf. \cite{brog95})  using the EKH restoration algorithm (\cite{EKH}) and a skynoise reduction technique developed by us (see Appendix~A).  The imaging of each requested field was obtained by combining partially overlapping restored on-the-fly maps.  

The main beam was measured and found to have a {\it HPBW} size of $\sim 11\arcsec$ using Uranus and Mars.  The absolute pointing of the telescope was found to be accurate to within $\sim 5\arcsec$.  The data were all taken in winter but with heterogeneous weather conditions. The zenith atmospheric optical depth varied between 0.1 and 0.5 and the skynoise level was low to high (correlation factor from 0.2 to 1). The resulting rms noise in individual maps  is $\sigma=10-150~\mjb$, reduced to $\sigma \sim 20~\mjb$ after reduction of the skynoise (see Appendix~A).  The final mosaics all have $\sigma < 20~\mjb$ rms with a median rms of $\sigma \sim 15~\mjb$. Uranus and Mars were also used for flux calibration and the overall absolute calibration uncertainty is estimated to be $\sim 20\%$.

\subsection{SiO(2-1) observations}\label{s:obs_sio}

Pointed observations in the SiO v=0 J=2$\to$1 transition were performed for 40 millimeter continuum sources listed in Table~\ref{t:densecores} in September 2003 and July 2004 with the IRAM 30~m telescope. We used the A100 (B100) SIS receiver and the VESPA autocorrelator with a frequency resolution of 40~kHz that is equivalent to $0.135~\kms$ at 86.847~GHz. The system temperature for both observing runs was between 110 and 130~K and the average rms for each data set is 0.03~K on a main beam brightness temperature scale, taking an efficiency $\eta_{\mbox{\tiny MB}}$ of 0.78. The angular resolution of the telescope at 87~GHz is $29\arcsec$. All observations were performed in position-switching mode with off-positions a few arcminutes away in azimuth. Pointing was checked regularly and was accurate to within $3\arcsec$. Standard calibration sources were observed regularly and were consistent within 10\%. The data were reduced with the IRAM software for spectral lines (CLASS), applying first order baselines and averaging individual spectra with rms weighting.

\section{Analysis of the 1.2~mm mapping of Cygnus~X}\label{s:structure}

\begin{figure*}
\centering
\includegraphics[angle=0,width=16cm]{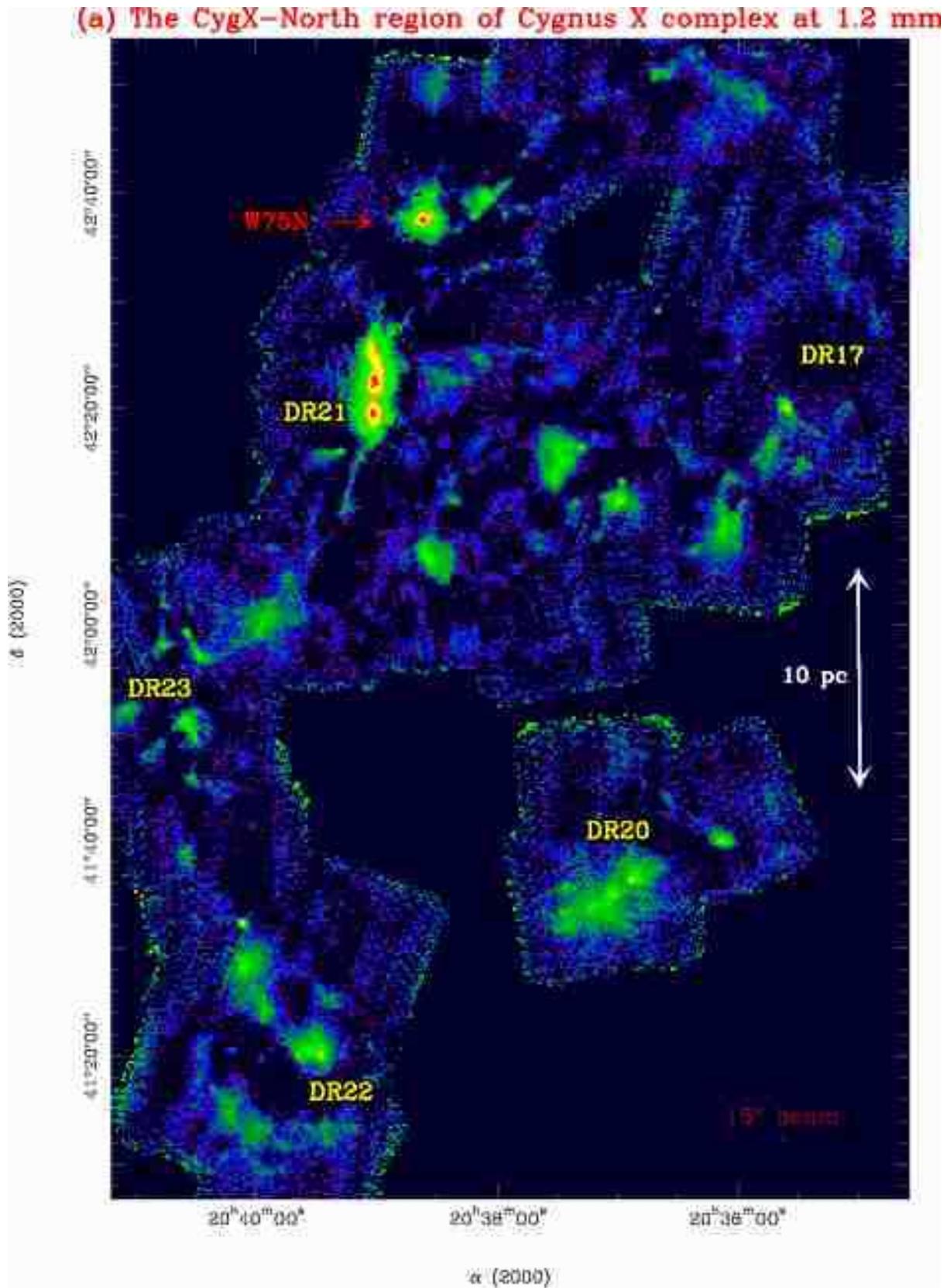}
\caption[]{Millimeter continuum imaging of the Cygnus~X molecular cloud complex obtained with the MAMBO and MAMBO-2 cameras installed at the IRAM 30~m telescope.  These 1.2~mm maps have been smoothed to an effective angular resolution of $15\arcsec$, allowing a sensitivity of 0.1--5~pc cloud structures.  The main radio sources (\cite{DR66}) and a few well-known sources are indicated as reference marks.  {\bf a} The CygX-North region (see Fig.~\ref{f:extinct} for its location): maximum flux is $\sim 8\,500~\mjb$ (color scale is saturated beyond $500~\mjb$) and rms noise level is $\sigma = 10-20~\mjb$.}
\label{f:mambo}
\end{figure*}

\setcounter{figure}{1}
\begin{figure*}
\centering
\includegraphics[angle=180,width=15.5cm]{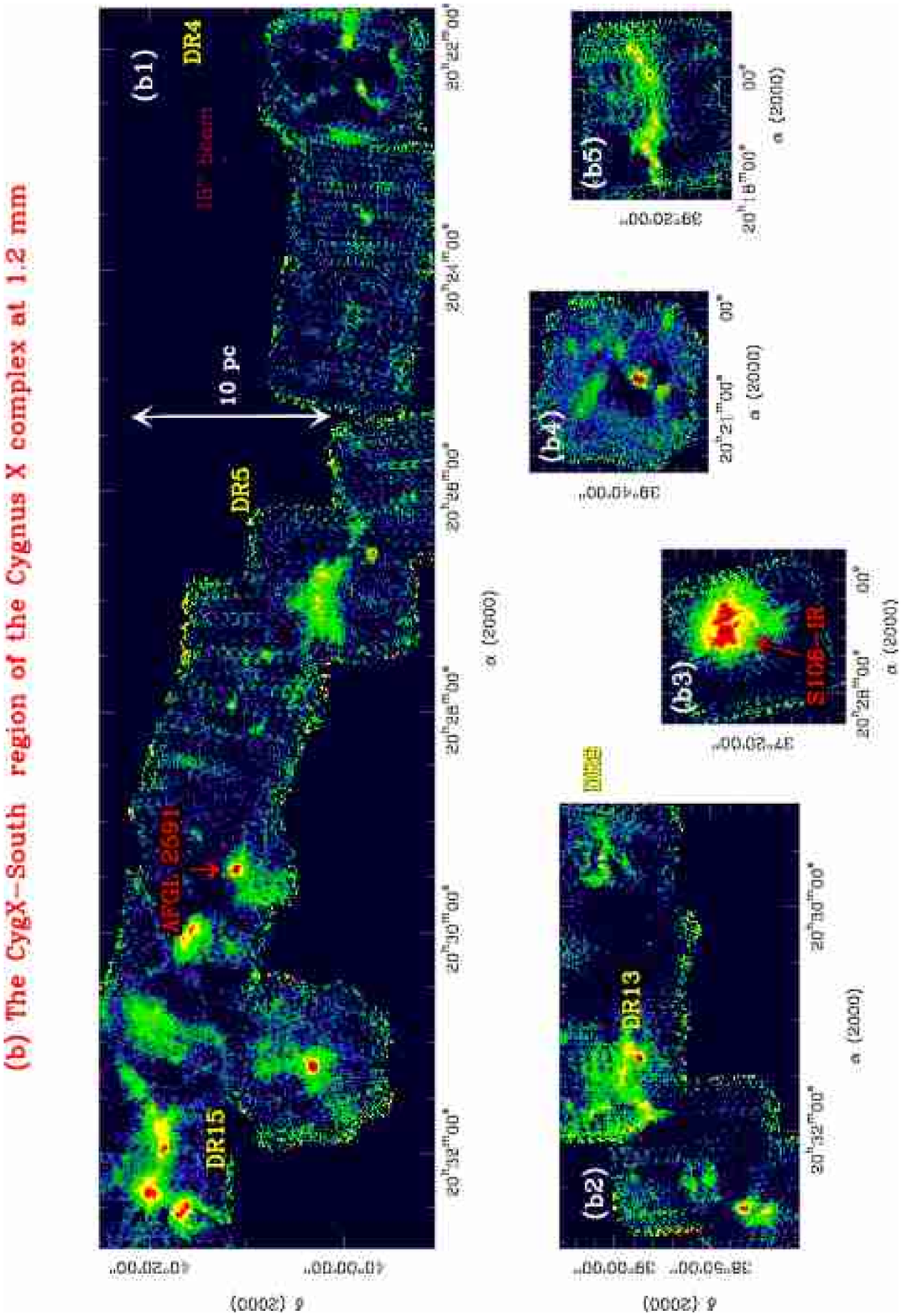}
\caption[]{(continued) {\bf b} The CygX-South region: maximum flux is $\sim 8\,500~\mjb$ (color scale is saturated beyond $500~\mjb$) and rms noise levels are $9-18~\mjb$ in {\bf b1}, $\sim 14~\mjb$ in {\bf b2}, $\sim 14~\mjb$ in {\bf b3}, $\sim 13~\mjb$ in {\bf b4}, and $\sim 15~\mjb$ in {\bf b5}. Note that the MAMBO-2 maps associated with DR12 and DR6 in Fig.~\ref{f:extinct} are ignored since they do not display any compact and dense gas.}
\end{figure*}

\subsection{Dust continuum images}\label{s:spatdyn}

The 1.2~mm continuum images are presented in Figs.~\ref{f:mambo}a-c with zooms in the left-hand parts of Figs.~\ref{f:msxnorth}--\ref{f:msxsouth} (see also Figs.~\ref{f:msxnorth_app}--\ref{f:msxnw_app} which are only available electronically).  The dense molecular gas sometimes follows filamentary structures (well observed in the $^{13}$CO(1-0) images of Schneider et al. in prep.) and displays many compact ($\sim 0.1$~pc) fragments.  We make a complete census of these 1.2~mm fragments in Sect.~\ref{s:census1} and a similar but less complete census of larger-scale ($\sim 1$~pc) cloud structures in Sect.~\ref{s:census2}.

The MAMBO-2 mosaics are sensitive to spatial scales ranging from $\sim 0.09$~pc (corresponding to an {\it HPBW} of $11\arcsec$ at 1.7~kpc) to $\sim 5$~pc (i.e. the mean azimuth extent of individual maps $\sim 10\arcmin$ at 1.7~kpc). The resulting  spatial dynamic  is $\sim 55$,  larger than that  of other (sub)millimeter dust continuum observations made with SCUBA at the JCMT, SIMBA at the SEST, or BOLOCAM at the CSO.  Therefore, large fast-mapping images with MAMBO-2 are currently the best tool to study the density structure of molecular clouds with high spatial resolution. Given the dynamic range in the MAMBO-2 image (within a $11\arcsec$ beam, cold dust emission ranges from $\sigma \sim 15~\mjb$ to $6\,300~\mjb$), the Cygnus~X complex is probed from $N\htwo \sim 9\times 10^{21}$~cm$^{-2}$  to $2\times 10^{24}$~cm$^{-2}$, or similarly from $A_\mathrm{V} \sim 10$~mag to $2100$~mag (i.e. with a column density dynamic of more than 200). The above estimate uses Eq.~1 from \cite{MAN98} with dust temperatures of 20~K and dust opacities of $\kmm=0.005$ and $0.01~\cmg$ for the lower and higher density cloud structures, respectively.

Due to the spatial filtering of cloud structures larger than clumps (i.e. $> 5$~pc, see Appendix~A) in MAMBO-2 images, the total mass detected in millimeter continuum emission is 10 to 20 times smaller ($37\times 10^3~\msun$ in CygX-North and $26\times 10^3~\msun$ in CygX-South) than that measured in the CO survey of \cite{schn06}.

\setcounter{figure}{1}
\begin{figure}
\centering
\includegraphics[width=9cm]{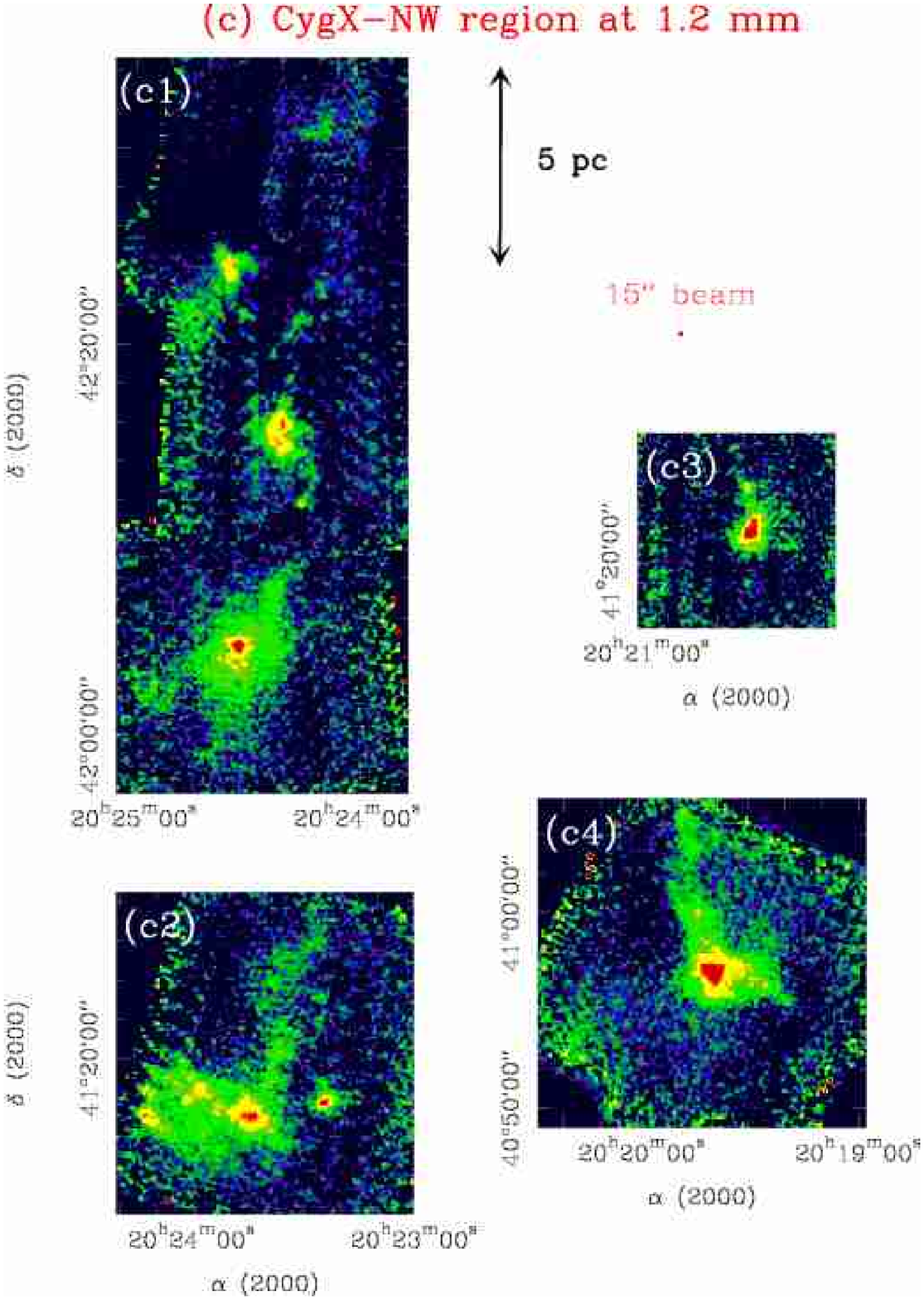}
\caption[]{(continued) {\bf c} The CygX-NW region: maximum flux is $\sim 800~\mjb$ (color scale is saturated beyond $500~\mjb$) and rms noise levels are $9-16~\mjb$ in {\bf c1}, $\sim 13~\mjb$ in {\bf c2}, $\sim 18~\mjb$ in {\bf c3}, and $\sim 12~\mjb$ in {\bf c4}.}
\end{figure}

\subsection{Census of compact cloud fragments in Cygnus~X}\label{s:census1}

Since our main goal is to investigate the best potential sites of high-mass star formation, we first focus on the small-scale  cloud fragments observed in our MAMBO-2 images. As expected, larger-scale ($\sim 1$~pc) cloud structures generally have a moderate volume-averaged density ($\sim 5\times 10^3~\cmc$ for a 1.2~mm peak flux of $5\sigma = 75~\mjb$ over a diameter of 1~pc, when a 20~K dust temperature and a $\kmm=0.01~\cmg$ dust mass opacity are assumed, cf. Eqs.~\ref{e:mass} and \ref{e:density}) and will therefore be less likely to form high-mass stars in the near future. We apply the source extraction technique developed by Motte et al. (2003) which uses a multi-resolution analysis (\cite{SM06}) and the Gaussclumps program (\cite{SG90}; \cite{kram98}). We first identify 129 compact ($< 1$~pc) cloud fragments above the median $5\sigma = 75~\mjb$ level in the MAMBO-2 maps, where spatial scales larger than 1~pc have been filtered out (see e.g. Fig.~\ref{f:extraction}b).  We then precisely derive the 1.2~mm characteristics of each fragment by fitting a 2D-Gaussian in the image where its local background (or surrounding large-scale emission) has been entirely removed (see Appendix~B for more details).
 
The compact fragments extracted here are marked and labeled in the left-hand parts of Figs.~\ref{f:msxnorth}--\ref{f:msxsouth} and Figs.~\ref{f:msxnorth_app}--\ref{f:msxnw_app}. Table~\ref{t:densecores}, which is only available electronically, lists the 129 fragments discovered in Cygnus~X and gives their 1.2~mm basic characteristics as estimated above. The fragments are listed in increasing RA coordinate for each region. They are numbered (Col.~1) and named after their 2000 equatorial coordinates (Col.~2). Their 1.2~mm peak flux density (Col.~3), their {\it FWHM} size (Col.~4), and integrated flux (Col.~5) are measured assuming a Gaussian shape. In Col.~9, we list previous names given to strong millimeter peaks of the DR21 filament (cf. \cite{CGC93}), as well as the name of nearby benchmark sources (e.g. \cite{car00}).

\subsection{Large-scale cloud structures in Cygnus~X}\label{s:census2}

To complement the study performed in Sect.~\ref{s:census1}, we here make a census of the large-scale ($\sim 1$~pc) cloud structures of Cygnus~X. We first smooth the MAMBO-2 maps to a $55\arcsec$ resolution, which roughly corresponds to $\sim 0.5$~pc and is the resolution of a N$_2$H$^+$ survey performed at the FCRAO (Schneider et al. in prep.). We then identify 40 large-scale cloud fragments above the median $5\sigma = 75~\mjb$ level and fit them by a 2-D Gaussian. Note that the skynoise reduction technique we applied may have filtered out part of the large-scale emission (for more details see Appendix~A). The present analysis is therefore less secure than that performed in Sect.~\ref{s:census1} for the small-scale fragments of the Cygnus~X molecular clouds. Table~\ref{t:clumps}, which is only available electronically, lists the 40 large-scale clumps found in Cygnus~X and gives their 1.2~mm basic characteristics (in Cols.~1-6) with the same convention as in Table~\ref{t:densecores}. In addition, Col.~7 lists the compact fragments (if any) harbored by each larger-scale structure.

\subsection{Mass and density derived from millimeter continuum maps}\label{s:mass} 

The mass and density of the compact cloud fragments ($< 1$~pc) and large-scale structures ($\sim 1$~pc)  of Cygnus~X are determined from their dust continuum parameters given in Tables~\ref{t:densecores} and \ref{t:clumps}. These tables list the mass (Col.~6 vs 5) and volume-averaged density (Col.~7 vs 6) estimated for each identified dense core and clump using methods and assumptions described below.

Before we estimate the mass of each cloud fragment, we should ensure that the millimeter sources identifed in Sects.~\ref{s:census1} and \ref{s:census2} are indeed density structures.  According to previous studies of high-mass star forming regions (see e.g. \cite{MSL03}), the main source of contamination of 1.2~mm fluxes is the free-free emission from \hii regions. Unfortunately, the published centimeter wavelength free-free images of Cygnus~X (e.g. \cite{WHL91}) lack the necessary angular resolution to estimate the amount of likely contamination of each 1.2~mm source. Furthermore, the spectral index of the free-free emission is unknown, precluding the extrapolation of its flux at 1.2~mm. Observations with high-density tracers such as molecular lines at the IRAM 30~m, Effelsberg, and FCRAO telescopes, as well as 350~$\mu$m continuum with SHARC~II at the CSO, suggest that the vast majority of the 1.2~mm compact and larger-scale fragments are definitive density structures traced by dust emission rather than free-free centimeter sources (e.g. Schneider et al. in prep.). Fifteen millimeter sources of Table~\ref{t:densecores} are suspected to contain \hii regions (cf. Sect.~\ref{s:hii}); their associated masses and densities are therefore upper limits.

We believe that the millimeter continuum emission of cloud fragments identified in Sects.~\ref{s:census1} and \ref{s:census2} is mainly thermal dust emission, which is largely optically thin. For any given dust properties and gas-to-dust ratio, the 1.2~mm fluxes are thus directly related to the total (gas~$+$~dust) mass of the fragments.  For present mass estimates, we use the 1.2~mm integrated fluxes ($\sint$) of Tables~\ref{t:densecores}--\ref{t:clumps} without correction for any free-free contamination, and derive the mass (\msmm, given in Tables~\ref{t:densecores}--\ref{t:clumps}), as follows:
\begin{eqnarray}\label{e:mass}
\msmm & =    & \frac{\sint\; d^{2}}{\kmm \; \Bmm} \nonumber\\
      &\simeq& 5.3 \ \msun \times 
              \left(\frac {\sint}{\mbox{100~mJy}}\right) 
              \left(\frac {d}{\mbox{1.7~kpc}} \right)^2 \nonumber\\
      &      & \times \left(\frac {\kmm}{0.01\,\cmg}\right)^{-1}
                     \left(\frac {\tdust}{20~\K}\right)^{-1},
\label{eq:mass}
\end{eqnarray}
where $\kmm$ is the dust opacity per unit mass column density at 1.2~mm, $\Bmm$ is the Planck function for a dust temperature \tdust, and where we assume a distance of 1.7~kpc.
 
The dust mass opacity (including dust properties and gas-to-dust mass ratio) is likely to vary with density, temperature, and the evolutionary state of the emitting medium (\cite{HMS95}).  Models of dust in low-mass protostellar cores (e.g. \cite{OH94}) suggest that a value of $\kmm=0.01~\cmg$ is well suited for cool ($10-30~\K$) and high-density ($n\htwo \sim 10^5~\cmc$) cloud fragments.  This value agrees with the cross-comparisons of dust emission surrounding an UC{\hii} region with its CO ice absorption and gas emission (\cite{vdT02}).  We therefore choose a dust opacity per unit (gas~$+$~dust) mass column density of $\kmm=0.01~\cmg$ for compact cloud fragments. We use this value for all the dense fragments because the average dust properties of these $\sim 0.1$~pc structures should not change drastically when they contain a massive protostar. We also adopt $\kmm=0.01~\cmg$ for larger-scale ($\sim 1$~pc) clumps, despite the fact that a lower value is generally used for lower-density starless cloud fragments (e.g. Ward-Thompson et al. 1999). We estimate that the absolute value taken for the dust mass opacity is uncertain by a factor of 2.

The temperature to be used in Eq.~(\ref{eq:mass}) is the mass-weighted dust temperature of the cloud fragments, whose value could be determined from gray-body fitting of their spectral energy distributions. Such measures are not yet available but we recently measured the ammonia rotational temperatures of the Cygnus~X dense cores (Wyrowski et al. in prep.). The ratios of NH$_3$ (1,1) and (2,2) transitions observed with the Effelsberg 100~m telescope ($40\arcsec$ beam) give cold temperatures ($\sim 15$~K) for the material located within a 0.33~pc diameter of a millimeter peak. To estimate the temperature averaged on $\sim 0.1$~pc scales, we assume  a classical $\rho(r) \propto r^{-2}$ density law and inner heating ($T(r) \propto r^{-0.4}$), in agreement with small-scale fragments being highly centrally condensed and generally protostellar (cf. Sects.~\ref{s:highdens}--\ref{s:potential}). We thus estimate that $\sim 0.1$~pc fragments should have mass-weighted dust temperatures in the range of 15--25~K. This does not preclude any hotter or colder components from being embedded in the $\sim 0.1$~pc fragments; temperature gradients will be discussed in companion papers (Schneider et al. in prep.; Wyrowski et al. in prep.). For simplicity, we assume $\tdust=20~\K$ in Eq.~(\ref{eq:mass}) for all the cloud fragments of Table~\ref{t:densecores}. We  add estimates made with $\tdust=40~\K$ for \hii regions (identified in Sect.~\ref{s:hii}) and the brightest ($\lbol > 5\times 10^3 \lsun$) infrared protostars of Cygnus~X: IRAS~20343+4129, studied by \cite{beut02} which corresponds to CygX-N5, and LkHA~225S which coincides with CygX-NW5. To avoid biasing our statistical results, we hereafter use the estimates made with a homogeneous 20~K  temperature. The dust temperature of large-scale clumps is more difficult to constrain since it may vary according to the presence or absence of strong external heating. We estimate that the mass-weighted dust temperature may vary from $\tdust=10$~K to 20~K, and choose $\tdust=15~\K$ as an averaged value to use in Eq.~(\ref{eq:mass}) for the large-scale ($\sim 1$~pc) clumps of Table~\ref{t:clumps}.

The mass estimates in Tables~\ref{t:densecores}--\ref{t:clumps} are globally uncertain by a factor of 2 due to uncertain dust emissivity. The individual masses can vary by $\pm 30\%$ relative to each other when dust temperatures vary from 15 to 25~K, or by $\pm 50\%$ for the 10 to 20~K temperature range (see also Eq.~\ref{eq:mass}).

The volume-averaged densities listed in Tables~\ref{t:densecores}--\ref{t:clumps} and used in Tables~\ref{t:scale}--\ref{t:lifetime} are estimated as follows:
\begin{eqnarray}\label{e:density}
<n\htwo> =    & \frac{ \msmm}{\frac{4}{3}\;\pi\;\times\;\mbox{\rm \it FWHM}^{\;3}}, \label{eq:density}
\end{eqnarray}
where $\msmm$ is the mass derived by Eq.~(\ref{eq:mass}) and {\it FWHM} is the full width at half maximum, determined by Gaussian fits. Using a radius equal to one {\it FWHM} in Eq.~(\ref{e:density}) allows one to accurately determine the volume-averaged density because the flux (and thus the mass) measured within such an aperture corresponds to $>98\%$ of the integrated flux (respectively total mass) of Gaussian cloud structures. The often-used, beam-averaged peak density would of course be higher, but is less relevant when estimating physical constraints such as the free-fall time.

\section{Signpost of stellar activity toward the Cygnus~X cloud fragments}\label{s:nature}

We investigate the spatial coincidence of the compact cloud fragments identified in Sect.~\ref{s:census1} with mid-infrared, SiO, centimeter free-free, and maser emission (see Sects.~\ref{s:infrared}--\ref{s:radio}). Large-scale cloud structures identified in Sect.~\ref{s:census2}, which are associated with neither compact fragments displaying signposts of stellar activity nor bright mid-infrared sources, qualify as starless clumps as indicated in Col.~7 of Table~\ref{t:clumps}.

\begin{figure*}
\vskip -0.2cm
\centerline{\hskip -0.1 cm \includegraphics[angle=270,width=10.9cm]{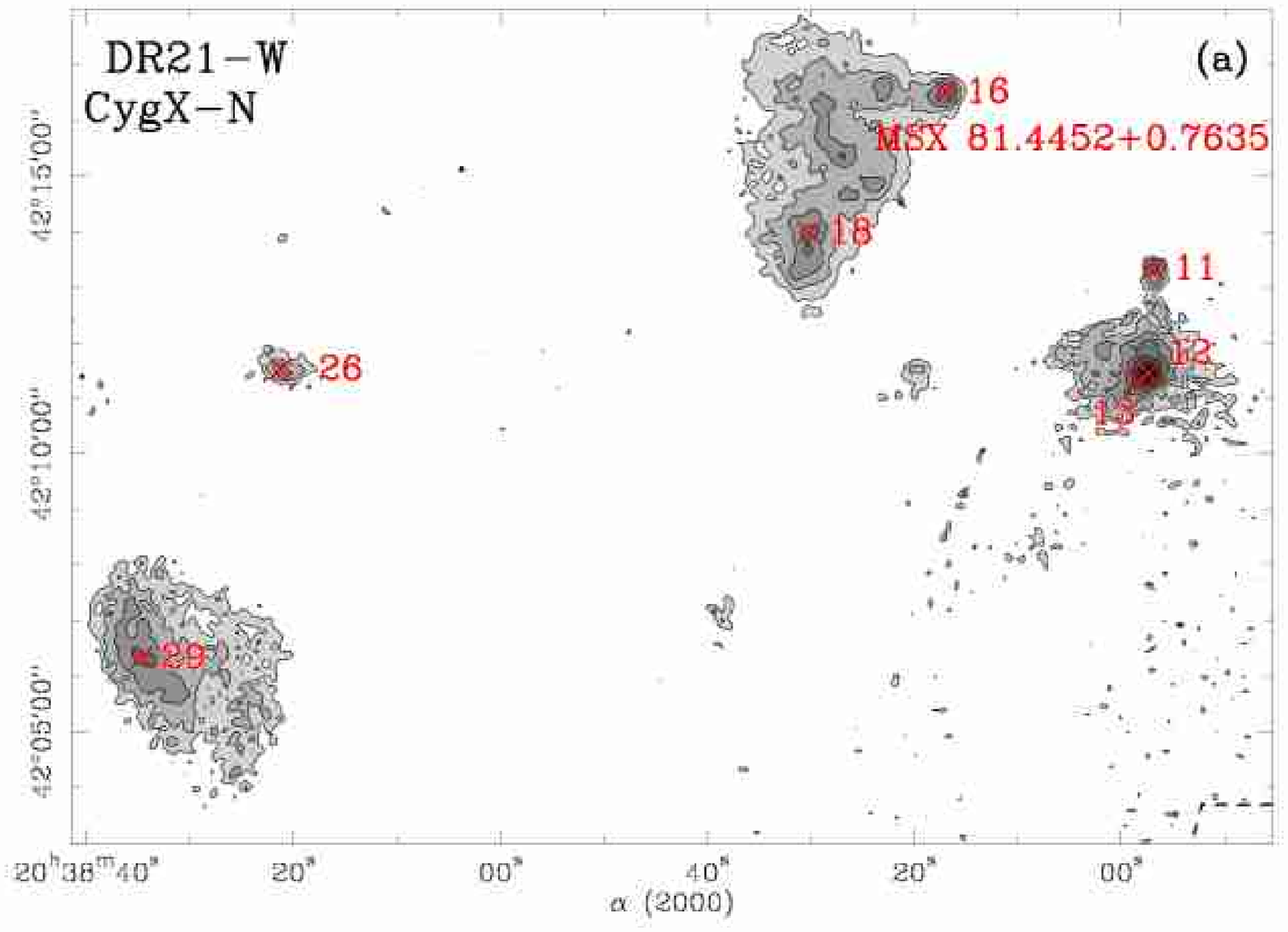}
\hskip -1.8 cm \includegraphics[angle=270,width=10.9cm]{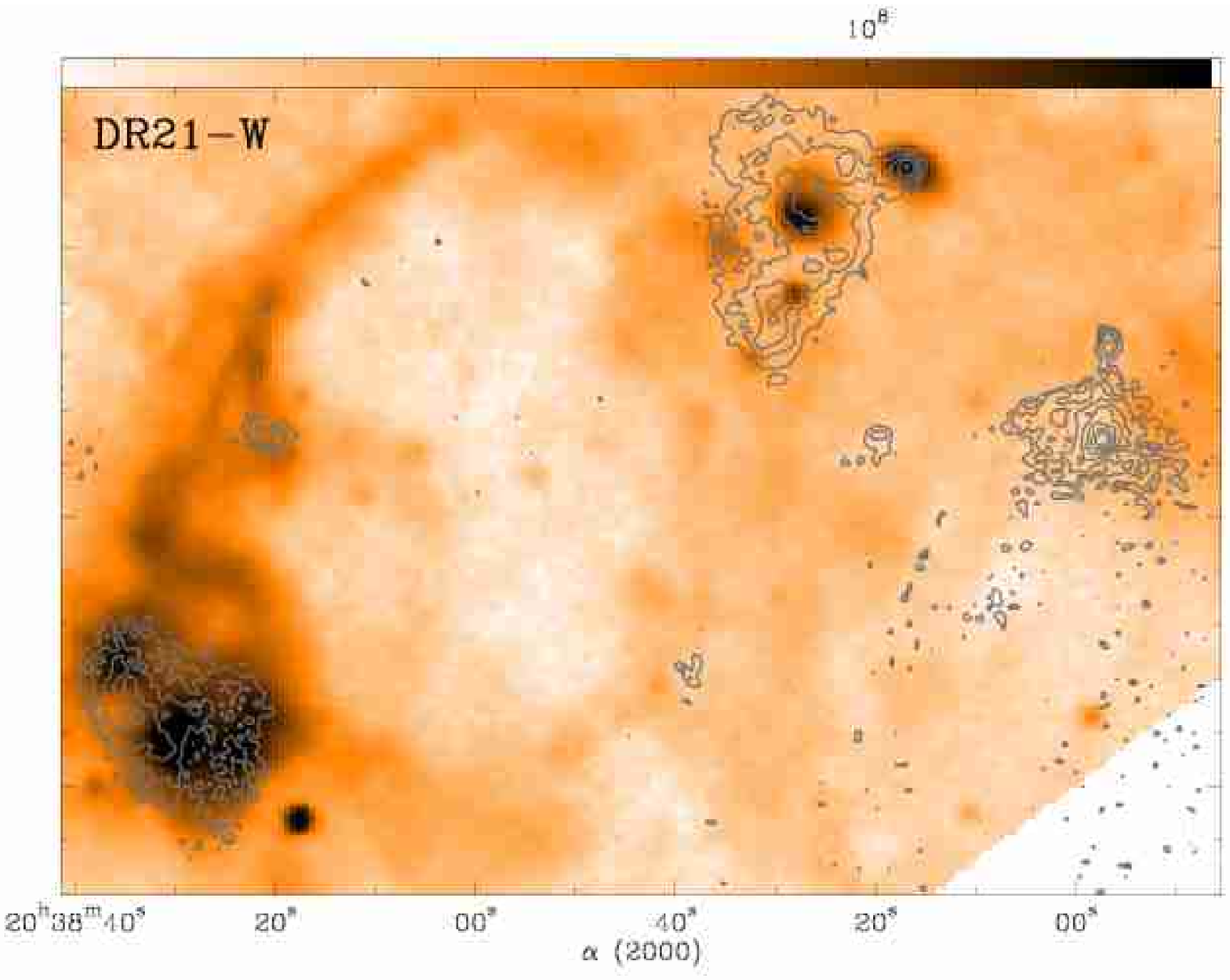}}
\vskip -0.7cm 
\centerline{\includegraphics[angle=0,width=6.89cm]{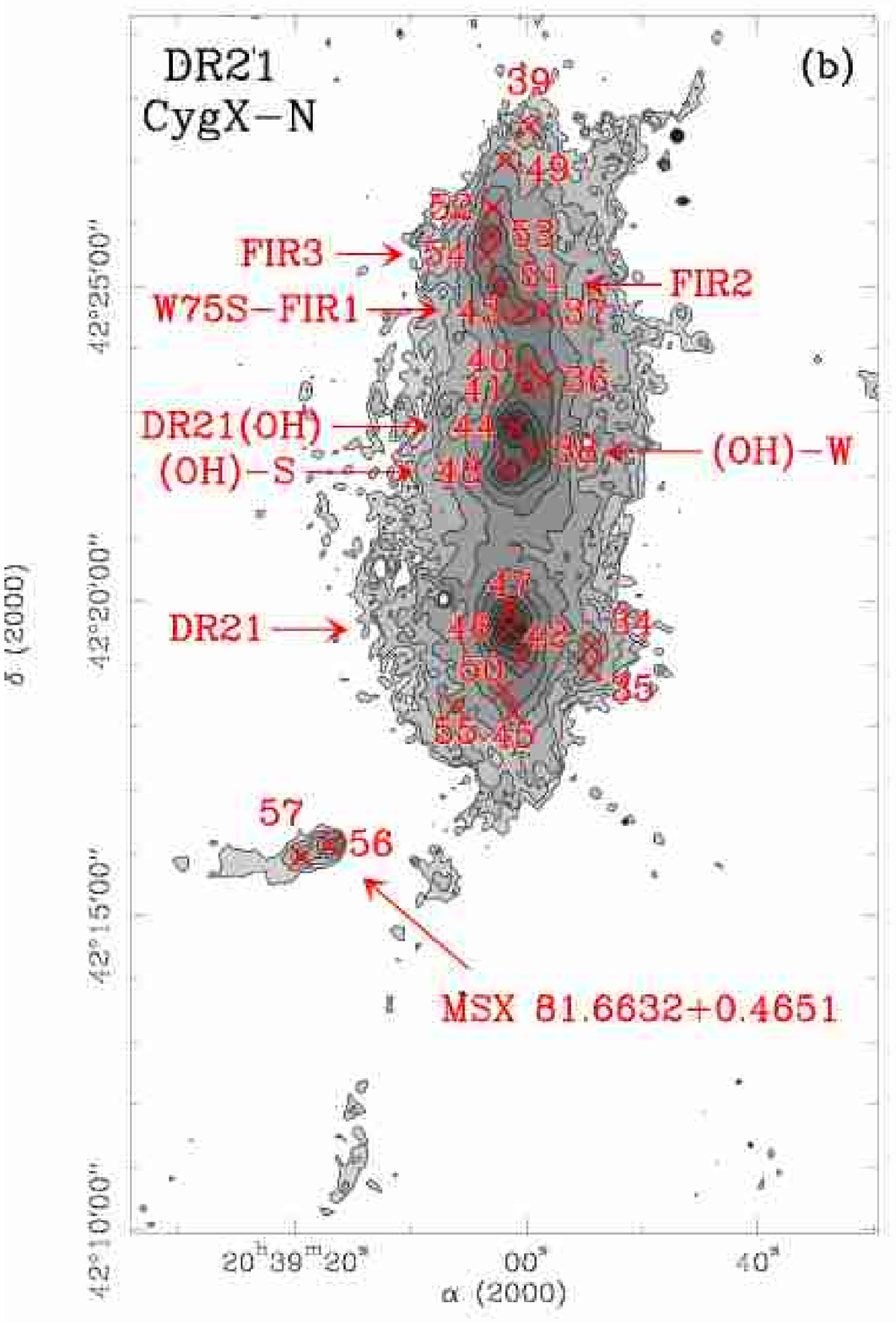}
\hskip -1.6cm\includegraphics[angle=0,width=6.89cm]{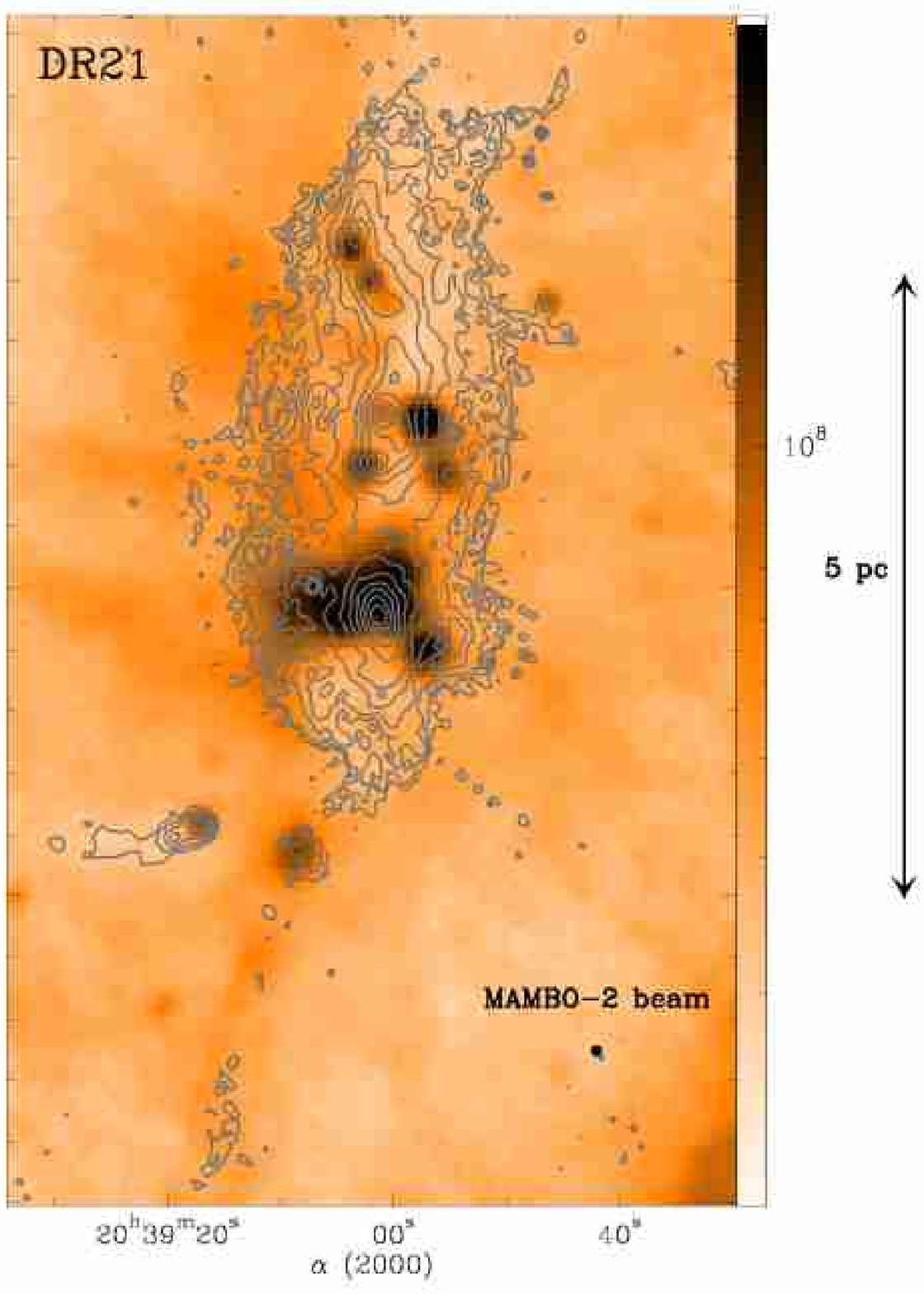}}
\vskip -2.2cm 
\centerline{\includegraphics[angle=0,width=6.4cm]{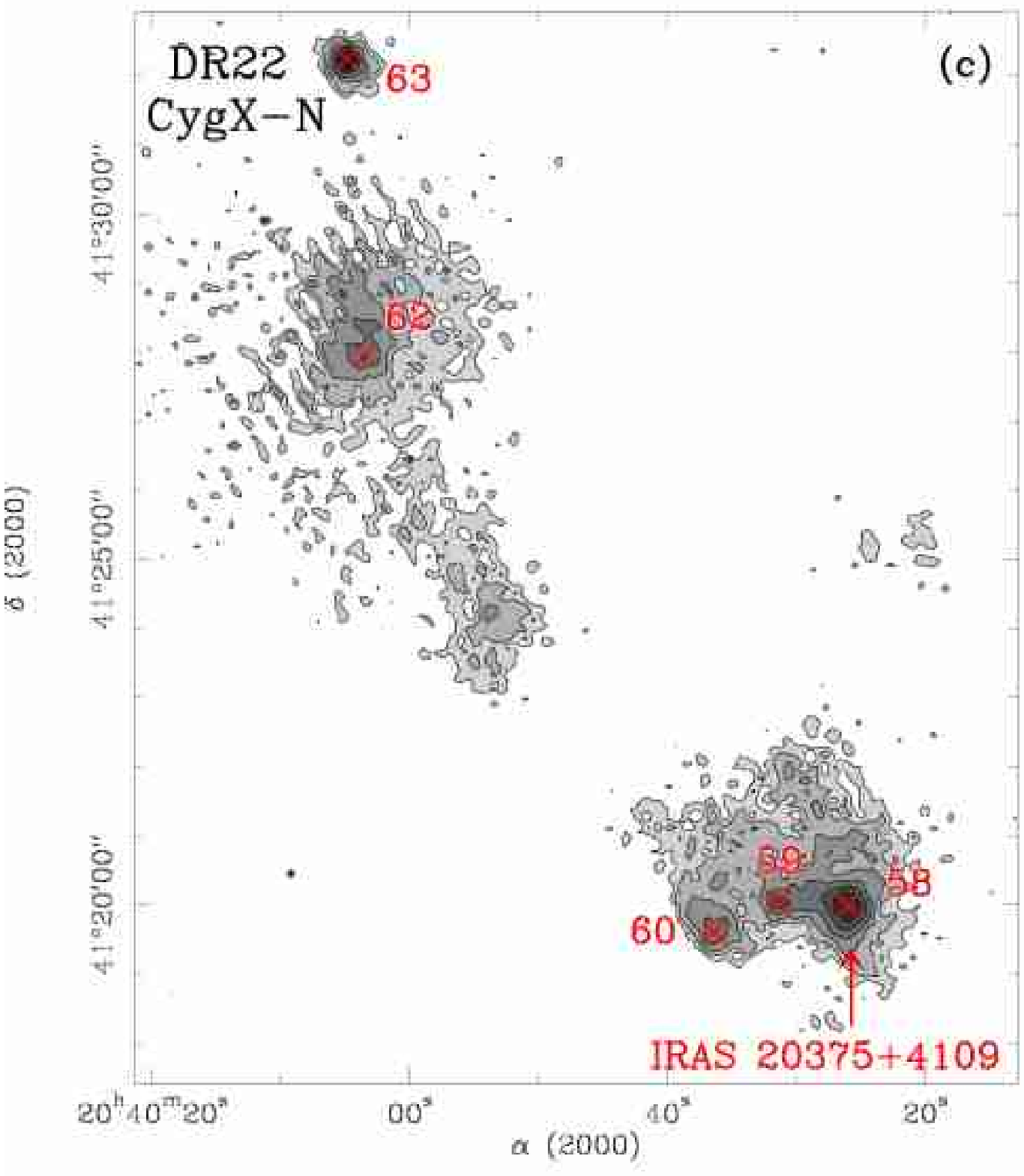}
\hskip -0.8cm \includegraphics[angle=0,width=6.4cm]{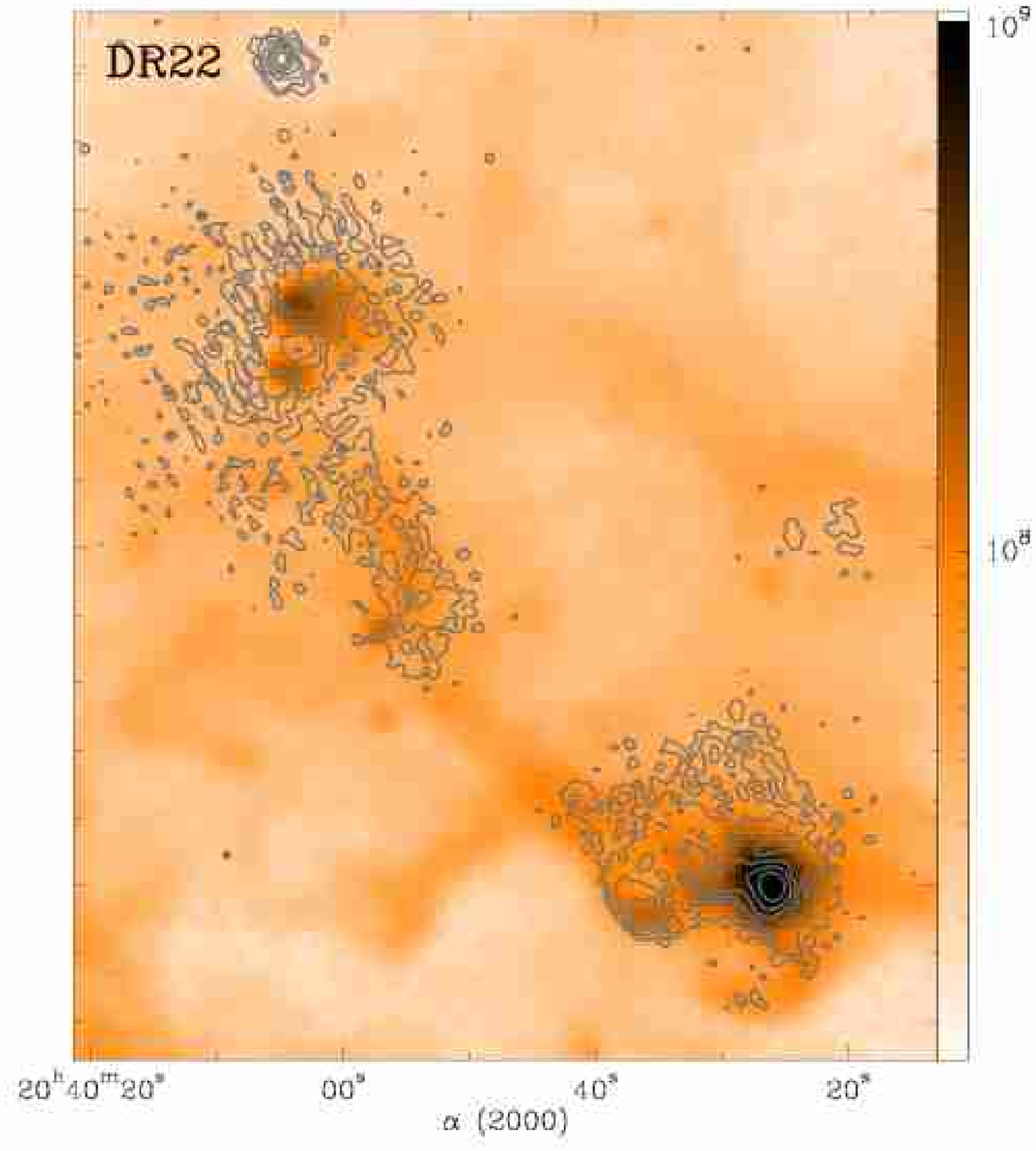}}
\vskip -0.8cm
\caption[]{MAMBO maps of CygX-North ({\bf left}: gray-scale and contours, {\bf right}: contours overlaid on 8~$\mu$m images obtained by \emph{MSX} and converted to Jy sr$^{-1}$) extracted from Fig.~\ref{f:mambo}a. Regions shown are west of DR21  ({\bf a}), around DR21 ({\bf b}) and DR22 ({\bf c}). The 1.2~mm and 8~$\mu$m images have $11\arcsec$ and $20\arcsec$ angular resolutions, respectively. The compact cloud fragments discovered in MAMBO images (see Table~\ref{t:densecores}) are labeled and marked by crosses in the gray-scale plot. The infrared sources which coincide with a MAMBO cloud fragment are also indicated. Contour levels are logarithmic and go from 40 to $800~\mjb$ in {\bf a} and {\bf c}, and from 40 to $4\,800~\mjb$ in {\bf b}.}
\label{f:msxnorth}
\end{figure*}

\begin{figure*}
\vskip -1.6cm
\centerline{\includegraphics[angle=0,width=5.46cm]{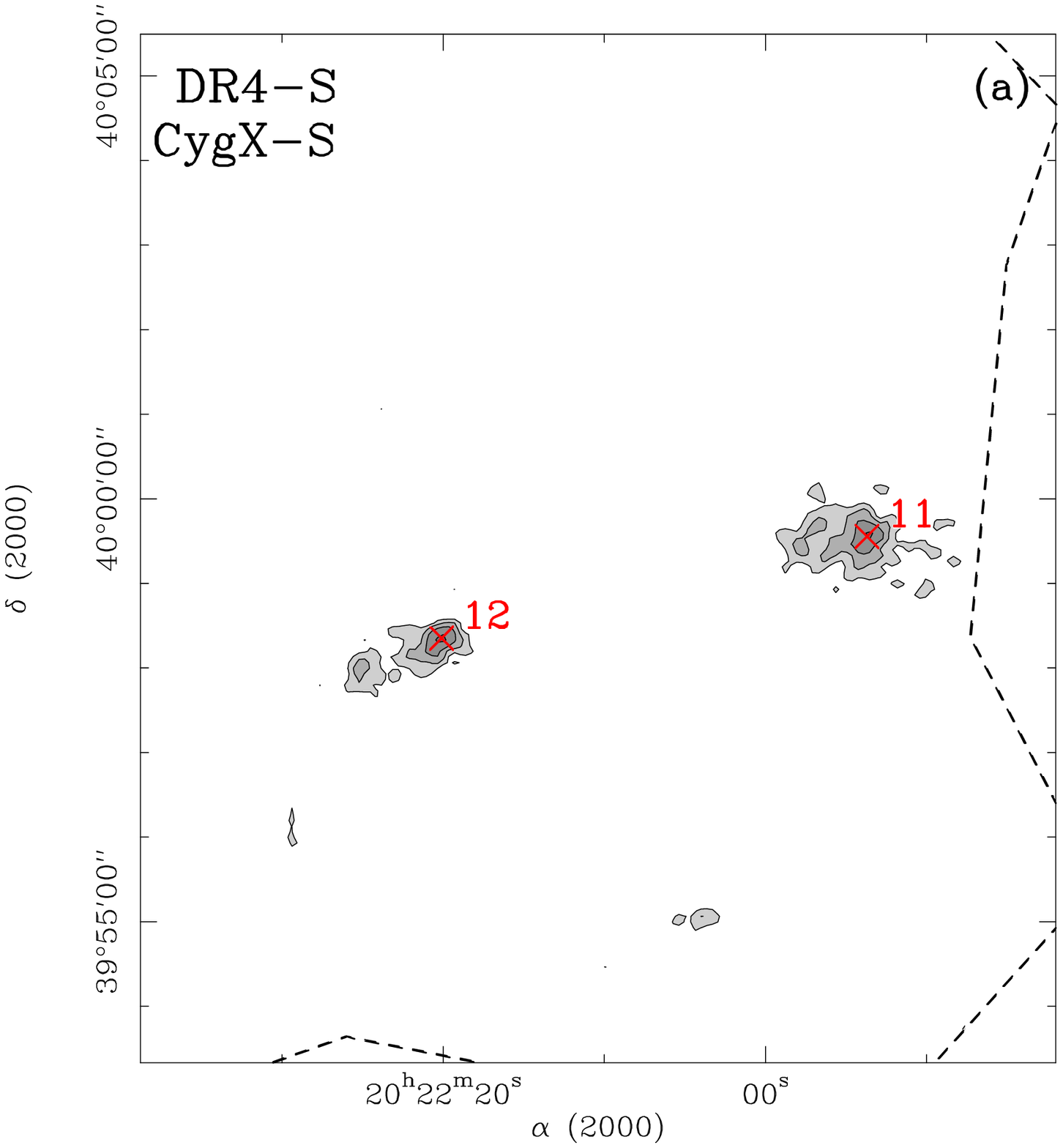}
\hskip -0.7cm \includegraphics[angle=0,width=5.46cm]{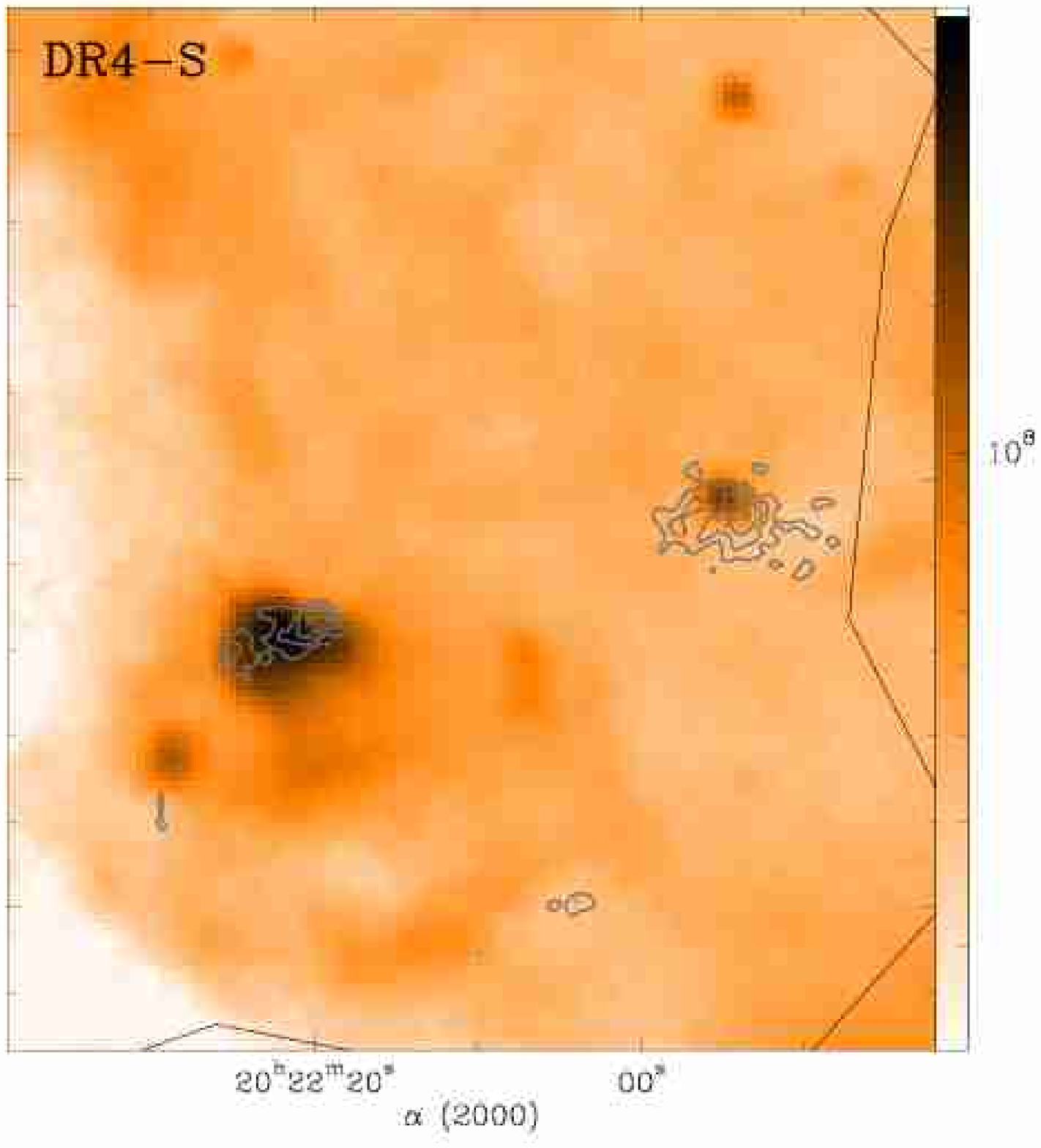}}
\vskip -1.6cm
\centerline{\includegraphics[angle=0,width=5.0cm]{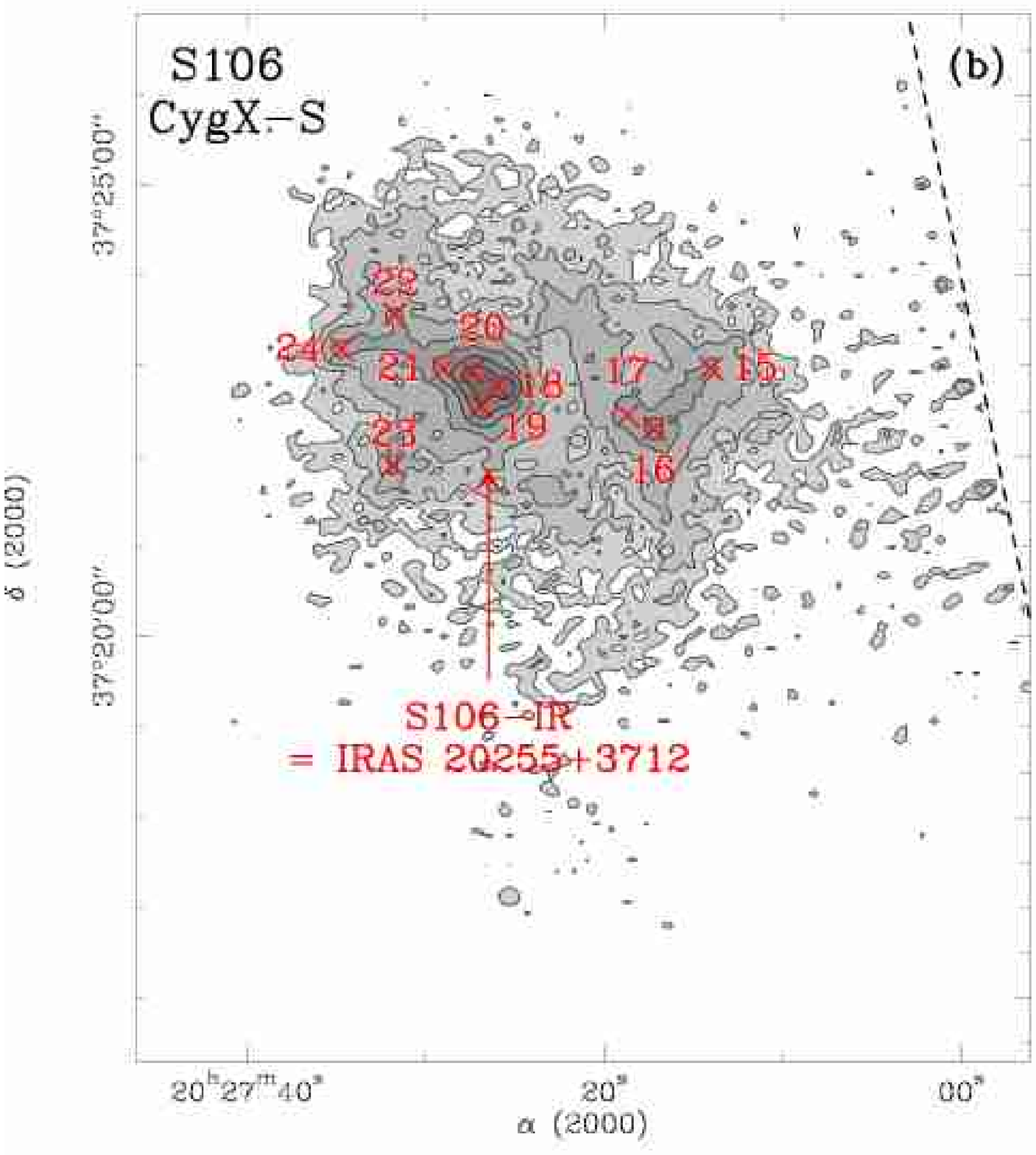}
\hskip -0.1cm \includegraphics[angle=0,width=5.0cm]{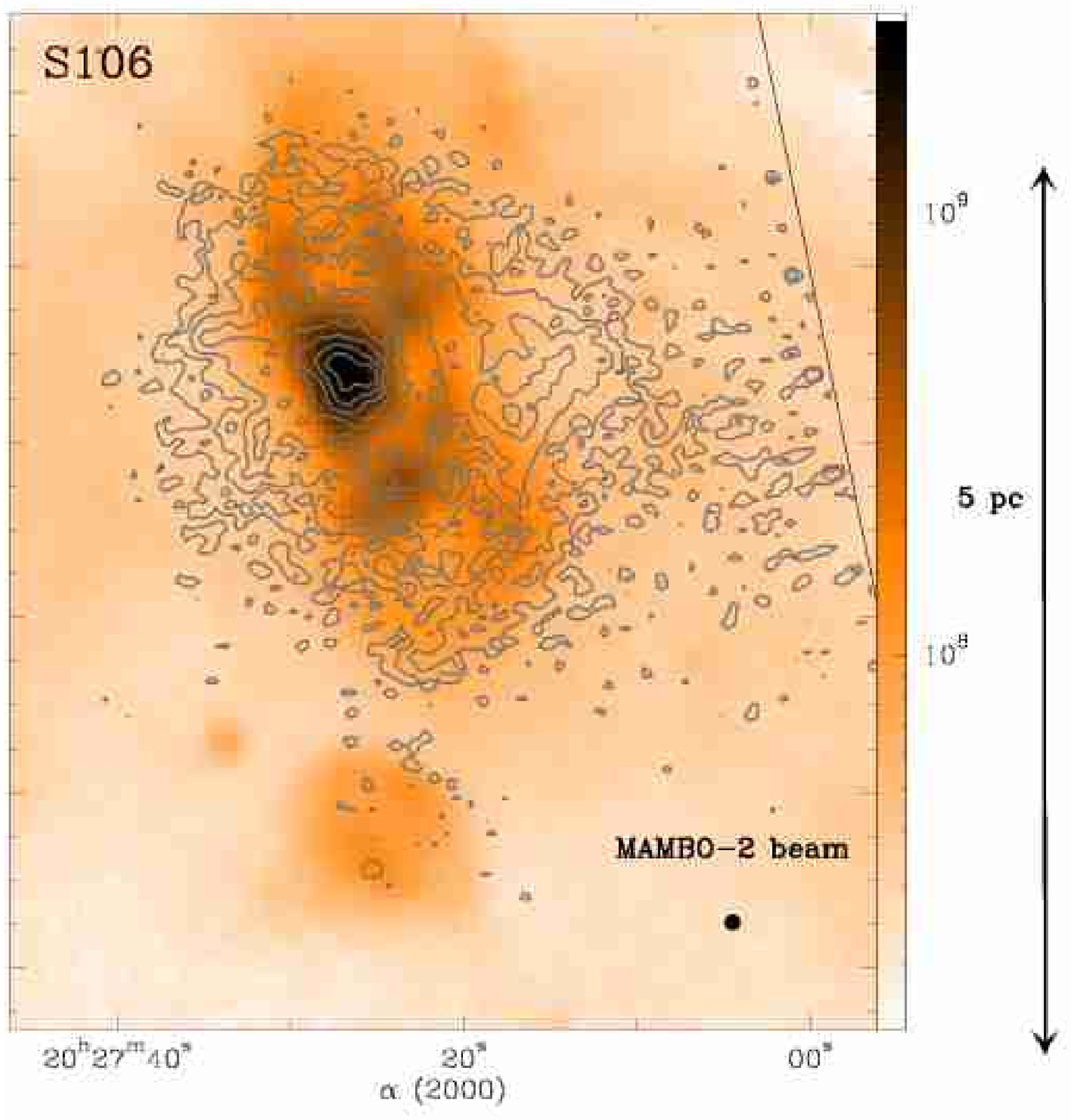}}
\vskip -0.7cm
\centerline{\includegraphics[angle=270,width=7.66cm]{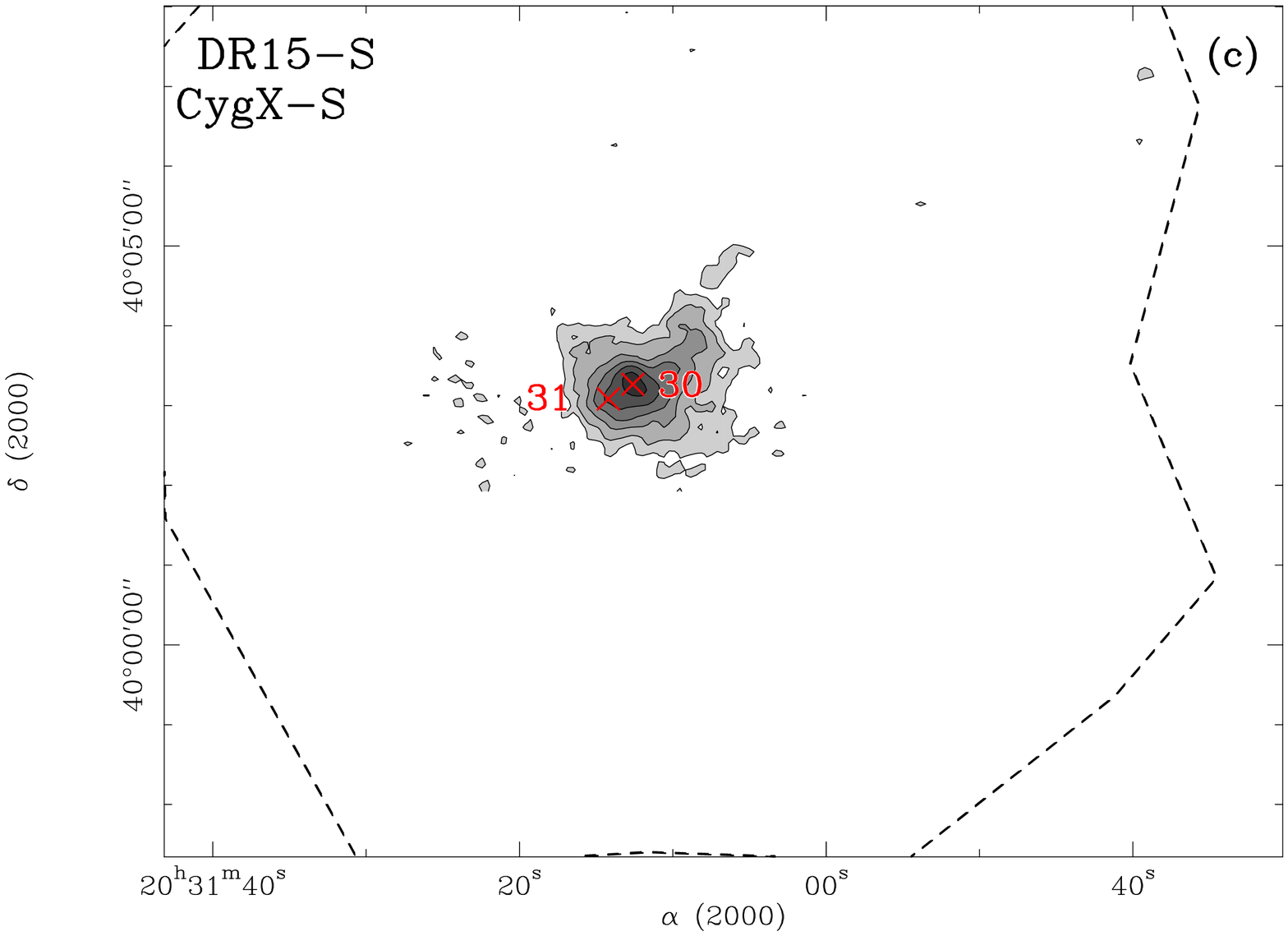}
\hskip -1.6cm \includegraphics[angle=270,width=7.66cm]{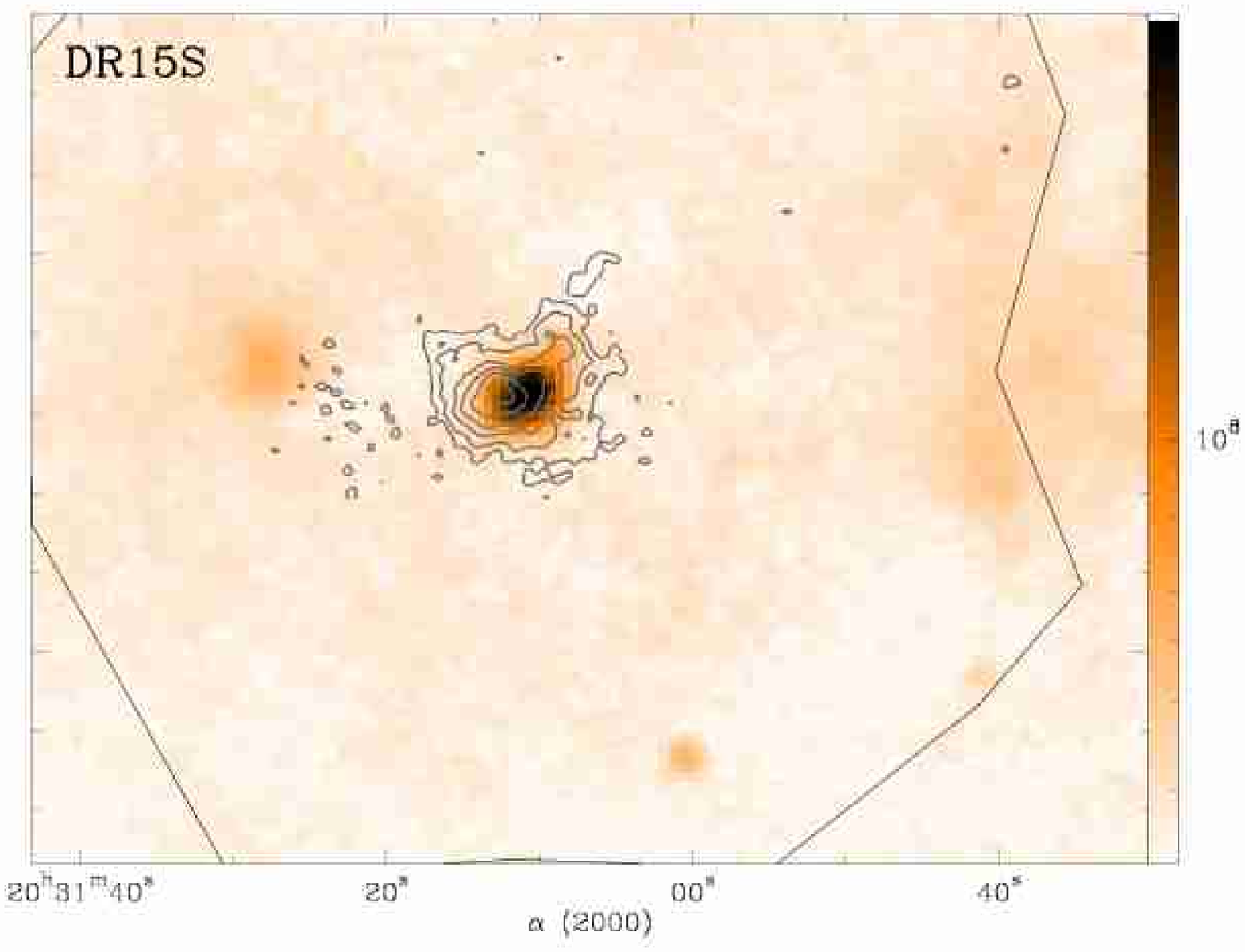}}
\vskip -0.8cm
\centerline{\includegraphics[angle=270,width=10.14cm]{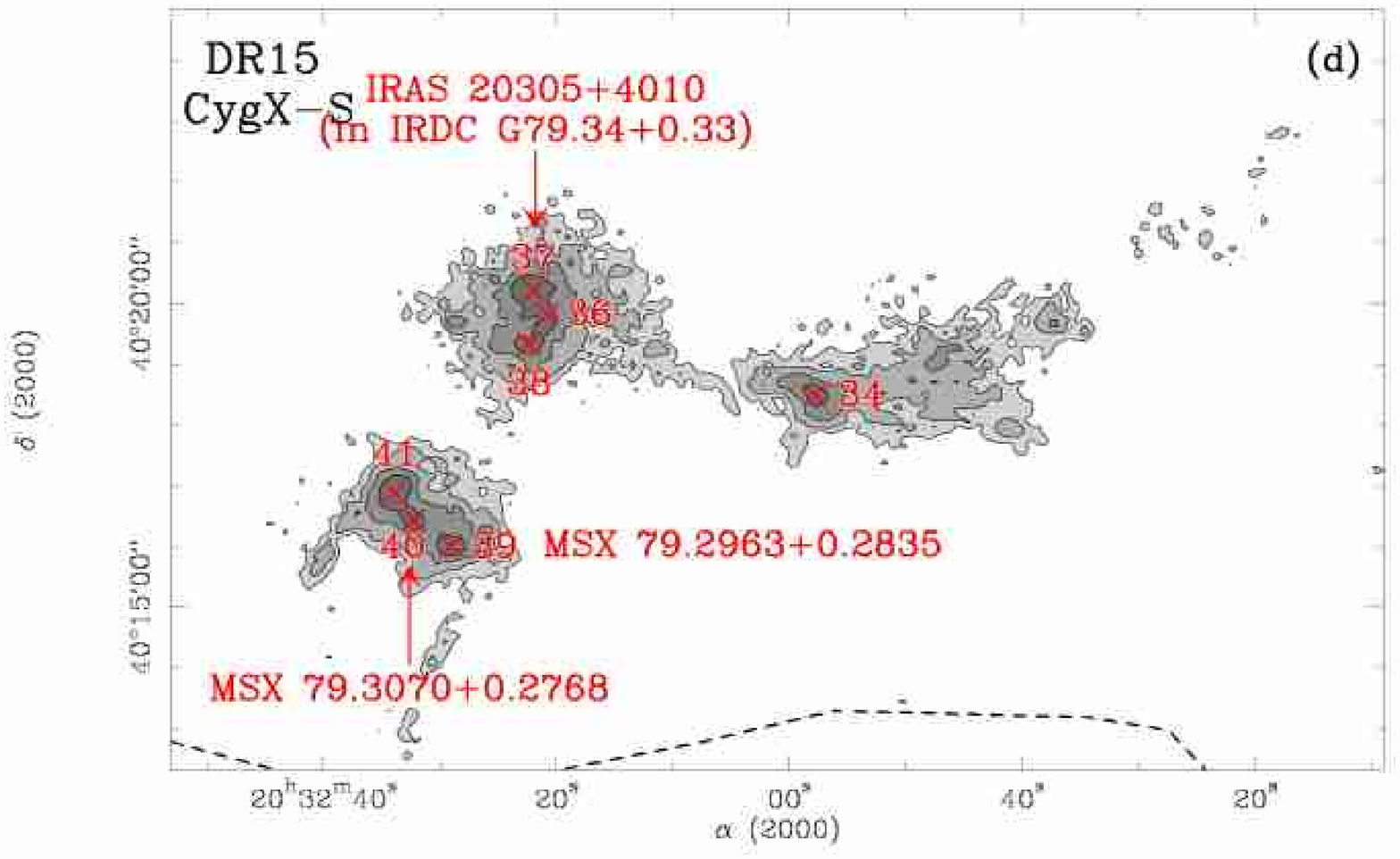}
\hskip -1.6cm  \includegraphics[angle=270,width=10.14cm]{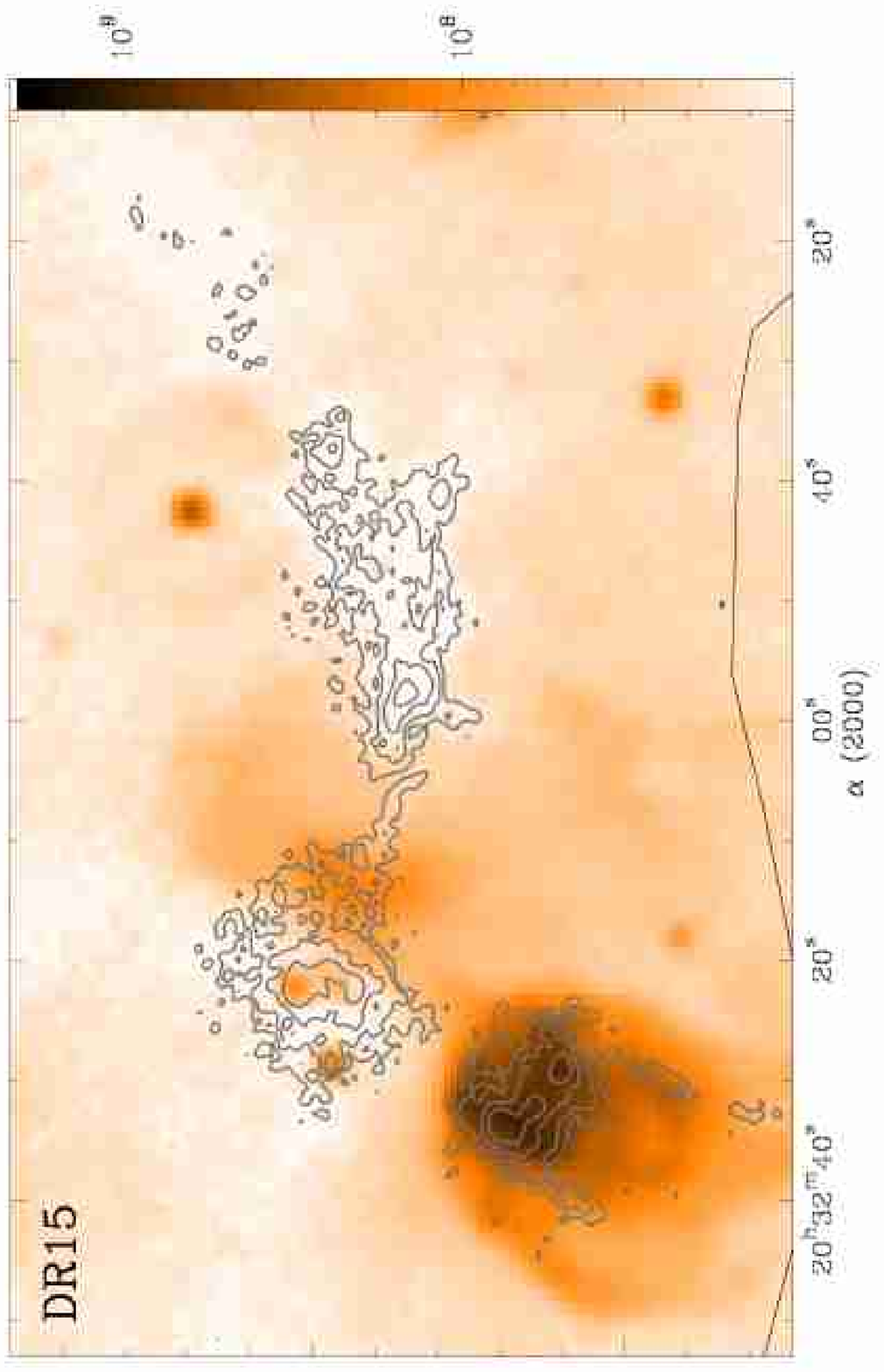}}
\caption[]{Same convention as Fig.~\ref{f:msxnorth} for MAMBO maps of CygX-South extracted from Fig.~\ref{f:mambo}b. Regions shown are south of DR4  ({\bf a}), around S106 ({\bf b}), south of DR15  ({\bf c}), and around DR15 ({\bf d}). Contour levels are logarithmic and go from 40 to $200~\mjb$ in {\bf a} and {\bf d}, from 60 to $1\,200~\mjb$ in {\bf b}, and from 40 to $400~\mjb$ in {\bf c}.}
\label{f:msxsouth}
\end{figure*}

\subsection{Coincidence with mid-infrared point sources}\label{s:infrared} 

When searching for the signature of embedded stellar embryos, the cross-correlation of compact millimeter sources with infrared catalogs is crucial. Among the near- to mid-infrared surveys that are currently available and complete for the Cygnus~X complex, we choose that performed by the \emph{MSX}\footnote{
The \emph{Midcourse Space Experiment} provides images observed at 8, 12, 15, and $21~\mu$m with a $20\arcsec$ resolution (\cite{egan01}).} 
satellite; its angular resolution ($20\arcsec$) is far better than that provided by \emph{IRAS} and its wavebands (from 8 to $21~\mu$m) are better suited than those of 2MASS. The \emph{Spitzer} satellite provides better angular resolution and sensitivity but the mapping of Cygnus~X with the \emph{Spitzer}/MIPS camera is far from being complete, and published results from the \emph{Spitzer}/IRAC camera generally do not cross-correlate well with (sub)millimeter sources (see \cite{davi07} and the counterexample ERO1/CygX-N57 of \cite{mars04}).

To search for coincidence with mid-infrared point sources, we carefully examine the structure of the \emph{MSX} emission at the exact location of each compact millimeter source (see right-hand side of Figs.~\ref{f:msxnorth}--\ref{f:msxsouth} and Figs.~\ref{f:msxnorth_app}--\ref{f:msxnw_app}). A millimeter source is defined to be detected by \emph{MSX} if it coincides with pointlike emission at both 8 and $21~\mu$m. Slightly more extended  \emph{MSX} emission is considered for three developed \hii regions (cf. Sect.~\ref{s:hii}): DR21 (associated with CygX-N46 and N47), S106-IR (associated with CygX-S18, S19 and S20) and IRAS~20306+4005 (associated with CygX-S39 and S40). Given the resolution of \emph{MSX} images, \emph{MSX} point sources have similar sizes to those of compact cloud fragments detected by MAMBO-2. The detection, at both 8 and $21~\mu$m, allows the rejection of peaks of infrared emission created by small grains at the interface of photo-dissociation regions and molecular clumps (e.g. \cite{abe02}). The latter are actually externally heated sources whose peaks shift with wavelength and sometimes disappear at $21~\mu$m. We add four more sources which are only detected at $21~\mu$m: CygX-N30 (W75N(B)), CygX-N46 (DR21) and N47 (DR21-D), whose $8~\mu$m emission is extinguished by their high-column density, and CygX-N65, which is very weak. The positional accuracy requested between the compact sources of MAMBO-2 and \emph{MSX} is  $10\arcsec$ (i.e. the sum of the maximum pointing errors of both images), which corresponds to 0.08~pc at 1.7~kpc. Such a narrow association is made possible by the small size ($\sim 0.1$~pc) of the compact cloud fragments extracted from the MAMBO-2 images. In fact if we consider looser associations, many  \emph{MSX} point sources lie on and possibly within clumps, at $\sim 0.2$~pc from the density peaks which we identified as compact fragments (e.g.  Figs.~\ref{f:msxnorth}a-b). We reject these loose associations, unlike previous studies that considered, for example, that CygX-S12 (respectively CygX-S30) and IRAS~20205+3948 (respectively IRAS~20293+3952) coincide (cf. Figs.~\ref{f:msxsouth}a and \ref{f:msxsouth}c, see also Beuther et al. 2002). In fact, these \emph{MSX} point sources lying within dust clumps sometimes coincide with extremely red objects detected by \emph{Spitzer}/IRAC (EROs, \cite{mars04}, \cite{davi07}), confirming that (sub)millimeter and infrared surveys generally identify different population of stars in terms of age and/or mass.

Our cross-correlations are illustrated in the right-hand parts of Figs.~\ref{f:msxnorth}--\ref{f:msxsouth} and 
Figs.~\ref{f:msxnorth_app}--\ref{f:msxnw_app}. Their results are given in Col.~8 of Table~\ref{t:densecores}. When there is  a coincidence with a \emph{MSX} point source, Col.~8 gives its $21~\mu$m flux as taken from the \emph{MSX} catalog (see Note c for exceptions) and Col.~9 gives the \emph{IRAS} or \emph{MSX} name. Otherwise, when the MAMBO-2 source is not detected or seen in absorption at $8~\mu$m, Col.~8 of Table~\ref{t:densecores} mentions ``--'' or ``Abs''.

With the above selection criteria, 36 compact cloud fragments of Cygnus~X (i.e. less than $1/3$ of the sample) are associated with mid-infrared emission detected by \emph{MSX}. The other 93 fragments are undetected down to $\sim 0.15$~Jy at 8~$\mu$m and $\sim 3$~Jy at 21~$\mu$m, which are the completeness levels of the \emph{MSX} catalog (MSX~C6) in Cygnus~X (see \cite{egan01}; \cite{Bontemps06}). Some of these sources are even seen  in absorption against the diffuse mid-infrared background. Interestingly, some of the Cygnus~X fragments which are undetected by \emph{MSX} harbor  a weak 24~$\mu$m source in \emph{Spitzer} images. To preserve the homogeneity of our study, we do not use this information here.  

\begin{figure*}[htbp]
\centering
\includegraphics[angle=0,width=16.8cm]{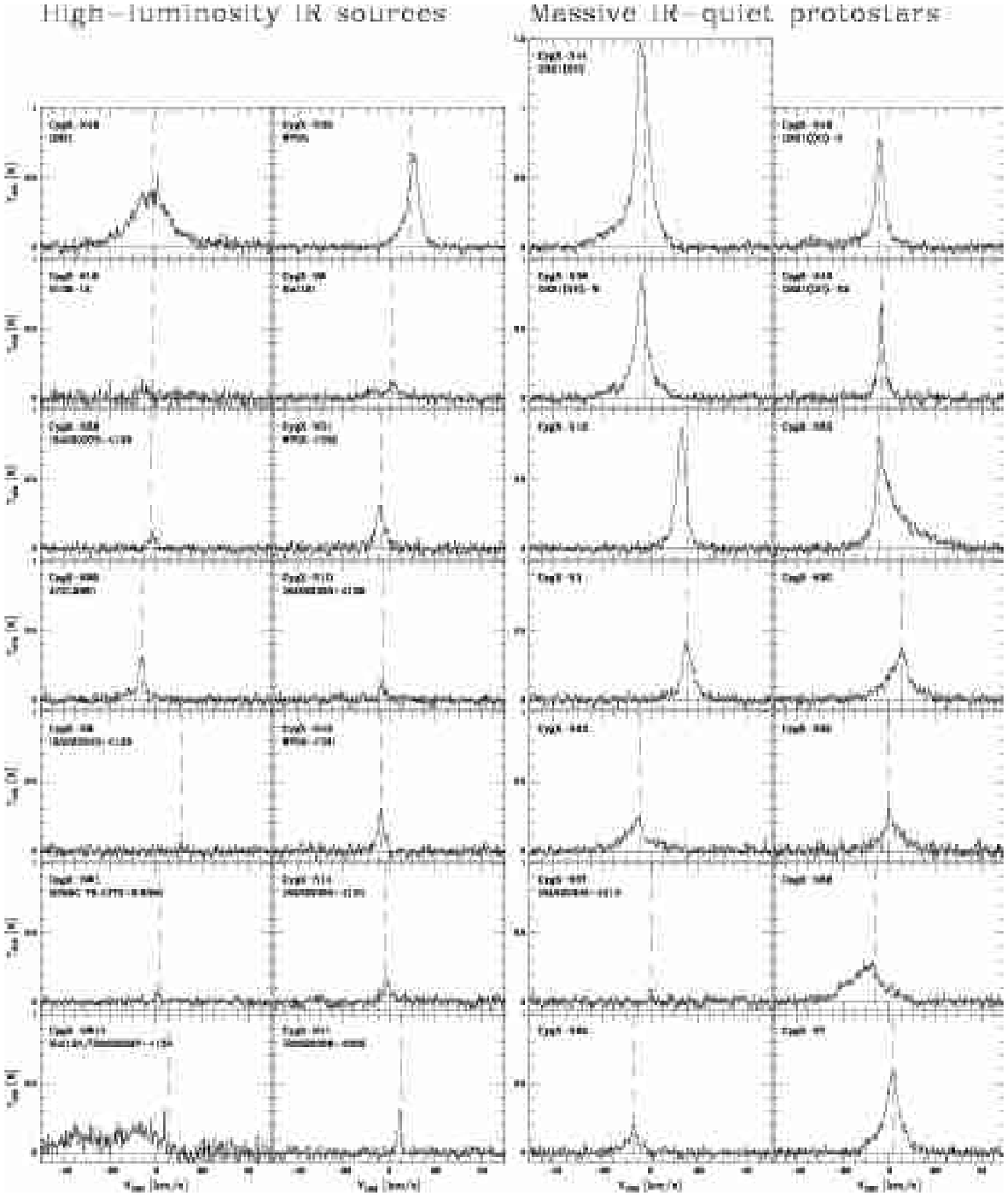}
\caption{SiO(2-1) lines observed toward the 28 most massive ($> 40~\msun$) cloud fragments of Cygnus~X: 14 high-luminosity infrared sources on the left and 14 infrared-quiet protostellar cores on the right, both ordered by decreasing $\msmm$ from top to bottom. The  local velocity at rest of each cloud fragment, as measured by  optically-thin tracers, is indicated by a dashed line.}
\label{f:sio}
\end{figure*}

\setcounter{table}{2}
\begin{table*}[htbp]
\caption[]{SiO(2-1) observations of the most massive dense cores of Cygnus~X compared with the brightest SiO protostellar sources at $< 500$~pc}
\label{t:sio}
\centering
\begin{tabular}{lcccccl}
\hline
Source	& \msmm		& $T_\mathrm{SiO}$	& $\Delta\,v_\mathrm{\it FWHM}$ 	& $\Delta\,v_\mathrm{outflow}$	& $\int T_\mathrm{SiO}~dv$ & Possible nature\\
name	& (\msun)		& (K) 			& (\kms) 			& (\kms)					& (K \kms) 			&\\
\hline
\hline
N46, DR21			& $< 609-949$ 	& 0.38 	& 16.9 	& 64.8 	& $8.6 \pm 0.42$ 	& UC\hii region\\
N30, W75N(B)			& $<361-563$	& 0.65 	& 5.4 	& 25.8 	& $4.7 \pm 0.16$ 	& hyper compact \hii region?\\
N44, DR21(OH)		& 446 		& 1.45 	& 6.0 	& 43.0 	& $13.4 \pm 0.25$ 	& {\bf massive IR-quiet} protostellar core\\
S18, S106-IR			& $< 145-226$ & $< 0.10$& - 		& - 		& $< 0.15$		& compact \hii region\\
N48, DR21(OH)-S		& 197		& 0.78 	& 4.0 	& 49.7 	& $5.4 \pm 0.26$ 	& {\bf massive IR-quiet} protostellar core\\
S8, Mol~121S			& $< 121-188$ & 0.10 	& 4.6		& 22.1 	& $0.96 \pm 0.18$ 	& UC\hii region\\
N58, IRAS~20375+4109 	& $< 92-144$	& 0.086 	& 3.8 	& 7.1 	& $0.34 \pm 0.10$ 	& compact \hii region\\
N38, DR21(OH)-W 		& 138 		& 0.86 	& 4.6 	& 37.5 	& $6.8 \pm 0.25$ 	& {\bf massive IR-quiet} protostellar core\\
N51, W75S-FIR2 		& 125 		& 0.29 	& 3.6 	& 12.7 	& $1.5 \pm 0.19$ 	& {\bf high-luminosity IR} protostellar core\\
S26, AFGL2591 		& $< 80-124$	& 0.30 	& 2.8 	& 18.7 	& $1.5 \pm 0.19$ 	& UC\hii region\\
N40, DR21(OH)-N2		& 106 		& 0.63 	& 2.6 	& 14.4 	& $2.3 \pm 0.18$ 	& {\bf massive IR-quiet} protostellar core\\
N12 					& 86 			& 0.87 	& 4.0 	& 26.5 	& $4.7 \pm 0.21$ 	& {\bf massive IR-quiet} protostellar core\\
N53 					& 85 			& 0.80 	& 6.4 	& 57.6 	& $8.9 \pm 0.31$ 	& {\bf massive IR-quiet} protostellar core\\
N3 					& 84 			& 0.40 	& 4.8	 	& 21.5 	& $2.6 \pm 0.21$ 	& {\bf massive IR-quiet} protostellar core\\
S30 					& 82 			& 0.32 	& 9.3 	& 35.2 	& $3.9 \pm 0.26$ 	& {\bf massive IR-quiet} protostellar core\\
N10, IRAS~20350+4126 	& $< 52-81$	& 0.12 	& 2.4 	& 4.7         & $0.35 \pm 0.10$     & UC\hii region?\\
N6, IRAS~20343+4129 	& 44-68 		& $< 0.07$& - 		& - 		& $< 0.10$ 		& {\bf high-luminosity IR} protostellar core\\
N43, W75S-FIR1 		& 66 			& 0.25 	& 3.4 	& 9.1 	& $0.94 \pm 0.15$ 	& {\bf high-luminosity IR} protostellar core\\
NW1, IRAS20178+4046 	& $<38-59$	& 0.05 	& 2.6 	& 3.4 	& $0.15 \pm 0.06$ 	& UC\hii region?\\
N63 					& 58 			& 0.23 	& 7.2 	& 36.5 	& $2.6 \pm 0.23$ 	& {\bf massive IR-quiet} protostellar core\\
S32 					& 54 			& 0.29 	& 6.2 	& 24.6 	& $2.4 \pm 0.28$ 	& {\bf massive IR-quiet} protostellar core\\
N14, IRAS~20352+4124 	& 50 			& 0.16 	& 2.4 	& 7.1 	& $0.61 \pm 0.13$ 	& {\bf high-luminosity IR} protostellar core\\
NW14, Mol124 			& 32-50		& 0.16 	& 41.6 	& 64.4       & $5.8 \pm 0.76$       & {\bf high-luminosity IR} protostellar core\\
S37, IRAS~20305+4010 	& 45 			& 0.05 	& - 		& 3.2 	& $0.09 \pm 0.09$ 	& {\bf massive IR-quiet} protostellar core\\
N68 					& 44 			& 0.27 	& 13.7 	& 43.8 	& $4.6 \pm 0.25$ 	& {\bf massive IR-quiet} protostellar core\\
S41, IRAS~20306+4005 	& 27-42 		& 0.30 	& 2.0 	& 8.2         & $0.61 \pm 0.09$     &  {\bf high-luminosity IR} protostellar core\\
N65 					& 41 			& 0.19 	& 3.8 	& 16.9 	& $1.06 \pm 0.16$ 	& {\bf massive IR-quiet} protostellar core\\
S7 					& 40 			& 0.58 	& 6.0 	& 28.4 	& $5.1 \pm 0.23$ 	& {\bf massive IR-quiet} protostellar core\\
\hline
NW2 				& 36 			& 0.10 	& - 		& 2.0 	& $0.18 \pm 0.09$ & IR-quiet protostellar core\\
S43 					& 35 			& 0.16 	& 2.2 	& 6.1 	& $0.59 \pm 0.13$ & IR-quiet protostellar core\\
N24 					& 34 			& 0.09 	& 4.7 	& 5.1 	& $0.39 \pm 0.10$ & IR-quiet protostellar core\\
NW5, LkHA 225S 		& 22-34 		& 0.045 	& - 		& 14.9 	& $0.69 \pm 0.26$ & infrared protostellar core\\
N60 					& 30 			& 0.08 	& 2.0 	& 4.5 	& $0.20 \pm 0.10$ & IR-quiet protostellar core\\
N16, MSX 81.4452+0.7635 & 29 		& $< 0.09$& - 		& - 		& $< 0.12$ 	      & infrared protostellar core\\
N28 					& 27 			& 0.09 	& 1.8 	& 3.5 	& $0.25 \pm 0.09$ & IR-quiet protostellar core\\
N62 					& 27 			& 0.10 	& 2.6 	& 8.5 	& $0.42 \pm 0.11$ & IR-quiet protostellar core\\
S16 					& 24			& 0.10 	& 2.6 	& 14.5 	& $0.51 \pm 0.13$ & IR-quiet protostellar core\\
N18 					& 23 			& $< 0.08$& - 		& - 		& $< 0.11$ & ?\\
N59 					& 20 			& $< 0.08$& - 		& - 		& $< 0.11$ & ?\\
NW9, IRAS~20216+4107 & 19 		& $< 0.10$& - 		& - 		& $< 0.15$ & infrared protostellar core\\
\hline
\hline
L1448-MM @ 1.7 kpc 	& - 		& - 	& - 		& $\sim140$ & $\sim 1.4^{\rm (a)}$ & low-mass class~0 protostar\\
L1157-MM @ 1.7 kpc 	& - 		& - 	& - 		& $\sim25$ & $\sim 1.8^{\rm (b)}$ & low-mass class~0 protostar\\
Orion-IRc2 @ 1.7 kpc 	& - 		& - 	& - 		& $\sim74$ & $\sim 9.2^{\rm (c)}$ & high-mass protostar\\
\hline
\end{tabular}
\begin{list}{}{}
\item[ $^{a}$ ]{Intensity which has been distance-corrected (extrapolation from 300~pc to 1.7~kpc) using values from Table~2 of \cite{nisi07} and excluding R4 which would fall outside the beam.}
\item[ $^{b}$ ]{Intensity which has been distance-corrected (extrapolation from 440~pc to 1.7~kpc) using values from Table~3 of \cite{nisi07} and summing up B0 and B1 components.}
\item[ $^{c}$ ]{Intensity which has been distance-corrected (extrapolation from 450~pc to 1.7~kpc) using the spectrum of Fig.~1 in Ziurys \& Friberg (1987).}
\end{list}
\end{table*}

\subsection{Survey for SiO emission}\label{s:sio}

Silicon monoxide (SiO) emission is an excellent tracer of shocked gas, usually associated with molecular outflows (e.g. \cite{ZF87}; \cite{MPBF92}; \cite{schi97}). SiO may also have been detected in shocks toward hot cores (e.g. \cite{HFM01} and references therein). We have surveyed for SiO($2-1$) emission by making single pointings toward almost\footnote{
Three sources (CygX-N32, N47, and S20) are included in beams targeting other fragments (CygX-N30, N46, and S18 respectively). Two sources (CygX-N69 and CygX-S10) were not observed because they are among the lowest-density fragments.}
all the most massive ($\msmm > 40~\msun$) fragments of Table~\ref{t:densecores} and a  few less massive fragments. Figure~\ref{f:sio} and Fig.~\ref{f:sio_app} (only provided electronically) display the SiO($2-1$) lines for all the observed fragments, ordered by decreasing $\msmm$. In Table~\ref{t:sio} and for each 1.2~mm compact fragment (Col.~1), we give the peak line temperature (Col.~3), the line width at half maximum (Col.~4), the full line width at the base (Col.~5), and the integrated intensity (Col.~6) of the SiO(2-1) emission. For direct comparison, we list the SiO line parameters scaled to the Cygnus X distance of three nearby protostars which are among the brightest known SiO outflow sources. 

Among the 28 most massive fragments whose SiO emission is displayed in Fig.~\ref{f:sio}, only two (CygX-S18, and CygX-N6) are not detected. This implies a very high detection rate of 93~\% ($>79~\%$ when considering the five high-mass fragments which have not been observed), with a typical detection level of $\sim 0.15~$K$\,\kms$ (see Table~\ref{t:sio}). This detection rate stays high even when considering the observation of the $20-40~\msun$ dense cores listed in Table~\ref{t:sio}. Down to $\sim 25~\msun$, the compact fragments of Table~\ref{t:sio} that do not have any SiO emission are all associated with infrared point sources: CygX-S18/S106-IR, CygX-N6/IRAS~20343+4129, and CygX-N16/MSX~81.4452+0.7635. This suggests that all massive fragments down to at least $40~\msun$, and maybe $25~\msun$, drive SiO outflows and/or contain hot cores. 

In Fig.~\ref{f:sio}, we show the emission spectra separately for the high-luminosity infrared sources ($\ge 10^3~\lsun$) and the massive infrared-quiet cores (cf. their definitions given in Sects.~\ref{s:HL-IR} and \ref{s:definition}). The two groups  obviously have different behavior. All infrared-quiet cores are detected in SiO(2-1) and most of them are particulary bright; the averaged peak line temperature and integrated intensity are 0.55~K and 4.6~K$\,\kms$, respectively. In contrast, only a small fraction of the high-luminosity infrared sources show a strong SiO line; the averaged peak line temperature and integrated intensity are 0.22~K and 1.9~K$\,\kms$, respectively. Since only two high-luminosity infrared sources (CygX-N48/DR21 and CygX-N30/W75N) are strong SiO emitters, the contrast is even larger when comparing the median values: 0.7~K$\,\kms$ for the high-luminosity infrared sources, and 3.1~K$\,\kms$ (i.e. 4.3 times more) for the infrared-quiet cores. The SiO lines of infrared-quiet cores are also 3 to 4 times brighter than the brightest SiO outflows of the nearby low-mass protostars (see Table~\ref{t:sio}).

Almost all the SiO lines shown in Figs.~\ref{f:sio} and \ref{f:sio_app} display outflow wings with line width at the base of emission that can reach up to $\sim 60~\kms$. Among the ten brightest SiO sources, the average of the line width at the base is as high as $41~\kms$, thus clearly suggesting outflow shocks. A few SiO lines might also indicate the presence of at least two emission components: a wide base (due to outflow shocks) and a narrower component which may indicate a contribution of shocks inside a hot core and/or lower velocity outflows. The best examples of two-component lines are those detected toward CygX-S26/AFGL~259, CygX-N44/DR21(OH), and CygX-N48/DR21(OH)-S, while the best example of single-component lines suggestive of a unique powerful outflow are CygX-N46/DR21, CygX-N53, CygX-S30, and CygX-N68.

\subsection{Coincidence with radio free-free and maser emission}\label{s:radio} 

We have used the SIMBAD\footnote{SIMBAD is the reference database for identification and bibliography of astronomical objects, developed and maintained by CDS, Strasbourg.} 
database to search for complementary signposts of stellar activity: centimeter free-free emission and OH, H$_2$O, and CH$_3$OH masers. SIMBAD gives inhomogeneous information on our sample of 129 compact cloud fragments, since it compiles results of Cygnus~X surveys performed  with limited angular resolution and high-angular resolution studies made for well-known sources. 

Within Table~\ref{t:densecores}, fifteen dense cores coincide with a small-diameter ($\le 10\arcsec$) source of radio emission detected at 2, 3.6, 6, 11, or 21~cm (e.g. \cite{KCW94}; \cite{zoot90}). We cannot check the nature of the detected centimeter emissions with the current database, but it most likely corresponds to free-free emissions emitted by \hii regions. Furthermore, several 1.2~mm continuum sources coincide with OH, H$_2$O, and/or CH$_3$OH maser sources (e.g. \cite{BE83}; \cite{vald01}; \cite{PMB05}). Most of these dense cores are associated with centimeter continuum and/or mid-infrared emission, but one (CygX-N53) is neither a centimeter nor a \emph{MSX} source. In Col.~9 of Table~\ref{t:densecores}, we indicate when some centimeter free-free and/or maser emission is detected toward a Cygnus~X dense core.

\begin{table*}[htbp]
\caption[]{Mean properties of the Cygnus~X clumps and cores compared with millimeter cloud structures of a few reference studies}
\label{t:scale}
\centering
\begin{tabular}{|l|ccc|ccc|c|}
\hline
& HMPOs	& IRDCs	& CygX clumps & Cygnus~X	& $> 40~\msun$ CygX &Nearby, low-mass	& $\rho$~Oph\\
& clumps	& clumps	& (all / starless)	& dense cores	& dense cores	&dense cores
	& condensations\\
\hline
\hline
{\it FWHM}~$^{\rm a}$ size (pc)
	& 0.5		& 0.5		& 0.68 / 0.78	& 0.10	& 0.13	& 0.08 	& 0.007\\
Mass$^b$ \msmm (\msun)	
	& 290	& 150	& 1000 / 780 	& 24 		& 91 		& 4.7 	& 0.15 \\
$<n\htwo>$$^c$ (\cmc)
	& $8.5\times 10^3$ 	& $5.9\times 10^3$ 	& $1.4\times 10^4$ / $6.7\times 10^3$ 	& $1.1\times 10^5$ 	& $1.9\times 10^5$ & $3.5\times 10^4$	& $1.9 \times 10^6$ \\
Number of sources
	& 69$^e$ 	& 190	& 40 / 10		& 128	& 33		& 22          & 60 \\
References$^d$  & (1) & (2) & (3) & (3) & (3) & (4), (5) & (5)\\
Distance (kpc) & 0.3-14  & 1.8-7.1 & 1.7 & 1.7 & 1.7 & 0.14-0.44 & 0.14\\
\hline
\end{tabular}
\begin{list}{}{}
\item[ (a) ] {Deconvolved {\it FWHM} sizes derived from a 2D-Gaussian fit, except for HMPOs and IRDCs for which sizes are taken from Table~1 of  \cite{beut02} (mean estimates are taken when the distance is unknown) and Table~3 of Rathborne et al. (2006)}.
\item[ (b) ]  {Mass consistently estimated from the 1.2~mm integrated flux using Eq.~(1). The mass values of \cite{beut02} have been recalculated using $\kmm=0.01~\cmg$ but those of nearby, low-mass dense cores and $\rho$~Oph condensations are kept while they assume $\kmm=0.005~\cmg$.}
\item[ (c) ]  {Volume-averaged density homogeneously recalculated using Eq.~(\ref{eq:density}) for every list of sources.}
\item[ (d) ]  {References: (1) \cite{beut02}; (2) Rathborne et al. (2006); (3) this paper; (4)  \cite{WMA99};
(5) \cite{MAN98}.}
\item[ (e) ]  {We limit our comparison to the main HMPOs components, which are the most likely progenitors of high-mass stars.}
\end{list}
\end{table*}

\begin{figure*}[htbp]
\includegraphics[angle=270,width=9.cm]{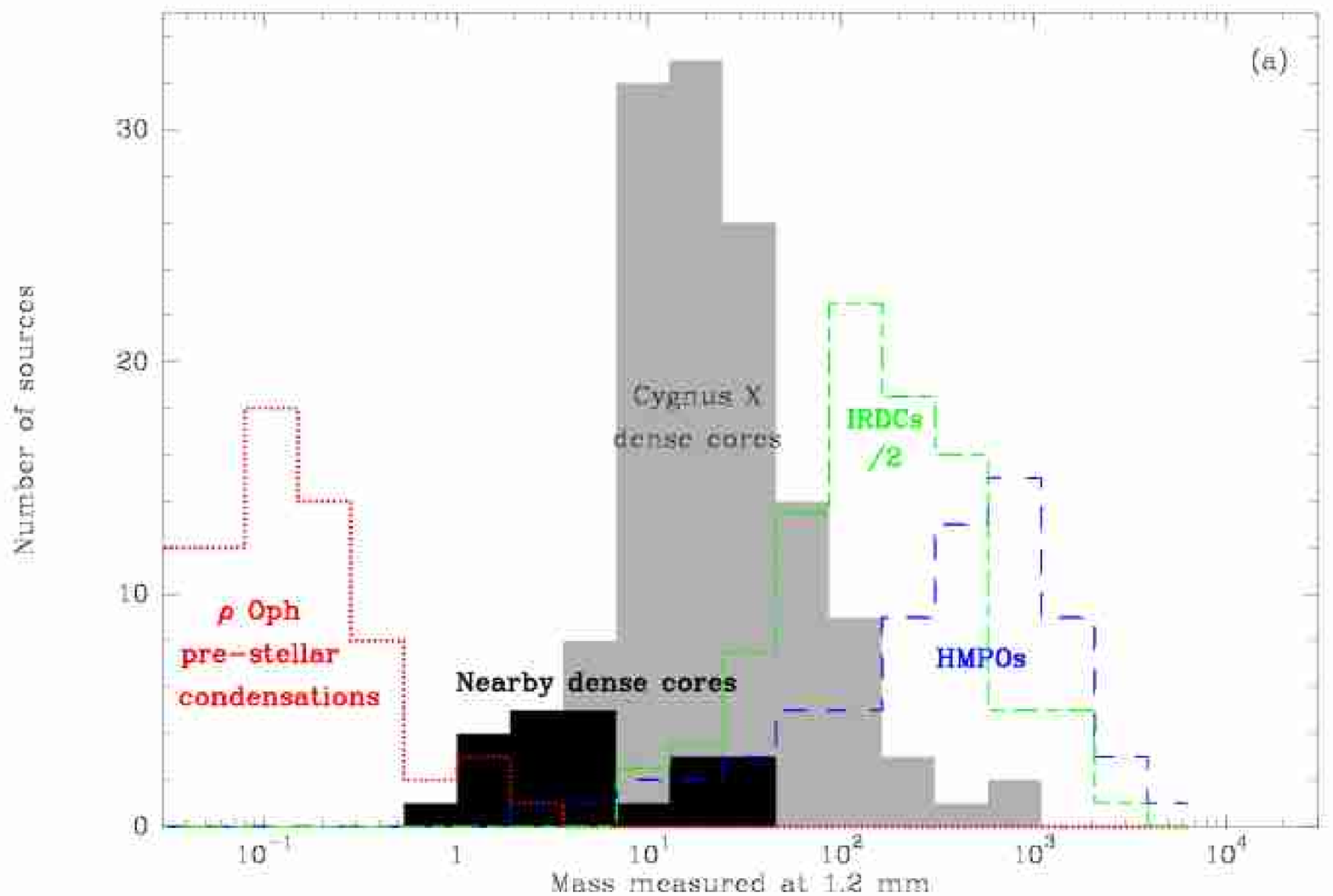} 
\includegraphics[angle=270,width=9.cm]{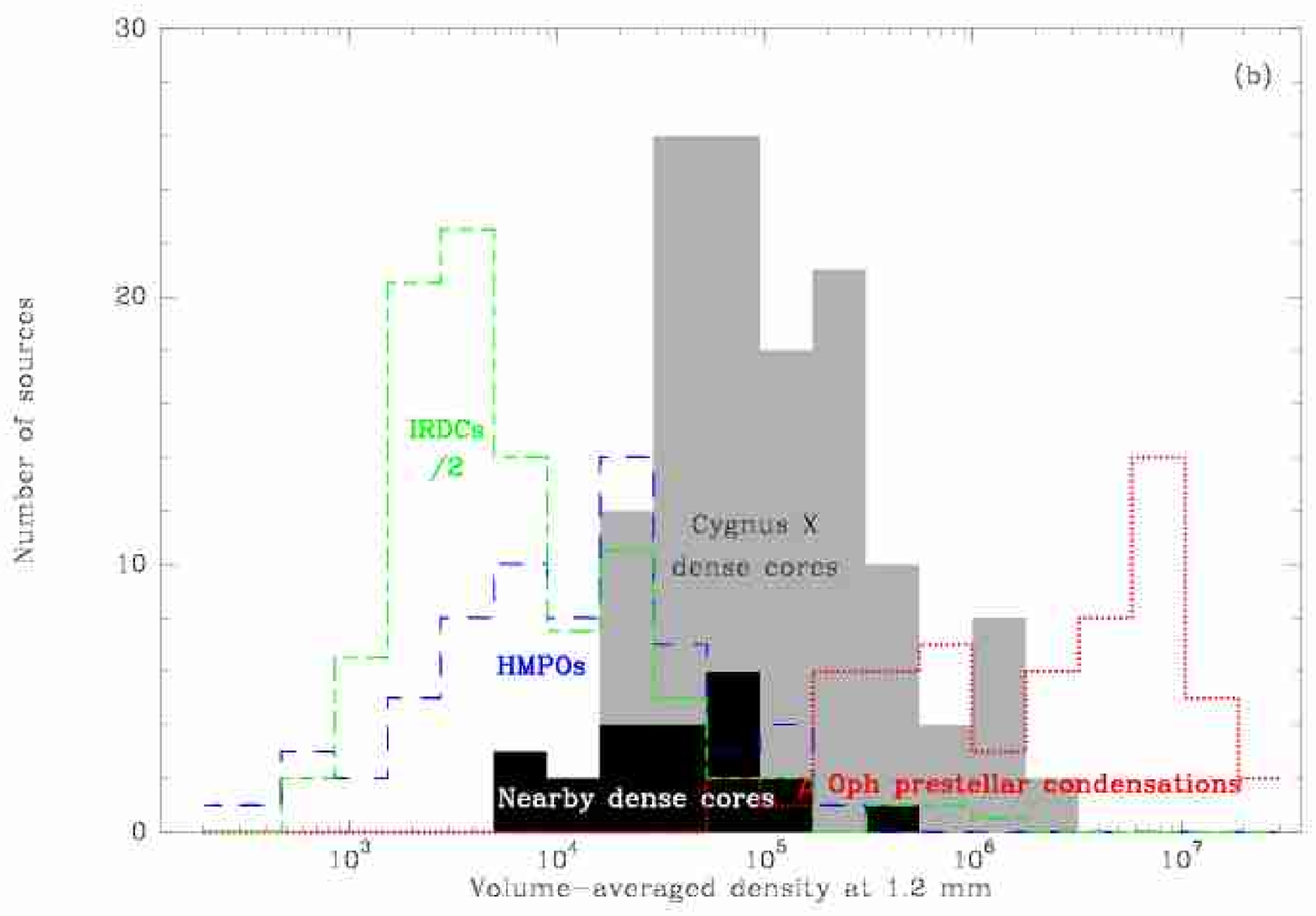}
\caption[]{Distribution of mass ({\bf a}) and volume-averaged density ({\bf b}) of the Cygnus~X dense cores compared to those of clumps hosting high-luminosity protostars (main component of HMPOs, Beuther et al. 2002) and clumps referred to as IRDCs (\cite{rath06}). The Cygnus~X histograms are also compared to density spectra of nearby dense cores (Motte et al. 1998; Ward-Thompson et al. 1999) and $\rho$~Oph condensations (Motte et al. 1998).}
\label{f:histo}
\end{figure*}

\section{Discussion}\label{s:discussion}

\subsection{A sample of high-density cores}\label{s:highdens}

Our 1.2~mm continuum imaging of the Cygnus~X molecular cloud complex gives a complete view of the structure of its highest-column density portions, with spatial scales ranging from 0.03~pc (deconvolved from the beam) to 3~pc. According to the terminology used for nearby molecular clouds (e.g. \cite{WBMK00}), it covers scales typical for dense cores ($\sim 0.1$~pc) and clumps ($\sim 1$~pc). The 129 compact cloud fragments identified in Cygnus~X have sizes ranging from 0.03 to 0.3~pc with a mean size of $\sim 0.1$~pc (see Sect.~\ref{s:census1} and Table~\ref{t:densecores}). They should therefore be called ``dense cores''. The 40 larger-scale structures of Table~\ref{t:clumps} have typical sizes of $\sim 0.7$~pc; we will call them ``clumps''. In Table~\ref{t:scale}, we give the mean properties of the Cygnus~X dense cores and clumps and compare them with those of a few representative studies (Motte et al. 1998; Ward-Thompson et al. 1999; Beuther et al. 2002; \cite{rath06}). In Fig.~\ref{f:histo}a-b, we plot the mass and density distributions of these 1.2~mm sources samples.

The typical size of the Cygnus~X dense cores  is 5 to 10 times smaller than the average cloud structures identified by single-dish studies of high-mass star-forming regions (see Table~\ref{t:scale} and also \cite{RW05}; \cite{hill05}). Using the terminology mentioned above, HMPOs (for high-mass protostellar objects, Beuther et al. 2002) and IRDCs (for infrared dark clouds, e.g. \cite{rath06}) are actually ``clumps''. We are here able to reach the scale of dense cores because Cygnus~X is one of the closest high-mass star-forming regions, our MAMBO-2 survey has a good spatial resolution, and we deliberately focus on high-density cloud structures. As a consequence of their small sizes, the Cygnus~X dense cores have masses ($4-950~\msun$, see Table~\ref{t:densecores}) which are up to one order of magnitude smaller than those obtained by \cite{beut02} and Rathborne et al. (2006) (see Table~\ref{t:scale} and Fig.~\ref{f:histo}a). In contrast, Cygnus~X dense cores have densities averaged over their {\it FWHM} size ($1\times 10^4-2 \times 10^6~\cmc$) which are more than \emph{one order of magnitude higher} than those of HMPOs and IRDCs (cf. Table~\ref{t:scale} and Fig.~\ref{f:histo}b). As interesting illustrations, the Cygnus~X HMPOs called IRAS~20343+4129 and IRAS 20293+3952 (Beuther et al. 2002) are twice as large and 7 times less dense than the dense cores we have identified within them (see CygX-N5, CygX-N6 and CygX-S30 in Table~\ref{t:densecores}).

On average, the Cygnus~X dense cores are a factor of 5 more massive and a factor of 3 denser than nearby dense cores identified by \cite{MAN98} and \cite{WMA99} (see Table~\ref{t:scale} and Figs.~\ref{f:histo}a-b). The latter sample includes isolated pre-stellar cores located at 140-440~pc from the Sun, and all the protocluster dense cores of the $\rho$~Ophiuchi main cloud which are either pre-stellar or already forming low-mass stars. Statistically, studying the Cygnus~X region gives access to dense cores which are one to two orders of magnitude more massive than those observed in nearby regions. The 33 most massive compact cloud fragments of Cygnus~X (i.e. with masses larger than $40~\msun$, see Table~\ref{t:sio}) already have 19 times more mass than the dense cores of low-mass star-forming regions (cf. Table~\ref{t:scale}). Thus, they do not have any equivalent in the nearby star-forming regions and represent good candidate sites for forming intermediate- to high-mass stars. 

In terms of average density, Cygnus~X dense cores are intermediary cloud structures between nearby dense cores and condensations in the $\rho$~Ophiuchi main cloud (cf. Table~\ref{t:scale} and Fig.~\ref{f:histo}b). This result suggests that the efficiency of mass transfer from the gas reservoir to the star(s) is high. Indeed, the star formation efficiency estimated in the $\rho$~Ophiuchi dense cores is $\sim 31\%$ over $10^6$~yr (\cite{bonte01}), and the good agreement in $\rho$~Oph between the pre-stellar condensation mass spectrum and the stellar IMF suggests an efficiency of $50\%-100\%$ (Motte et al. 1998; \cite{bonte01}). While the low-mass starless condensations are believed to be the direct progenitors of single stars, the Cygnus~X dense cores will probably form small groups of stars. The large mass of the Cygnus~X dense cores, compared to that of nearby dense cores and $\rho$~Oph condensations, argues for the formation of intermediate- to high-mass stars (cf. Fig.~\ref{f:histo}a). With their high density and small size, the Cygnus~X dense cores can probably be seen as the inner part of clumps/protoclusters over which the stellar IMF does not globally apply. Therefore, we may expect that the $>40~\msun$ dense cores of Cygnus~X have a high probability of forming $10-20~\msun$ of stars, including at least one high-mass star. The detection of maser emission toward a few moderate-mass ($\sim 20~\msun$) dense cores that remain weak at mid-infrared wavelength (cf. Wyrowski et al. in prep.) suggests that the $>40~\msun$ limit for a Cygnus~X dense core to host a high-mass protostar is reasonable.

\subsection{The high potential of Cygnus~X dense cores to form massive stars}\label{s:potential}

The Cygnus~X molecular complex has already formed generations of high-mass stars, since it contains several OB associations (\cite{uyan01}) and numerous \hii regions (\cite{WHL91}). We here discuss its ability to form high-mass stars in the near future by making a census of embedded \hii regions (Sect.~\ref{s:hii}), high-luminosity protostellar cores (Sect.~\ref{s:HL-IR}), and massive infrared-quiet cores which are bright in SiO (Sect.~\ref{s:sioprotostar}). We summarize in Table~\ref{t:lifetime} the number of objects found at various stages of the high-mass star formation process.

\subsubsection{Embedded \hii regions}\label{s:hii}

Within our sample of high-density cores, we identify 15 embedded \hii regions from their bright infrared and centimeter continuum emissions (see Cols.~8-9 of Table~\ref{t:densecores}). Most of them are already recognized as compact or ultra-compact \hii regions (\cite{DR66}; \cite{hasch81}; \cite{KCW94}; \cite{mol98b}; \cite{trin03}). Five others (CygX-N47/DR21-D, CygX\--N58\-/IRAS~20375+4109, CygX-S18/S106-IR, S20, CygX\--S27\-/MSX~77.9550+0.0058) coincide with radio centimeter sources from the 1.4~GHz Galactic Plane Survey (\cite{zoot90}) or old centimeter maps (\cite{harr73}; \cite{piph76}). As expected, observations of high-density tracers such as CS, NH$_3$ line and 350~$\mu$m continuum made toward these 1.2~mm sources suggest some free-free contamination (e.g. Motte et al. in prep.). 

Despite harboring an \hii region, most of these massive young stellar objects are embedded and some of them (e.g. CygX-N46/DR21, CygX-N30/W75(N), CygX-S/AFGL~2591) even show high levels of ongoing star formation activity, such as maser emission and/or powerful outflow shocks (see Tables~\ref{t:densecores} and \ref{t:sio} and Figs.~\ref{f:sio} and \ref{f:sioflux}). In agreement with these \hii regions being compact, ultra-compact, or even hyper-compact, our 1.2~mm study is sensitive to young and thus ``embedded \hii regions''.

\begin{figure}[htbp]
\includegraphics[angle=270,width=9.cm]{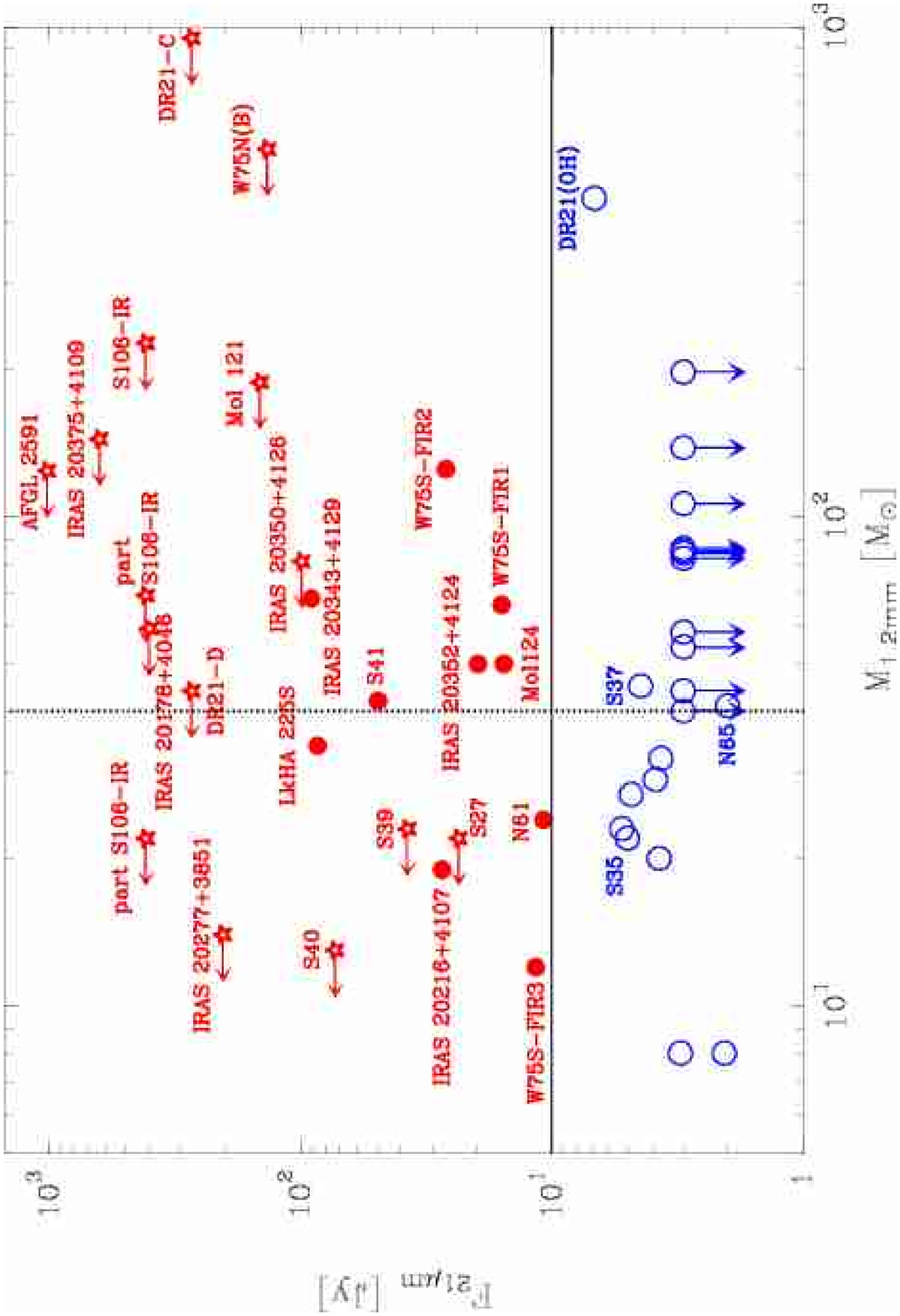} 
\caption[]{Separating the high-luminosity ($> 10^3~\lsun$) sources from infrared-quiet cores using their $21~\mu$m flux (limit set to 10~Jy). In Cygnus~X, 15 UC\hii regions (red star-like markers) and 10 infrared protostars (red filled circles) have high luminosity. A few infrared-quiet cores (blue open circles) are weak \emph{MSX} sources (like DR21(OH)) but most of them remain undetected (those more massive than $40~\msun$ are plotted here).}
\label{f:lum}
\end{figure}

\subsubsection{``High-luminosity infrared sources''}\label{s:HL-IR}

Dense cores which are luminous at infrared wavelengths are usually considered to be the best candidates to be (or to host) high-mass protostars (e.g. Wood \& Churchwell 1989). Following a similar philosophy, we qualify as ``high-luminosity infrared sources'' those Cygnus~X dense cores with bolometric luminosity larger than $10^3~\lsun$, which corresponds to that of a B3 star on the main sequence. This luminosity converts into a \emph{MSX} $21~\mu$m flux of $\sim 10$~Jy, assuming that the luminosity of high-mass protostars is dominated by their mid- to far-infrared luminosity and their average colors are defined by \cite{WC89} (see their Table~1). In practice, we estimate the total flux in the \emph{IRAS} bands using the simplified formulae\footnote{
$F_{IRAS}= 1.75 \times 10^{-13}$~W$\,$m$^{-2} \times (S_{12}/0.79 + S_{25}/2 + S_{60}/3.9 + S_{100}/9.9)$ with $S_{12}$, $S_{25}$, $S_{60}$, $S_{100}$ expressed in Jy.}
of \cite{caso86} and the \cite{WC89} colors to express this total flux as a function of the single \emph{IRAS} $25~\mu$m flux. We finally convert the $25~\mu$m flux into an expected \emph{MSX} flux at $21~\mu$m by an interpolation between 25 and $12~\mu$m using the Wood \& Churchwell color. For $d = 1.7$~kpc, we finally estimate a typical \emph{MSX} flux at $21~\mu$m of $\sim 10$~Jy$\, \times\, (L_{IRAS}/ 10^3~\lsun)$ for a B3 protostar. Figure~\ref{f:lum} displays the $21~\mu$m flux detected toward Cygnus~X dense cores as a function of their mass. With a 10~Jy limit in Fig.~\ref{f:lum}, we detect all the embedded \hii regions plus 10 sources which are ``high-luminosity infrared protostellar cores''. Three of the latter are associated with OH, H$_2$O, and/or CH$_3$OH maser emission: CygX-N43/W75S-FIR1, CygX-N44/DR21(OH), and CygX-NW5/LkA~225S. 

The high-luminosity infrared sources of the present Cygnus~X survey are mostly \emph{IRAS} sources fulfilling the \cite{WC89} criteria for UC\hii regions or objects in an earlier stage of high-mass star formation. Many of them were already identified as high-mass protostellar sources: e.g. W75S-FIR1 and -FIR2 (see \cite{CGC93}) and IRAS~20343+4129 and IRAS~20216+4107 (Beuther et al. 2002).

\subsubsection{Strong SiO emission as a probe for high-mass protostars}\label{s:sioprotostar}

In contrast to high-luminosity infrared sources, massive infrared-quiet cores are not all expected to host high-mass protostars. It is therefore interesting and somewhat surprising to find that all of them are bright SiO emitters (cf. Table~\ref{t:sio} and Fig.~\ref{f:sio}). As shown below, their SiO(2-1) lines are well interpreted as due to a powerful outflow that is probably driven by one high-mass protostar.

\begin{figure}
\includegraphics[angle=270,width=9.cm]{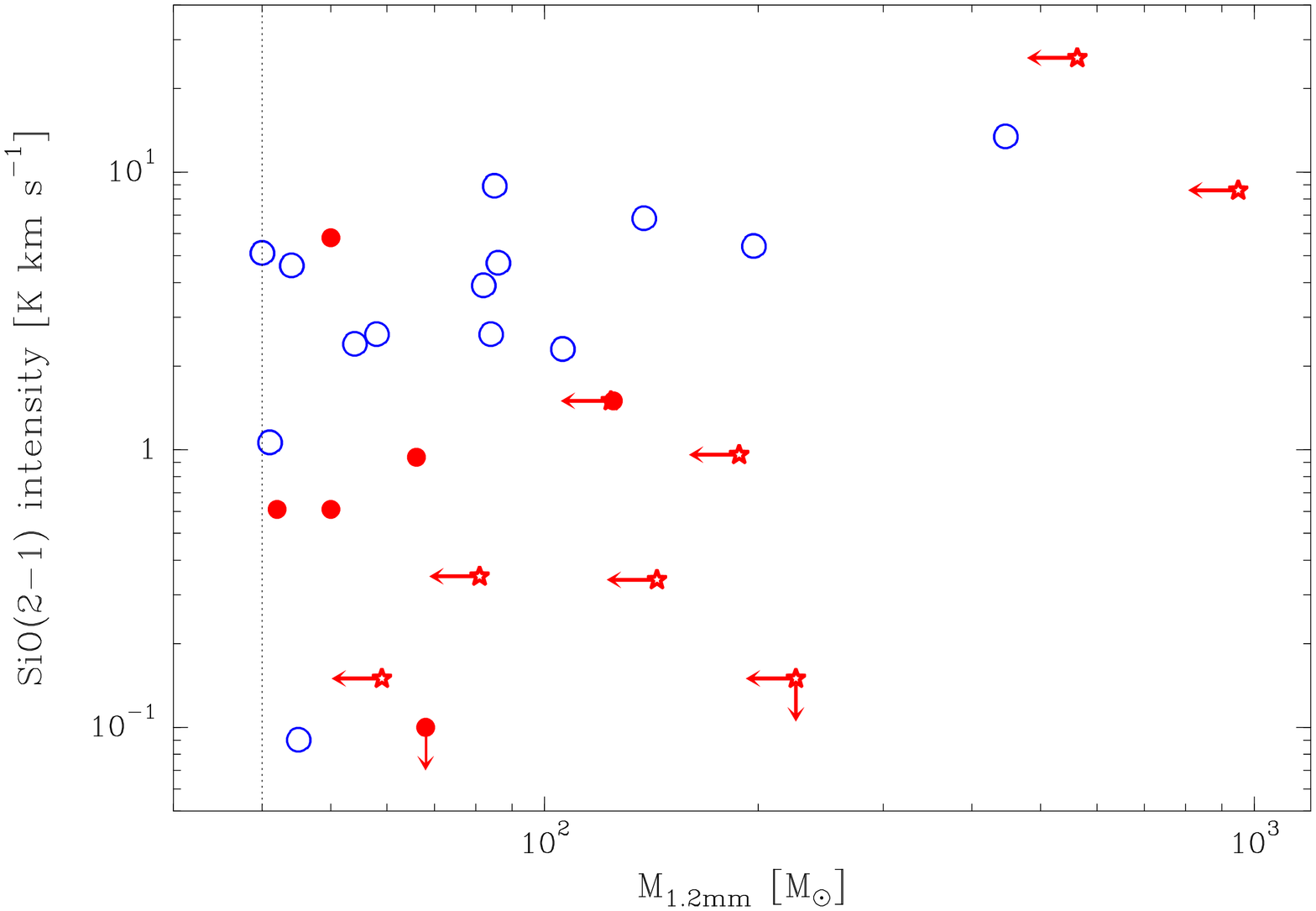}
\caption[]{Integrated intensity of SiO(2-1) detected toward the Cygnus~X dense cores more massive than $40~\msun$, as a function of their mass. Infrared-quiet cores (blue open circles) are definitely stronger in SiO than the high-luminosity protostellar cores (red filled circles) and UC\hii (red star-like markers).}
\label{f:sioflux}
\end{figure}

As shown in Fig.~\ref{f:sioflux}, which presents the SiO intensity of the Cygnus~X dense cores versus their mass, the SiO lines detected for infrared-quiet cores are several times brighter than those observed for the high-luminosity sources. As an example, the well-recognized UC\hii region CygX-S26/AFGL~2591 is weaker (in SiO) than a completely unknown source such as CygX-N53. Since (SiO) outflows are tracing the accretion process (see \cite{BATC96}), this result suggests that massive infrared-quiet cores are in a more active accretion phase than high-luminosity infrared sources. Furthermore, the SiO emission of massive infrared-quiet cores is several times stronger than the most extreme SiO shocks associated with nearby low-mass class~0 protostars (see e.g. Table~\ref{t:sio}).  This suggests that the infrared-quiet cores host either a clear single SiO outflow driven by a high-mass protostar, or a group of SiO outflows as extreme as those of L1448-MM or L1157-MM. Quantitatively the mean SiO brightness of the massive infrared-quiet cores in Cygnus~X would require the summation of two low-mass class~0s driving extreme SiO outflows (and up to 5 in the case of CygX-N53). This is rather improbable, since we evaluate the occurence of such SiO outflows to be of the order of $6\%$ (2 out of 35 class~0s known at $d<500\,$pc; e.g. \cite{AWB00}). Statistically, the summation of several tens of low-mass class~0s would be required to explain the bright SiO emissions observed here. The SiO outflows of these infrared-quiet cores better resemble that of the high-mass protostar Orion-IRc2 (see \cite{ZF87} and Table~\ref{t:sio}). These intensity comparisons suggest that the infrared-quiet protostellar cores of Cygnus~X are much more active (in star formation) than nearby low-mass protostars, and that a large fraction of them should host high-mass analogs of class~0s.

The profile of the SiO lines provides us with useful additional constrains. Half of the infrared-quiet protostellar cores in Table~\ref{t:sio} (CygX-N53, N63, N68, S30, S32 and possibly N3, N65, S7) show clear and smooth outflow wings in their SiO line profiles. These profiles are so clean that it can be taken as a good indication that the SiO emission is dominated by a single SiO outflow which, in turn, has to be powered by a high-mass protostar. Four infrared-quiet protostellar cores (CygX-N44/DR21(OH), CygX-N48/DR21(OH)-S, CygX-N38/DR21(OH)-W, and CygX-N12) have a profile which is more indicative of two emission components: a strong narrow line plus a line with large wings (see Sect.~\ref{s:sio}). This can arise from two different SiO outflows which would both still be powerful, or from a single outflow composed of two main velocity patterns in the SiO shocks. An alternative interpretation is that, besides the outflow, some shocks associated with a hot core would be detected as a narrow SiO line component. 

Therefore, the $>40~\msun$ dense cores of Cygnus~X all have a high probability of containing (at least) one high-mass protostar, whether or not they are bright infrared sources. Our selection of  extreme-density cores has thus proven to be very efficient in locating sites of future formation of high-mass stars.

\subsection{``Massive infrared-quiet protostars'': definition and lifetime}\label{s:IR-Q}

\subsubsection{Definition}\label{s:definition}

We define as ``massive infrared-quiet protostellar cores'' the Cygnus~X dense cores that are more massive than $40~\msun$, weak at infrared wavelengths (i.e. $<10$~Jy at $21~\mu$m), but present some signposts of stellar activity. Table~\ref{t:densecores} provides a list of 33 dense cores more massive than  $40~\msun$ (see also Table.~\ref{t:scale})  and Fig.~\ref{f:lum} allows the identification of ``high-luminosity infrared sources'' (see also Sect.~\ref{s:HL-IR}). The  17 remaining dense cores should either be infrared-quiet protostellar cores or starless cores. Among them, four are already identified as protostellar cores by their weak \emph{MSX} detection and/or their association with masers: CygX-N44/DR21(OH), CygX-N65, CygX-S37/IRAS~20305+4010, and CygX-N53. The majority of the 17 massive infrared-quiet cores  have been surveyed in SiO(2-1); 14 positions are listed in Table~\ref{t:sio} and CygX-N32 is included in the beam targeting CygX-N30. The brightness and line shape of their SiO emission prove that the infrared-quiet cores are all protostellar in nature and that they probably host a high-mass protostar (see Sect.~\ref{s:sioprotostar}).

\begin{table*}[htbp]
\caption[]{Massive young stellar objects found in Cygnus~X  at various stages of the high-mass star formation process}
\label{t:lifetime}
\centering
\begin{tabular}{|l|cc|cc|cc|}
\hline
& Starless	& Pre-stellar	& Massive IR-quiet		& High-luminosity	& \hii 	& OB stars \\
& clumps	& cores		& protostars	& IR protostars		& regions	& \\
\hline
\hline
Number in Cygnus~X
	& 10	& $\le 1$	& 17	& 25	& $\sim 800$	& $2600\pm 1000$ \\
References &	this paper & this paper & this paper & this paper &  \cite{WHL91} & \cite{knod00}\\
Statistical lifetime$^a$
	& $\sim 7\times 10^3$~yr 	& $\le 8 \times 10^2$~yr 	& $\sim 1.3\times 10^4$~yr 	& $\sim1.9 \times 10^4$~yr & $\sim 6\times 10^5$~yr	& $(2\pm 1) \times 10^6$~yr \\
$<n\htwo>$ 
	& $6.7\times10^3~\cmc$ & $\sim 1.5\times 10^5~\cmc$ 	& \multicolumn{2}{c|}{$1.7\times 10^5~\cmc$} & & \\
Free-fall time$^b$
	& $4\times 10^5$~yr 	& $\sim 9\times 10^4$~yr 	& \multicolumn{2}{c|}{$8\times 10^4$~yr} & & \\
\hline
Number in Orion 	& ? & ? & $\sim 1$ & $\sim 1$ & $\sim 6$ & $\sim 81$ \\
\hline
\end{tabular}
\begin{list}{}{}
\item[ (a) ]  {Lifetime of massive starless clumps, pre-stellar cores, infrared-quiet protostars, high-luminosity infrared sources, and \hii regions measured relative to the known age of OB stars (cf. \cite{hans03}) using the census given in Line~3.}
\item[ (b) ] {Free-fall time measured from the mean values of the volume-averaged density given in Line~6: $t_{\mbox{\small free-fall}} = \sqrt{ \frac {3 \;\pi} {32 \; G \; <\rho>} }$.}
\end{list}
\end{table*}

\subsubsection{Lifetime of high-mass protostars}\label{s:lifetime}

Our study supports the existence of massive infrared-quiet protostars, overlooked by infrared-based surveys (see Figs.~\ref{f:lum}--\ref{f:sioflux}). It provides a sample of (mostly) new massive infrared-quiet protostellar cores which contain high-mass protostars. We still lack the necessary angular resolution to identify individual high-mass protostars, but the characteristics of each Cygnus~X dense core more massive than $40~\msun$ are believed to be shaped by those of a single embedded high-mass protostar (cf. Sects.~\ref{s:highdens} and \ref{s:sioprotostar}). With this assumption, we give the first lifetime estimates for this new population of high-mass protostars, assuming a constant star formation rate in Cygnus~X over the past 1-2~Myr.

Since our sample of Cygnus~X dense cores is complete for embedded, massive, young stellar objects, we can statistically estimate the relative lifetimes of massive infrared-quiet protostars and high-luminosity infrared sources (including infrared protostars and embedded \hii regions). The separation of infrared sources between embedded \hii regions and high-luminosity infrared cores is not final since centimeter continuum follow-up observations are necessary, and a low-level of centimeter free-free detection might be indicative of outflow shocks rather than \hii regions. In the following discussion, we therefore treat cores hosting high-luminosity infrared sources and possibly protostars as a single population. According to the numbers of high-luminosity infrared sources and massive infrared-quiet protostars given in Table~\ref{t:lifetime} (and taken from Sects.~\ref{s:hii}--\ref{s:HL-IR} and Sect.~\ref{s:definition}), the infrared-quiet protostellar phase of high-mass stars might last as long as their better-known infrared-bright protostellar (or embedded H\mbox{\sc ~ii}) phase. This result is in disagreement with the idea that high-mass protostars are very luminous from the very first stage of their evolution and thus that their infrared-quiet phase is very short-lived.

Table~\ref{t:lifetime} gives our current estimates of the number of massive young stellar objects forming and stars already formed in Cygnus~X. The number and age of OB stars in Cygnus~X are taken to be those of the Cyg~OB2 association (\cite{knod00}; \cite{hans03}), increasing uncertainties to take into account the other (smaller) OB associations which have different ages (\cite{uyan01}). The number of \hii regions is crudely given by the numbers of centimeter free-free sources observed in Cygnus~X (\cite{WHL91}). The statistical lifetimes of high-mass protostars derived relative to the number of OB stars give a time range of $(0.5-3.1) \times 10^4$~yr for the infrared-quiet protostellar stage and $(1.2-7.9)\times 10^4$~yr for the complete protostellar stage (see also Table~\ref{t:lifetime}). For the sake of comparison, the statistical lifetime measured for the high-mass protostellar phase in Cygnus~X ($\sim 3.2 \times 10^4$~yr) is one order of magnitude smaller than the typical age of nearby, low-mass class~I protostars (e.g. \cite{KH95}) and more closely resembles that of low-mass class~0s (cf. \cite{AWB00}). 

\begin{figure}[htbp]
\includegraphics[angle=270,width=9.1cm]{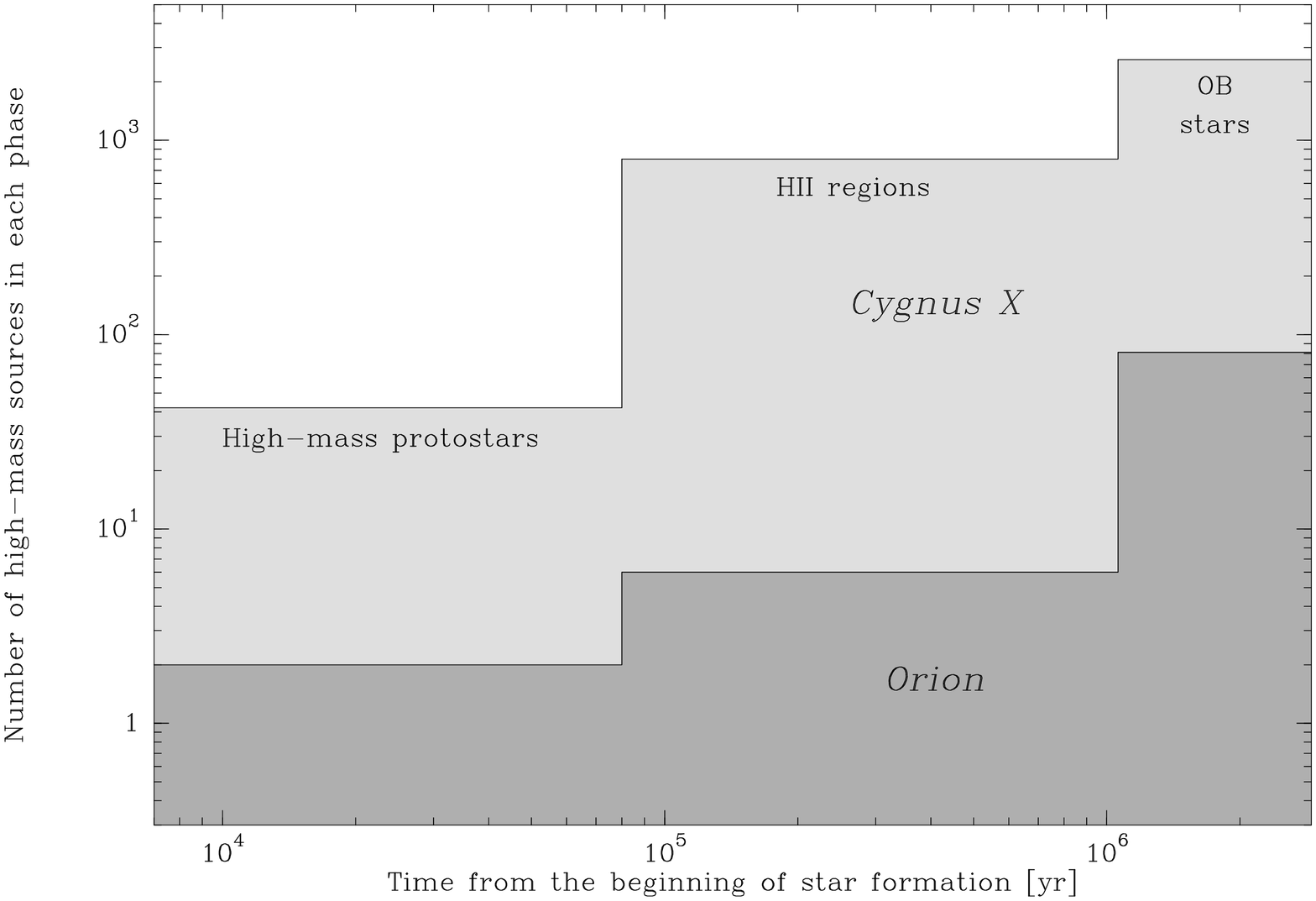}
\caption[]{Number of massive young stellar objects as a function of the time elapsed since the beginning of the protostellar collapse in Cygnus~X and Orion. The age estimates (also given in Sect.~\ref{s:lifetime}) are {\it not} statistical, in order to have the X-axis and Y-axis independent.}
\label{f:lifetime}
\end{figure}

Figure~\ref{f:lifetime} displays the number of high-mass protostars, \hii regions and OB stars (i.e. objects in the three main evolutionary phases) as a function of the time elapsed since the beginning of the protostellar collapse (i.e. since the formation of a hydrostatic stellar embryo). Absolute lifetimes of the high-mass protostellar phase (encompassing both the infrared-quiet and the high-luminosity infrared protostellar phases) as well as those of the high-mass starless phases can be roughly estimated from their free-fall dynamical timescales. Table~\ref{t:lifetime} gives the estimates we made using  the mean values of the volume-averaged density of massive dense cores observed in Cygnus~X. The most massive dense cores of Cygnus~X should thus collapse in a free-fall time of $\sim 8\times 10^4$~yr, which is taken to be the end of the high-mass protostellar phase in Fig.~\ref{f:lifetime}. The \hii regions phase should last for several times the typical lifetime of UC\hii regions ($\sim 10^5$~yr, cf. Wood \& Churchwell 1989), starting in Fig.~\ref{f:lifetime} at the end of the protostellar phase. The free-fall time of high-mass protostars in Cygnus~X agrees, within a factor of 3, with the statistical lifetimes of the complete population of high-mass protostars (including both infrared-quiet and high-luminosity infrared protostars) derived relative to the number of OB stars (see Table~\ref{t:lifetime}).

To compare Cygnus~X to Orion, Table~\ref{t:lifetime} makes the census of all massive young stellar objects known in Orion, and Fig.~\ref{f:lifetime} shows their distribution as a function of time. The number of high-mass stars (down to B3 spectral type) within all the Orion OB associations is measured by \cite{BGZ94}, while the \hii regions are associated with the well-known reflection nebulae M42, NGC~2024, NGC~1977, M43, NGC~2068, and NGC~2071. One source could qualify as a high-luminosity infrared protostar (Orion-IRC2) and another as a massive infrared-quiet protostar (OMC1-S). Interestingly, the massive young stellar objects forming, and stars formed, in Cygnus~X are 20 times more numerous than in Orion (see Table~\ref{t:lifetime}), in good agreement with the mass ratio (13) of the Cygnus~X and Orion cloud complexes.

\subsection{Where is the pre-stellar phase of massive star formation?}\label{s:pscore}

Among the 33 most massive ($\ge 40~\msun$) dense cores of Cygnus~X, no starless cores are found, with the possible exception of CygX-N69 which was not selected for SiO observations because of its relatively large size and corresponding low-density ($\sim 10^4~\cmc$).  This non-detection of starless cores means that high-mass pre-stellar cores, if they exist, are short-lived ($< 10^3$~yr, see Table~\ref{t:lifetime}). This is particularly surprising since the lifetime of pre-stellar cores and condensations in nearby, low-mass star-forming regions is more than two orders of magnitude longer ($1-4 \times 10^5$~yr, e.g. \cite{oni02}; \cite{KWA05}; \cite{andr07}). Below, we consider two interpretations: either the star formation activity of Cygnus~X is not continuous (see Sect.~\ref{s:burst}) or massive pre-stellar clumps are dynamically evolving into protostars (see Sect.~\ref{s:dynamic}).

\subsubsection{Star formation in Cygnus~X cannot be caused by a burst}\label{s:burst}

If the Cygnus~X molecular complex were experiencing a sudden burst of star formation, it might be due to the interaction with the \hii regions and/or OB associations, including the Cyg~OB2 cluster. In this case, we expect differences in the concentration of gas and in the population of young stellar objects in the CygX-North and CygX-South regions.

Evidence shows that the concentration of gas is different in CygX-North and CygX-South. The CO survey of \cite{schn06}, the present MAMBO-2 maps, and our small-scale analysis trace various cloud structures (see Sects.~\ref{s:spatdyn}--\ref{s:census1}). We can thus measure the gradual concentration of molecular clouds (0.5--50~pc scales), first into 0.1--5~pc scales, and then into 0.1~pc dense cores. In CygX-North ($280\times10^3~\msun$, cf. Schneider et al. 2006), the concentration factor is 8 for each step while in CygX-South ($480\times10^3~\msun$) it is 18 for each step. Therefore, when considering the mass involved in dense cores, the CygX-North region is 5 times more concentrated  (CO clouds concentrate into 0.1~pc cores with an efficiency of $\sim 1.5\%$) than CygX-South (concentration efficiency of $\sim 0.3\%$). This suggests that CygX-North could be 5 times more efficient in forming dense cores, and possibly forming stars, than CygX-South.  

In contrast, the relative numbers of ``high-luminosity infrared sources'' versus ``massive infrared-quiet protostellar cores'' are identical (within our large error bars) in both regions. Moreover, other generations of (massive) young stars have already formed nearby (see e.g. the \emph{MSX} sources at $\sim 0.2$~pc from the dense cores). The agreement between the free-fall lifetime and the statistical lifetime of high-mass protostars relative to OB stars also suggests that star formation has been continuous, at least over the past $\sim 10^6$~yr (see also Fig.~\ref{f:lifetime}). Therefore, it is highly improbable that Cygnus~X is currently experiencing a burst of star formation.

\subsubsection{Is the  star formation in Cygnus~X dynamic?}\label{s:dynamic}

The present study fails to find a single good candidate for being a pre-stellar dense core, i.e. a starless cloud structure with a $\sim 0.1$~pc size and a $\sim 10^5~\cmc$ volume-averaged density, which is gravitationally bound. This does not preclude, however, the existence of smaller and denser pre-stellar condensations ($\sim 0.01$~pc size and $\sim 10^6~\cmc$ density according to Motte et al. 1998, see Table~\ref{t:scale}) within the Cygnus~X protostellar cores. On the other hand, we have identified 10 clumps which may be starless ($\sim 0.8$~pc size and $7 \times 10^3~\cmc$ density, see Tables~\ref{t:clumps} and \ref{t:scale}). The Virial mass of these clumps ($\mvir=3\, R \sigma^2/G$, \cite{BMK92}) are estimated from the N$_2$H$^+$ linewidth measured at their peak location (Schneider et al. in prep.) and the {\it FWHM} size given in Table~\ref{t:clumps}; they suggest that most of these clumps are close to being gravitationally bound. 

From the numbers of starless clumps and pre-stellar cores detected in Cygnus~X and compared to the numbers of OB stars, we statistically estimate their lifetimes to be $\sim 10^4$~yr and $< 10^3$~yr, respectively (see Table~\ref{t:lifetime}). These lifetime values are considerably smaller (by factors of 60-110) than the free-fall times we estimated. They are also shorter than the lifetimes observed in nearby regions that form mostly low-mass stars, since starless cloud structures with volume-averaged densities of $\sim 10^3~\cmc$ and $\sim 10^4~\cmc$ have pre-stellar lifetimes of $\sim 10^6$~yr and $\sim 10^5$~yr (see Fig.~11 of \cite{KWA05} and references therein). Therefore, massive pre-stellar cores seem to be short-lived or even transient cloud structures compared to nearby low-mass pre-stellar cores, to high-mass protostars, and to massive starless clumps.

A dynamical process should thus be acting to create pre-stellar condensations from starless clumps. It should be supersonic to concentrate mass  from densities of $7 \times 10^3~\cmc$ to $2 \times 10^5~\cmc$ in a time as short as, or even shorter than, the free-fall timescale. Similarly short lifetimes are theoretically expected in molecular clouds where highly turbulent processes dominate and pre-stellar cores are magnetically supercritical (e.g. \cite{VS05}). Such dynamical processes are also necessary for the protostellar lifetime to last for only one free-fall time (see Table~\ref{t:lifetime}) and, consequently, for the accretion to be strong enough ($\sim10^{-3}~\msun$~yr$^{-1}$) to overcome the radiation pressure and form a high-mass star. In fact, the DR21 filament is observed to undergo a global supersonic collapse, possibly due to a strong external compression (\cite{chan93}; \cite{mott05}). This is encouraging but it remains to be verified whether such a dynamical process acts throughout the complex.

Taken together, the lifetimes measured here suggest a faster evolution (by a factor 10 at least) of the pre-stellar and protostellar phases of high-mass stars compared to those typically found for nearby, low-mass stars. Molecular clouds harboring high-mass star formations should thus be quantitatively different in terms of e.g. density and kinematics than those of our close neighborhood.

\section{Conclusions}\label{s:summary}

In order to improve our knowledge of the high-mass star formation process, we started an unbiased search for its earliest phases, i.e. the high-mass analogs of low-mass pre-stellar cores and class~0 protostars. One of the best targets for such a study is Cygnus~X, which we recently recognized as the richest molecular and \hii complex at less than 3~kpc (it is located at 1.7~kpc from the Sun). We have therefore made a complete 1.2~mm continnum survey of the Cygnus~X cloud complex and have performed SiO(2-1) follow-up observations of the best candidate progenitors of high-mass stars. Our main findings can be summarized as follows:

   \begin{enumerate}
      \item Our MAMBO-2 (1.2~mm, $11\arcsec$ resolution) imaging of the Cygnus~X molecular cloud complex gives a complete view of the cloud structures ranging from 0.03 to 5~pc, i.e. from dense cores to clumps (cf. the terminology used for nearby molecular clouds).  We performed a multi-resolution analysis to identify a few tens of massive large-scale ($\sim 0.7$~pc) clumps and extract 129 compact ($\sim 0.1$~pc) dense cores.
      \item The Cygnus~X dense cores are the high-density parts of massive clumps which resemble HMPOs or IRDCs. The dense cores have similar sizes ($\sim 0.1$~pc) but mass and volume-averaged densities ($4-950~\msun$, $\sim 1.1 \times 10^5~\cmc$) larger than those of nearby, pre-stellar dense cores. The most massive ($> 40~\msun$) and thus ``extreme-density'' cores of this large and unbiased sample are probable precursors of high-mass stars spanning, for the first time, the whole evolution from pre-stellar cores to ``embedded \hii regions''.
      \item We have used the \emph{MSX} 21~$\mu$m flux arising from our MAMBO-2 dense cores to identify the high-luminosity ($> 10^3~\lsun$) massive young stellar objects of Cygnus~X: 15 UC\hii regions and 10 ``high-luminosity infrared protostars''. Strikingly, half of the dense cores considered as the best candidate precursors of high-mass stars are found to be infrared-quiet (i.e. 17 Cygnus~X dense cores more massive than $40~\msun$ have $F_{21\mu m} < 10$~Jy).
      \item We have surveyed the most massive ($> 40~\msun$) dense cores of Cygnus~X in SiO(2-1) to search for shocked gas in molecular outflows and/or hot cores. The association of high-velocity SiO emission with \emph{all} massive infrared-quiet cores provides persuasive evidence that stars are already forming in these cores as outflows are tracing the accretion process. The brightness and shape of the SiO line compared to that of nearby, low-mass protostars suggest that these extreme-density cores may host ``massive infrared-quiet protostars''.
      \item Our unbiased survey of the massive young stellar objects in Cygnus~X shows that massive infrared-quiet protostars do exist,  and that their lifetime should be comparable to that of more evolved high-luminosity infrared protostars. By comparing the number of high-mass protostars and OB stars in the whole of Cygnus~X, we estimate a statistical lifetime of $3 \times 10^4$~yr for high-mass protostars. One order of magnitude smaller than the lifetime of nearby, low-mass protostars, such a value agrees, within a factor of 3, with the free-fall time of Cygnus~X dense cores.
      \item Our complete census of massive young stellar objects in Cygnus~X fails to discover the high-mass analogs of pre-stellar dense cores ($\sim 0.1$~pc, $> 10^4~\cmc$). Their corresponding lifetime ($< 10^3$~yr) is smaller than one free-fall time, in marked contrast with nearby, pre-stellar cores that typically live for 2--5 free-fall times. We propose that the starless, massive but lower-density clumps ($800~\msun$, $\sim 0.8$~pc, $\sim 7 \times 10^3~\cmc$) that we observe rapidly concentrate and collapse to form high-mass protostars.
      \item Lifetime measurements of the pre-stellar and protostellar phases of high-mass stars in Cygnus~X suggest a dynamical contraction for pre-stellar cores and a supersonic evolution for protostars. Highly turbulent processes throughout the Cygnus~X molecular cloud complex would be necessary in such a dynamical picture of the high-mass star formation process.
      \item Our Cygnus~X study suggests that far-infrared to sub-millimeter continuum imaging of entire star-forming complexes (such as those proposed with \emph{Herschel} by Motte, Zavagno, Bontemps et al.: the HOBYS\footnote{
``HOBYS: the \emph{Herschel} imaging survey of OB Young Stellar objects'' is a Guaranteed Time Key Programme jointly proposed by the SPIRE and PACS consortia, and the Herschel Science Centre.}
       Key Programme) will definitely contribute to a better knowledge of the high-mass star formation process during its earliest phases.
  \end{enumerate}

\begin{acknowledgements}
We are grateful to Axel Weiss, Roberto Neri and Clemens Thum for their help in taking and reducing the MAMBO data. We thank Henrik Beuther and Jill Ratherborne for providing electronic versions of their tables, as well as Friedrich Wyrowski for unveiling his Effelsberg results before publication. We are grateful to Tom Megeath for useful discussions on Orion and Arnaud Belloche for the suggestion of using volume-averaged densities.
\end{acknowledgements}

\newpage
~
\newpage

~\vskip 5cm
\centerline{\Huge Online Material}
\newpage
~
\newpage

\section*{Appendix~A: A skynoise reduction technique dedicated to extended sources}
\begin{figure}[h]
\centering
\vskip 0.2 cm
\includegraphics[angle=0,width=8cm]{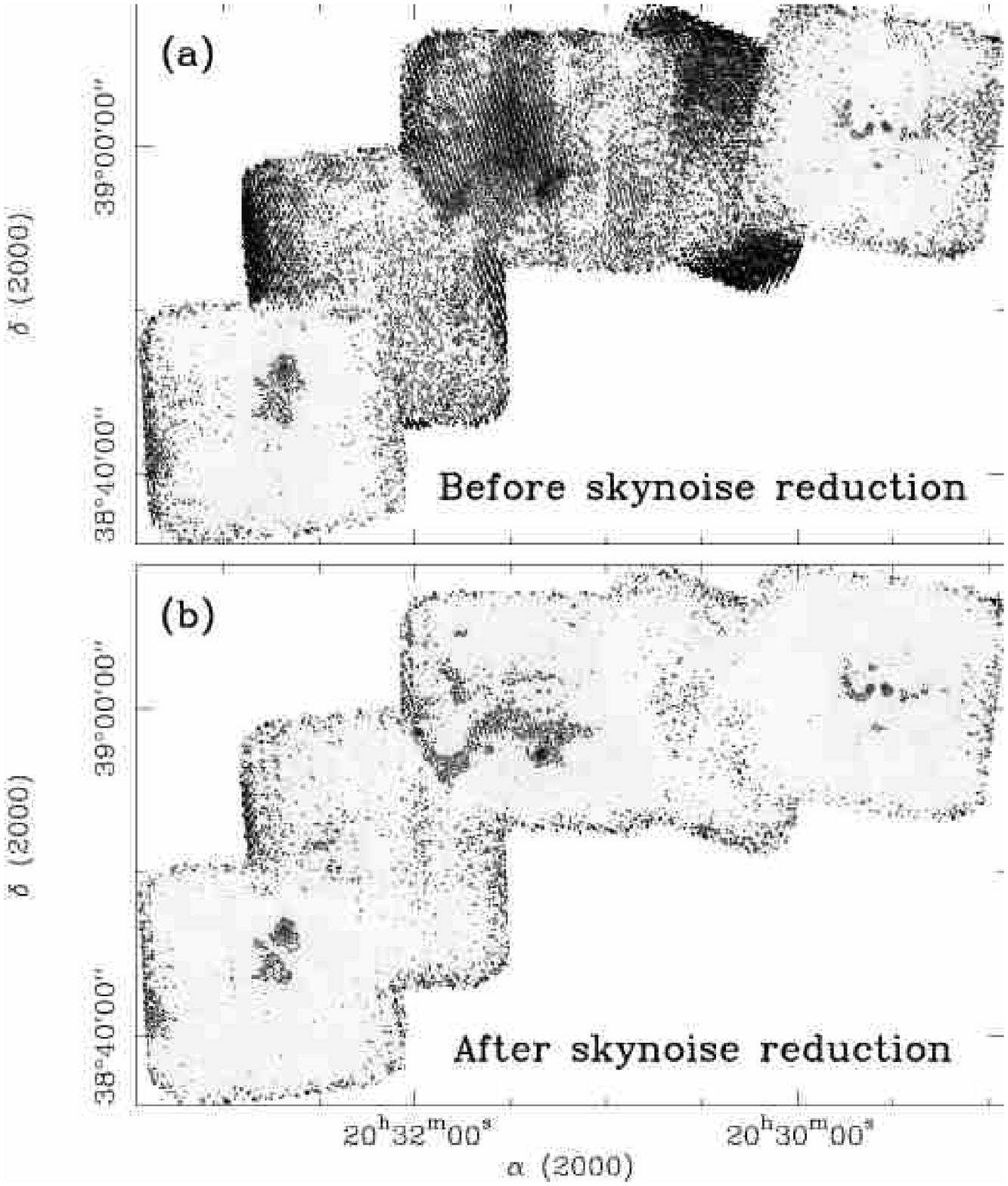}
\caption[]{1.2~mm map of the DR13 filament in the Cygnus~X molecular cloud complex obtained with MAMBO-2 at the IRAM 30~m telescope. Contour levels are -45~\mjb and then 45 to $315~\mjb$ with steps of 45~\mjb.
 {\bf(a)}  Mosaic built from individual maps taken with low to medium skynoise.  As a result, the rms noise level ranges from  $15~\mjb$ to $60~\mjb$.
{\bf(b)}  Mosaic built from individual maps processed by our skynoise reduction program.  The resulting noise level is relatively uniform with $\sigma = (14\pm 2)~\mjb$.
}
\label{f:skynoise}
\end{figure}

Ground-based (sub)millimeter continuum observations of astronomical sources are hampered by the emission of the atmosphere and its fluctuations, called ``skynoise".  The bulk of the atmospheric emission and its slow variations in time can be removed by the dual-beam scanning mode generally used for bolometer observations (the ``EKH method'', \cite{EKH}).  However, the residual atmospheric fluctuations generate an excess noise in the restored maps that can dominate the Gaussian noise (see e.g. Fig.~\ref{f:skynoise}a).  In the present study, more than half of the on-the-fly maps were observed with a skynoise level qualified as medium or high.  In our maps, medium skynoise implies $\sigma =30-60~\mjb$ and high skynoise $\sigma > 60~\mjb$, while the nominal rms  is $\sigma \sim 15~\mjb$.  This excess noise can theoretically be suppressed by using the fact that it is well correlated across the whole array.  Several skynoise reduction algorithms have been developed assuming that either the atmospheric emission is equal in all receivers at all times (``mean" or ``neighbor" approximation), or that  it is correlated on a few seconds of time with correlation coefficients decreasing when the angular separation of receivers increases (see e.g. NIC and MOPSIC IRAM softwares for bolometer-array data, \cite{brog95} and \cite{zylka98}).  While efficient for the detection of pointlike sources, such assumptions can have dramatic consequences on the representation of the extended emission.  Indeed, the large-scale structure of a source leads to a correlation of bolometer signals which is non-zero, and caution must be taken not to filter out most of the extended emission.

We have developed a program which uses a source model and the mean approximation to remove the skynoise while preserving, as much as possible, the structure of sources like molecular clouds (compare Figs.~\ref{f:skynoise}a-b). We describe below its four steps: \\
(1) The first step consists of creating a good source model.  A first-order data processing is done by reducing all individual maps and combining them with weights consistent with their intrinsic quality, i.e. accounting for both Gaussian noise and skynoise levels (see e.g. Fig.~\ref{f:skynoise}a).  This first-order mosaic is used as a model of the source after cutting its noisier parts (mostly map edges) and removing artefacts such as negative levels stronger than the rms.\\
(2) This source model is then subtracted from all the original scans.  To do so, we first simulate the on-the-fly dual-beam signals of each bolometer when the MAMBO-2 camera is mapping the source model with the observational geometry of the real scans (see \cite{MA01}, and the routine ``simulate'' in NIC). These simulated source signals are then removed from the original on-the-fly dual-beam signals recorded by each bolometer. At this point, and for each scan, the bolometer data streams should only contain Gaussian noise and skynoise.\\
(3) As a third step, the skynoise emission contained in each processed scan is estimated by using the ÔÔaveraging methodÕÕ.  While cruder than any ``correlation method'' (like that proposed in NIC), it allows the treatment of each observing point (taken every half a second) independently.  We often find the averaging method to be more efficient, suggesting that the skynoise power is strong at high frequency.\\
(4) Afterwards, the original scans are reduced for skynoise by subtracting, for each scan, the skynoise data stream estimated above from the on-the-fly dual-beam signals of each bolometer.  Finally, the individual scans, now treated for skynoise, are reduced with the default procedure and combined to create a better (second-order processed) mosaic (see e.g. Fig.~\ref{f:skynoise}b).

This process can evidently be iterated to improve the map sensitivity but a single iteration already reduces most of the skynoise and probably gives more secure results. The final individual maps have a noise rms of $20 \pm 5~\mjb$, $15~\mjb$ being the nominal rms in the absence of skynoise, and with a zenith atmospheric opacity of $\sim 0.2$.  Our skynoise reduction technique has thus improved the map sensitivity by a factor of $\sim 2$ on average, and up to a factor of 6 for the poorer quality original maps.

Our method relies on the quality of the source model and thus mainly on the redundancy of observations taken in every single pixel. The redundancy is currently $\sim 20$ for one single map and $\sim 40$ on average, since most parts were imaged several times. The mosaics shown in Figs.~\ref{f:mambo}a-c recover most of the extended emission up to $\sim 10\arcmin$, which is the azimuthal size of individual maps. In agreement, the Cygnus~X main region has recently been imaged  with the SHARC-II camera installed at the CSO telescope (Motte et al. in prep.). Those data generally confirm the cloud structures detected in our MAMBO-2 maps.  We acknowledge, however, that the extended emission is attenuated in a couple of sites where data are too noisy and not redundant enough to build a good source model (e.g. south of DR23).

\section*{Appendix~B: An extraction technique dedicated to compact sources}

Since stars are generally forming in the densest parts of molecular clouds, we have developed a source extraction technique aiming at identifying and extracting dense compact fragments that should be the best potential sites of star formation. 

\begin{figure}
\vskip 2cm
\hskip -0.7cm \includegraphics[angle=270,width=8cm]{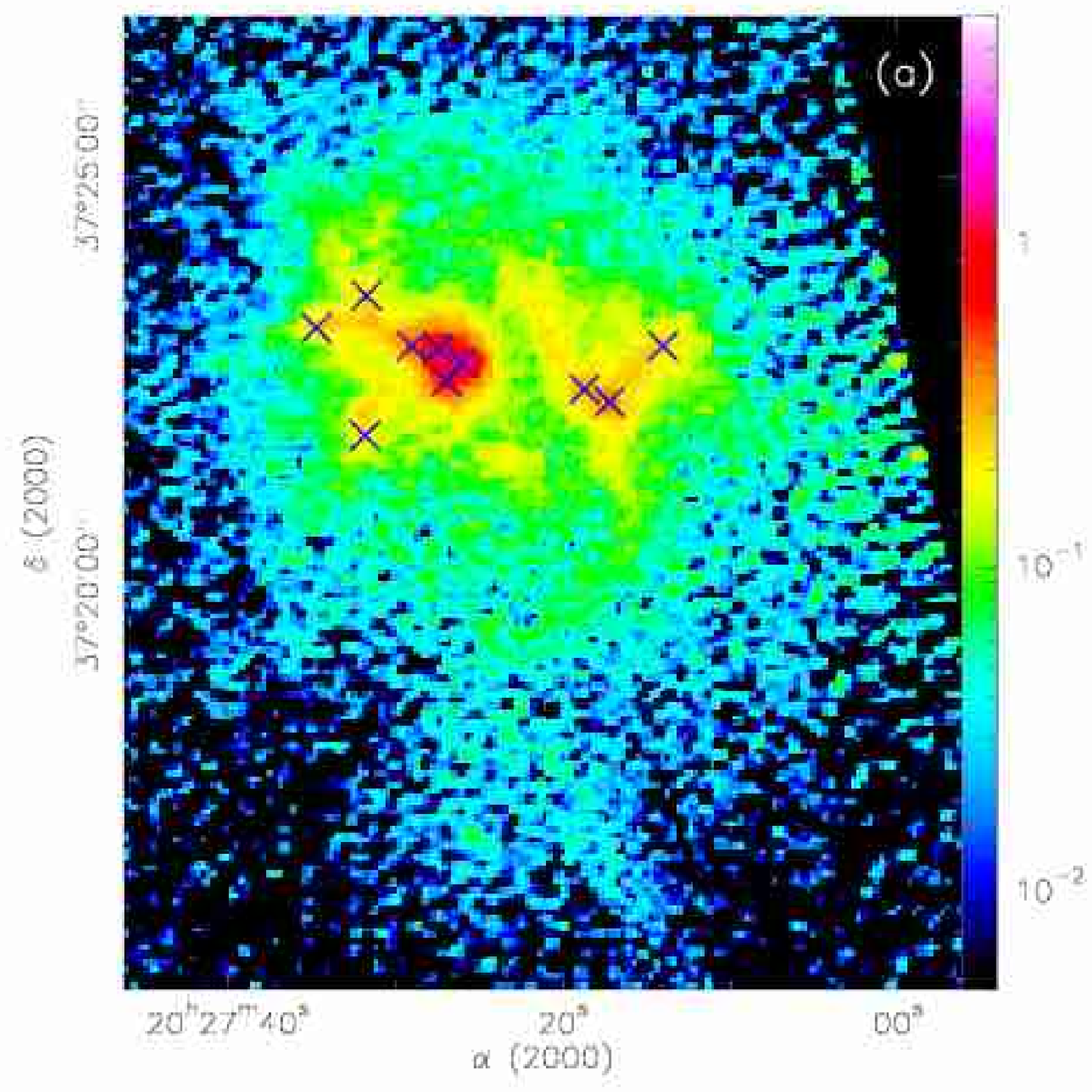}
\vskip -8cm\hskip 4.1cm\includegraphics[angle=270,width=8cm]{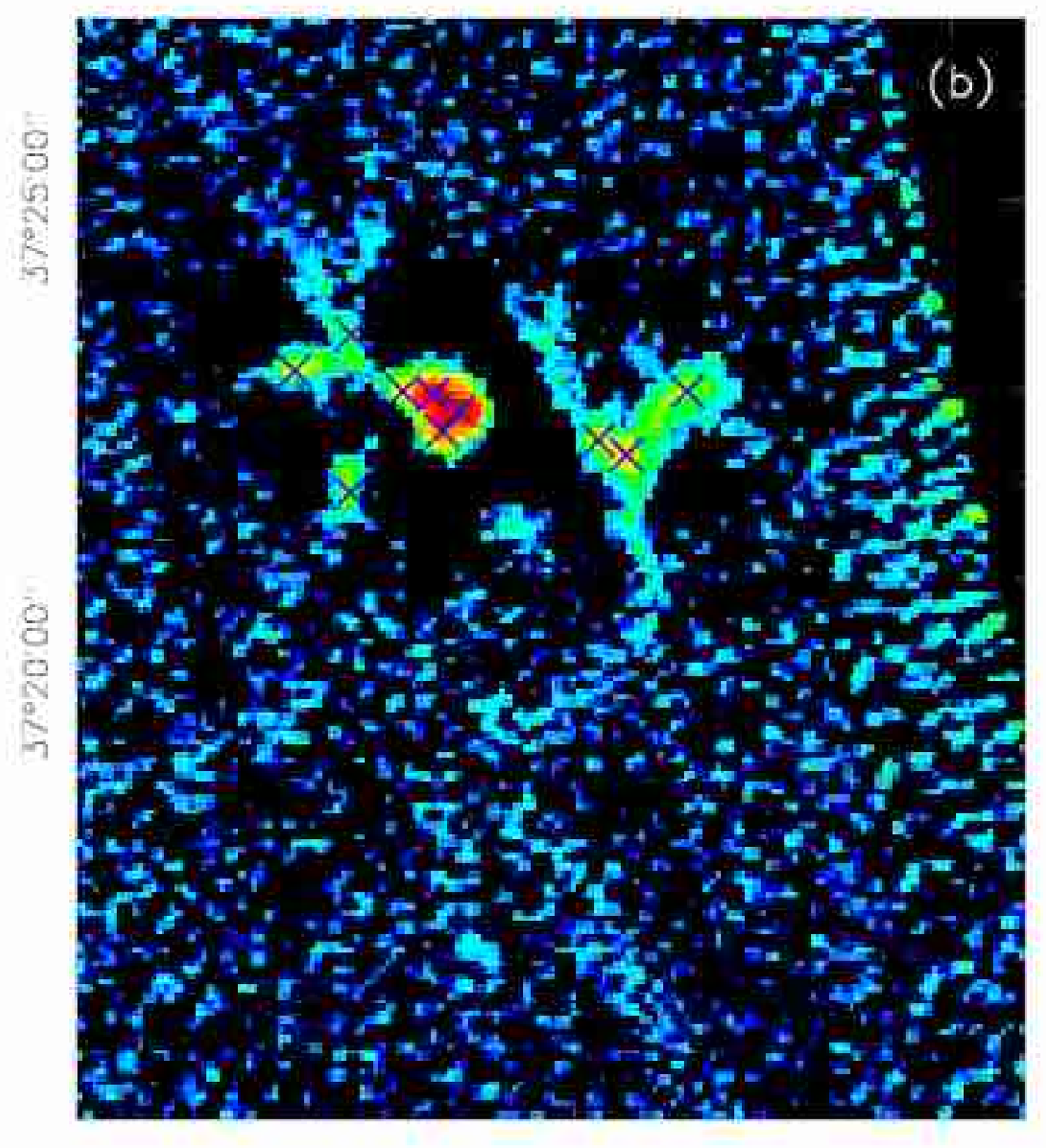}
\vskip -1.cm\hskip 4.1cm\includegraphics[angle=270,width=8cm]{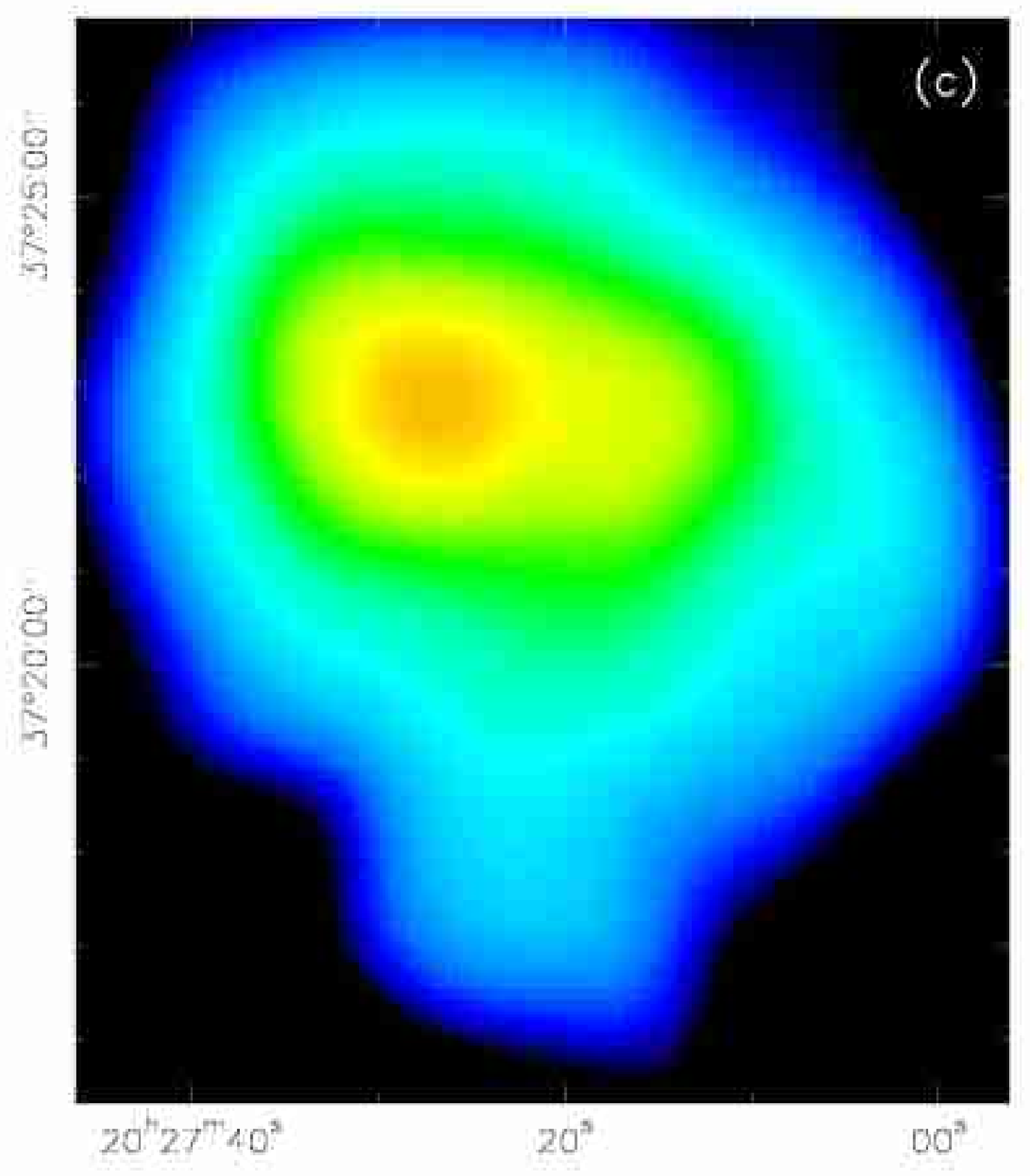}
\vskip -0.5cm
\caption[]{Separation of cloud structures from the MAMBO-2 map of S106 shown in ({\bf a}) into ({\bf b}) dense cores (i.e. compact structures with size scales smaller than $128\arcsec$) and ({\bf c}) their environment (i.e. cloud structure with a typical scale $>128\arcsec$). Dense cores extracted by the technique described in Appendix~B are indicated by crosses.}
\label{f:extraction}
\end{figure}

In spirit, this method is equivalent to an eye search of local density peaks followed by flux measurements with an aperture optimized for each source. We outline below the procedure for this automated method, which was originally developed by Motte et al. (2003):\\
(1) As a first step, one has to determine the cutoff lengthscale at which the sources of interest (here dense cores, cf. Fig.~\ref{f:extraction}b) separate from their surrounding (here clumps and clouds, cf. Fig.~\ref{f:extraction}c). This scale corresponds to twice the mean outer radius of the most compact objects one can distinguish on the MAMBO-2 map or the mean projected distance between them. This scale is typically about 5 to 10 times the {\it HPBW}.\\
(2) For further analysis, we only keep spatial scales smaller than the cutoff lengthscale determined at step~1 (here 1~pc). For this purpose, we use a multiresolution program based on wavelet transforms (MRE, cf. \cite{SM06}). This makes a series of filtering operations and provides ``views'' of the image at different spatial scales. We therefore sum\footnote{
The original image can be expressed as the sum of all wavelet views (also called planes) plus the smoothed image (last plane containing all remaining scales).} 
all wavelet planes of the original image up to that scale to create images like that shown in Fig.~\ref{f:extraction}b.\\
(3) As a third step, we use a version of the Gaussclumps program (\cite{SG90}; \cite{kram98}) which has been customized for continuum images (see \cite{MSL03} and \cite{mook04}) and optimized for compact sources\footnote{
Gaussclumps parameters for initial guesses are: {\it FWHM}$= 1.1 \times \mbox{\it HPBW}$, aperture $= 1.5 \times \mbox{\it HPBW}$}.
Gaussclumps is applied to the filtered image created at step~2. It identifies compact fragments above a $5~\sigma$ level (here $75~\mjb$) and gives first-guess parameters of a 2D-Gaussian fit. If the Gaussian shape is not ideally fitting compact fragments, it gives appropriate estimates of their size and mass even when their environment is not perfectly filtered out. Gaussclumps is known to be efficient at deblending compact sources. Noisy edges that usually pose problems (see e.g. \cite{kram98}) have been removed.\\
(4) Finally, the most critical and time consuming step is to improve the extraction of individual fragments from their surrounding background. The main goal is to avoid Gaussian fits being biased to a single lengthscale, which would be that determined at step~1. We therefore repeat steps~2 and 3 for the sources identified at step~3 and use a cutoff scale optimized for each fragment. We estimate the scale up to which one fragment is resolved from its surrounding to be twice the first-guess measure of its Gaussian {\it FWHM} (see step 3). Therefore, for each of the (here 129) sources, we create a final-guess filtered image and adjust its Gaussian fit parameters.

\setcounter{table}{0}
\begin{table*}[htbp]
\caption[]{Properties of dense cores (compact cloud fragments) detected in the Cygnus~X complex}
\label{t:densecores}
\centering
\begin{tabular}{rcrcrrccl}
\hline
& Fragment & \speak$^{\rm a}$ & {\it FWHM}~$^{\rm a}$ & \sint$^{\rm a}$  & \msmm$^{\rm b}$ & $<n\htwo>$$^{\rm c}$ & MSX$^d$ & Name, comments, and coincidence\\
& name & (mJy/ & (pc~$\times$~pc) & (mJy) & (\msun) & (\cmc) & 21~$\mu$m & with stellar activity signatures\\
&		& beam)	&		&		&		&	& (Jy) &\\
\hline\hline
& CygX-North region & & & & & & &\\ 
\hline
N1 & J203532.7+421953  & 120 & $0.15 \times 0.09$ 
       & 320  & 17 & $4.7 \times 10^4$ 	& -- &\\ 
N2 & J203534.0+422023  & 110 & $0.22 \times 0.07$ 
       & 380  & 20 & $4.0 \times 10^4$ 	& 3.74 & MSX 81.3039+1.0520\\ 
N3 & J203534.2+422005  & 720 & $0.11 \times 0.09$ 
       & 1580  & 84 & $3.6 \times 10^5$ 	& -- &\\ 
N4 & J203558.0+422431  & 90 & $0.06 \times 0.06$ 
       & 120  & $<6$ & $<1.2 \times 10^5$ 		& -- & probably not an embedded YSO$^e$\\ 
N5 & J203606.5+413958  & 160 & $0.12 \times 0.07$ 
       & 350  & 19 & 9.4$\times 10^4$ 	& -- & \\	
N6 & J203608.1+413958  & 260 & $0.23 \times 0.14$ 
       & 1280  & 44-68 & 3.2-5.0$\times 10^4$ 	& 90.88 & IRAS~20343+4129\\ 
N7 & J203615.7+425305  & 80 & $0.12 \times 0.10$ 
       & 190  & 10 & $3.2 \times 10^4$ 	& -- &\\ 
N8 & J203638.4+422911  & 170 & $0.17 \times 0.09$ 
       & 500  & 27 & $6.4 \times 10^4$ 	& Abs &\\ 
N9 & J203639.5+425113  & 100 & $0.23 \times 0.14$ 
       & 500  & 26 & $1.7 \times 10^4$ 	& Abs &\\ 
N10 & J203652.2+413623  & 590 & $0.14 \times 0.09$ 
       & 1530  & $<$52-81 & $<$1.5-2.4$\times 10^5$ 	& 99.33 & {\tiny IRAS~20350+4126, cm$^1$ free-free,}\\ 
       & & & & & & & & {\tiny maser$^2$ (CH$_3$OH/OH)}\\ 
N11 & J203656.8+421320  & 180 & $0.11 \times 0.06$ 
       & 340  & 18 & $1.4 \times 10^5$		& Abs &\\ 
N12 & J203657.4+421127  & 740 & $0.11 \times 0.09$ 
       & 1630  & 86 & $3.8 \times 10^5$ 	& Abs &\\ 
N13 & J203658.4+421118  & 90 & $0.04 \times 0.03$ 
       & 100  & 5 & $7.1 \times 10^5$ 		& Abs &\\ 
N14 & J203700.9+413457  & 500 & $0.10 \times 0.07$ 
       & 940  & 50 & $3.5 \times 10^5$ 	& 19.71 & IRAS~20352+4124\\
N15 & J203710.9+413339  & 80 & $0.09 \times 0.03$ 
       & 110  & 6 & $2.1 \times 10^5$ 		& -- &\\ 
N16 & J203716.9+421631  & 210 & $0.12 \times 0.12$ 
       & 550  & 29 & $7.5 \times 10^4$ 	& 3.87 & MSX 81.4452+0.7635\\ 
N17 & J203726.5+413319  & 100 & $0.15 \times 0.03$ 
       & 200  & 10 & $1.6 \times 10^5$ 	& -- &\\
N18 & J203730.2+421359  & 110 & $0.17 \times 0.13$ 
       & 430  & 23 & $2.7 \times 10^4$ 	& -- &\\ 
N19 & J203745.8+424343  & 80 & $0.09 \times 0.03$ 
       & 110  & 6 & $1.9 \times 10^5$ 		& -- &\\ 
N20 & J203747.4+423842  & 100 & $0.10 \times 0.03$ 
       & 150  & 8 & $2.1 \times 10^5$ 		& 2.06 & MSX 81.7947+0.9116\\ 
N21 & J203801.7+423933  & 100 & $0.16 \times 0.14$ 
       & 380  & 20 & $2.5 \times 10^4$ 	& -- &\\ 
N22 & J203803.4+423956  & 90 & $0.27 \times 0.11$ 
       & 420  & 22 & $1.9 \times 10^4$ 	& -- &\\ 
N23 & J203804.9+423317  & 80 & $0.22 \times 0.06$ 
       & 250  & 13 & $4.0 \times 10^4$ 	& -- &\\ 
N24 & J203805.8+423952  & 140 & $0.17 \times 0.17$ 
       & 650  & 34 & $3.0 \times 10^4$ 	& -- &\\  
N25 & J203810.4+423805  & 170 & $0.12 \times 0.11$ 
       & 450  & 24 & $6.5 \times 10^4$ 	& -- &\\ 
N26 & J203821.2+421131  & 100 & $0.20 \times 0.10$ 
       & 340  & 18 & $2.6 \times 10^4$ 	& -- &\\ 
N27 & J203831.6+423804  & 90 & $0.10 \times 0.10$ 
       & 200  & 11 & $4.1 \times 10^4$ 	& -- &\\ 
N28 & J203833.7+423944  & 140 & $0.19 \times 0.11$ 
       & 510  & 27 & $3.9 \times 10^4$ 	& -- &\\ 
N29 & J203834.6+420619  & 110 & $0.25 \times 0.13$ 
       & 540  & 29 & $2.0 \times 10^4$ 	& -- &\\ 
N30 & J203836.6+423730  & 3650 & $0.15 \times 0.10$ 
       & 10640  & $<$361-563 & $<$0.8-1.2$\times 10^6$ 	& 136 & {\tiny W75N(B), cm free-free$^3$,}\\
       & & & & & & & & {\tiny maser$^4$ (CH$_3$OH/H$_2$O/OH)}\\ 
N31 & J203836.9+423750  & 120 & $0.03 \times 0.03$ 
       & 130  & 7 & $1.2 \times 10^6$ 		& -- & close to \hii region W75N(A)\\ 
N32 & J203838.3+423733  & 540 & $0.16 \times 0.10$ 
       & 1630  & 86 & $1.7 \times 10^5$ 	& -- &\\ 
N33 & J203847.0+423800  & 100 & $0.08 \times 0.07$ 
       & 170  & 9 & $8.1 \times 10^4$ 		& -- &\\ 
N34 & J203854.3+421917  & 150 & $0.19 \times 0.07$ 
       & 430  & 23 & $6.1 \times 10^4$ 	& -- &\\ 
N35 & J203854.4+421855  & 90 & $0.13 \times 0.03$ 
       & 160  & 8 & $1.5 \times 10^5$ 		& -- &\\ 
N36 & J203858.5+422330  & 90 & $0.19 \times 0.03$ 
       & 210  & 11 & $1.2 \times 10^5$ 	& -- &\\ 
N37 & J203858.6+422435  & 80 & $0.03 \times 0.03$ 
       & 80  & 4 & $8.7 \times 10^5$ 		& -- &\\ 
N38 & J203859.4+422223  & 830 & $0.18 \times 0.09$ 
       & 2610  & 138 & $2.7 \times 10^5$ 	& -- & DR21(OH)-W\\ 
N39 & J203859.9+422732  & 70 & $0.16 \times 0.08$ 
       & 200  & 10 & $3.0 \times 10^4$ 	& -- &\\ 
N40 & J203859.8+422343  & 400 & $0.33 \times 0.08$ 
       & 2000  & 106 & $1.1 \times 10^5$ 	& -- & DR21(OH)-N2\\ 
N41 & J203900.2+422325  & 180 & $0.03 \times 0.03$ 
       & 180  & 10 & $1.9 \times 10^6$ 	& -- & DR21(OH)-N1\\ 
N42 & J203900.3+421907  & 90 & $0.05 \times 0.03$ 
       & 100  & 5 & $4.5 \times 10^5$ 		& -- &\\ 
N43 & J203900.6+422435  & 370 & $0.19 \times 0.10$ 
       & 1250  & 66 & $1.1 \times 10^5$ 	& 15.9 & W75S-FIR1, maser$^{5,6}$ (H$_2$O/OH)\\ 
N44 & J203901.0+422246  & 2970 & $0.14 \times 0.11$ 
       & 8430  & 446 & $1.0 \times 10^6$ 	& 6.78 & DR21(OH), \\
       & & & & & & & & maser$^{2,6}$ (CH$_3$OH/H$_2$O/OH)\\
N45 & J203901.1+421814  & 130 & $0.13 \times 0.08$ 
       & 300  & 16 & $6.4 \times 10^4$ 	& Abs &\\ 
N46 & J203901.4+421934  & 4200 & $0.19 \times 0.14$ 
       & 17930  & $<$609-949 & $<$5.9-9.1$\times 10^5$ 	& 272 & DR21, cm$^7$ free-free,\\
       & & & & & & & & maser$^8$ (H$_2$O)\\ 
N47 & J203901.4+421954  & 630 & $0.08 \times 0.03$ 
       & 840  & $<$28-44 & $<$1.2-1.9$\times 10^6$ 	& 272 & DR21-D, cm$^7$ free-free\\ 
N48 & J203901.5+422203  & 1160 & $0.17 \times 0.11$ 
       & 3720  & 197 & $3.5 \times 10^5$ 	& -- & DR21(OH)-S\\ 
N49 & J203902.0+422700  & 80 & $0.15 \times 0.03$ 
       & 160  & 8 & $1.2 \times 10^5$ 		& -- &\\ 
N50 & J203901.9+421835  & 80 & $0.24 \times 0.03$ 
       & 240  & 12 & $1.0 \times 10^5$ 	& -- &\\ 
N51 & J203902.4+422459  & 680 & $0.19 \times 0.10$ 
       & 2370  & 125 & $2.0 \times 10^5$ 	& 26.50 & W75S-FIR2\\ 
N52 & J203903.0+422615  & 180 & $0.16 \times 0.12$ 
       & 630  & 33 & $4.9 \times 10^4$ 	& -- &\\ 
N53 & J203903.2+422549  & 590 & $0.14 \times 0.09$ 
       & 1600  & 85 & $2.2 \times 10^5$ 	& -- & maser$^2$ (CH$_3$OH/H$_2$O)\\ 
N54 & J203903.6+422530  & 150 & $0.18 \times 0.08$ 
       & 450  & 24 & $6.0 \times 10^4$ 	& 10.85 & W75S-FIR3\\ 
N55 & J203906.1+421819  & 100 & $0.15 \times 0.05$ 
       & 220  & 11 & $8.0 \times 10^4$ 	& Abs &\\ 
N56 & J203916.9+421607  & 270 & $0.12 \times 0.08$ 
       & 600  & 32 & $1.4 \times 10^5$ 	& 3.68 & MSX 81.6632+0.4651\\ 
N57 & J203919.3+421556  & 130 & $0.09 \times 0.05$ 
       & 210  & 11 & $1.5 \times 10^5$ 	& Abs & ERO1\\ 
\hline
\end{tabular}
\end{table*}

\setcounter{table}{0}
\begin{table*}[htbp]
\caption[]{Continued}
\centering
\begin{tabular}{rcrcrrccl}
\hline
& Fragment & \speak$^{\rm a}$ & {\it FWHM}~$^{\rm a}$ & \sint$^{\rm a}$ & \msmm$^{\rm b}$ & $<n\htwo>$$^{\rm c}$ & MSX$^{\rm d}$ & Name, comments, and coincidence\\
& name & (mJy/ & (pc~$\times$~pc) & (mJy) & (\msun) & (\cmc) & 21~$\mu$m & with stellar activity signatures\\
&		& beam)	&		&		&		&	& (Jy) &\\
\hline
N58 & J203925.9+412001  & 430 & $0.23 \times 0.19$ 
       & 2730  & $<$92-144 & $<$4.2-6.6$\times 10^4$ 	& 634.13 & {\tiny IRAS~20375+4109, cm$^9$ free-free}\\ 
N59 & J203931.2+412003  & 200 & $0.12 \times 0.05$ 
       & 370  & 20 & $2.0 \times 10^5$ 	& --  &\\ 
N60 & J203936.2+411937  & 160 & $0.16 \times 0.13$ 
       & 560  & 30 & $4.2 \times 10^4$ 	& -- &\\ 
N61 & J203957.6+415911  & 100 & $0.12 \times 0.09$ 
       & 230  & 12 & $4.7 \times 10^4$ 	& 11.63 & MSX 81.5168+0.1926\\
N62 & J204003.6+412756  & 140 & $0.18 \times 0.12$ 
       & 500  & 27 & $3.6 \times 10^4$ 	& -- & {\tiny close to MSX 81.1109-0.1459}\\
       & & & & & & & & {\tiny and MSX 81.1225-0.1343}\\ 
N63 & J204005.2+413213  & 740 & $0.08 \times 0.05$ 
       & 1090  & 58 & $1.1 \times 10^6$ 	& Abs &\\ 
N64 & J204027.2+415651  & 100 & $0.12 \times 0.09$ 
       & 220  & 12 & $4.6 \times 10^4$ 	& Abs &\\ 
N65 & J204028.4+415711  & 230 & $0.16 \times 0.12$ 
       & 780  & 41 & $6.4 \times 10^4$ 	& 2.0 & \\
N66 & J204029.8+414933  & 110 & $0.15 \times 0.06$ 
       & 250  & 13 & $6.1 \times 10^4$ 	& Abs &\\ 
N67 & J204031.0+414957  & 100 & $0.08 \times 0.05$ 
       & 150  & 8 & $1.4 \times 10^5$ 		& Abs &\\ 
N68 & J204033.5+415903  & 390 & $0.10 \times 0.10$ 
       & 840  & 44 & $2.0 \times 10^5$ 	& Abs &\\ 
N69 & J204033.7+415059  & 110 & $0.47 \times 0.18$ 
       & 1280  & 68 & $1.2 \times 10^4$ 	& -- &\\ 
N70 & J204034.3+413845  & 80 & $0.12 \times 0.07$ 
       & 170  & 9 & $5.0 \times 10^4$ 		& Abs &\\ 
N71 & J204038.2+415259  & 80 & $0.21 \times 0.06$ 
       & 230  & 12 & $3.4 \times 10^4$ 	& -- &\\ 
N72 & J204045.7+415750  & 100 & $0.28 \times 0.13$ 
       & 540  & 29 & $1.7 \times 10^4$ 	& -- &\\ 
\hline\hline
& CygX-South region & & & & & & &\\  
\hline
S1 & J201643.3+392313  & 90 & $0.08 \times 0.05$ 
       & 140  & 8 & $1.1 \times 10^5$ 		& 3.08 & IRAS~20149+3913\\
S2 & J201648.5+392212  & 100 & $0.17 \times 0.06$ 
       & 260  & 14 & $5.4 \times 10^4$ 	& Abs &\\ 
S3 & J201658.6+392106  & 210 & $0.12 \times 0.08$ 
       & 440  & 23 & $1.2 \times 10^5$ 	& 5.27 & IRAS~20151+3911\\ 
S4 & J201737.0+392139  & 90 & $0.14 \times 0.10$ 
       & 230  & 12 & $3.1 \times 10^4$ 	& Abs &\\ 
S5 & J201745.3+392041  & 120 & $0.16 \times 0.09$ 
       & 340  & 18 & $4.1 \times 10^4$ 	& -- &\\ 
S6 & J202036.5+393821  & 120 & $0.03 \times 0.03$ 
       & 120  & 6 & $1.2 \times 10^6$ 		& -- &\\ 
S7 & J202038.0+393820  & 280 & $0.22 \times 0.03$ 
       & 760  & 40 & $3.5 \times 10^5$ 	& -- & \\	
S8 & J202038.8+393751  & 900 & $0.19 \times 0.12$ 
       & 3540  & $<$121-188 & $<$1.4-2.1$\times 10^5$ 	& 145.48 & {\tiny Mol121S, cm$^{10}$ free-free,}\\
       & & & & & & & & {\tiny maser$^{11}$ (H$_2$O)}\\ 
S9 & J202039.5+393738  & 120 & $0.03 \times 0.03$ 
       & 130  & 7 & $1.1 \times 10^6$ 		& -- &\\ 
S10 & J202044.4+393520  & 180 & $0.19 \times 0.14$ 
       & 760  & 40 & $4.1 \times 10^4$ 	& -- &\\ 
S11 & J202153.7+395933  & 110 & $0.15 \times 0.11$ 
       & 340  & 18 & $3.6 \times 10^4$ 	& -- &\\ 
S12 & J202220.1+395821  & 130 & $0.16 \times 0.08$ 
       & 350  & 18 & $5.5 \times 10^4$ 	& -- & close to IRAS~20205+3948\\
S13 & J202632.4+395721  & 160 & $0.15 \times 0.11$ 
       & 500  & 27 & $5.0 \times 10^4$ 	& 4.84 & MSX 78.3762+1.0191\\ 
S14 & J202634.8+395725  & 200 & $0.09 \times 0.05$ 
       & 330  & 17 & $2.4 \times 10^5$ 	& Abs &\\
S15 & J202714.0+372258  & 120 & $0.29 \times 0.12$ 	
       & 630  & 34 & $2.2 \times 10^4$ 	& Abs &\\ 
S16 & J202717.2+372218  & 260 & $0.11 \times 0.05$ 
       & 460  & 24 & $3.0 \times 10^5$ 	& -- &\\ 
S17 & J202718.7+372227  & 130 & $0.07 \times 0.06$ 
       & 210  & 11 & $1.4 \times 10^5$ 	& -- &\\ 
S18 & J202726.0+372246  & 960 & $0.21 \times 0.13$ 
       & 4270  & $<$145-226 & $<$1.2-2.0$\times 10^5$ 	& 413.63 & {\tiny S106-IR, cm$^{12}$ free-free,}\\
       & & & & & & & & {\tiny maser$^8$ (H$_2$O)}\\ 
S19 & J202726.9+372233  & 310 & $0.08 \times 0.03$ 
       & 420  & $<$14-22 & $<$6.1-9.5$\times 10^5$ 	& 413.63 & {\tiny part of S106-IR, cm$^{12}$ free-free}\\ 
S20 & J202727.4+372256  & 510 & $0.11 \times 0.11$ 
       & 1300  & $<$44-69 & $<$1.2-2.0$\times 10^5$ 	& 413.63& {\tiny part of S106-IR, cm$^1$ free-free}\\ 
S21 & J202729.0+372258  & 120 & $0.03 \times 0.03$ 
       & 130  & 7 & $1.2 \times 10^6$ 		& -- & \\ 
S22 & J202731.7+372334  & 80 & $0.13 \times 0.03$ 
       & 150  & 8 & $1.5 \times 10^5$ 		& -- &\\ 
S23 & J202731.8+372154  & 110 & $0.24 \times 0.03$ 
       & 310  & 16 & $1.2 \times 10^5$ 	& -- &\\ 
S24 & J202734.7+372311  & 140 & $0.18 \times 0.09$		
       & 420  & 22 & $4.9 \times 10^4$ 	& -- &\\ 
S25 & J202923.6+401110  & 120 & $0.05 \times 0.03$ 
       & 140  & 7 & $6.1 \times 10^5$ 		& -- &\\ 
S26 & J202924.8+401118  & 880 & $0.14 \times 0.10$ 
       & 2340  & $<$80-124 & $<$2.1-3.4$\times 10^5$ 	& 1023.40 & {\tiny AFGL~2591, cm$^{13}$ free-free,}\\
        & & & & & & & & {\tiny maser$^8$ (H$_2$O/OH?)}\\ 
S27 & J202931.8+390113  & 130 & $0.19 \times 0.09$ 
       & 420  & $<$14-22 & $<$2.7-4.2$\times 10^4$ 	& 23.55 & MSX 77.9550+0.0058,\\
       & & & & & & & & cm$^9$ free-free\\ 
S28 & J202936.8+390112  & 120 & $0.12 \times 0.08$ 
       & 270  & $<9-14$ & $<3.6-5.6 \times 10^4$ 	& 204.20 & {\tiny IRAS~20277+3851, cm$^1$ free-free}\\
S29 & J202958.3+401558  & 160 & $0.23 \times 0.12$ 
       & 710  & 37 & $3.4 \times 10^4$ 	& -- &\\ 
S30 & J203112.6+400316  & 360 & $0.16 \times 0.16$ 
       & 1560  & 82 & $7.7 \times 10^4$ 	& -- & close to IRAS~20293+3952\\  
S31 & J203114.2+400305  & 110 & $0.13 \times 0.07$ 
       & 220  & 12 & $6.5 \times 10^4$ 	& -- &\\ 
S32 & J203120.3+385716  & 250 & $0.21 \times 0.11$ 
       & 1010  & 54 & $6.0 \times 10^4$ 	& -- &\\ 
S33 & J203144.5+385639  & 80 & $0.14 \times 0.11$ 
       & 230  & 12 & $2.7 \times 10^4$ 	& -- &\\ 
S34 & J203157.8+401830  & 150 & $0.20 \times 0.12$ 
       & 610  & 32 & $3.4 \times 10^4$ 	& -- & in IRDC G79.3-P3\\ 
S35 & J203158.8+385836  & 200 & $0.12 \times 0.06$ 
       & 410  & 22 & $1.4 \times 10^5$ 	& 5.00 & \\ 
S36 & J203220.8+401949  & 80 & $0.07 \times 0.03$ 
       & 100  & 5 & $2.5 \times 10^5$ 		& -- & in IRDC G79.34+0.33\\ 
S37 & J203222.0+402012  & 240 & $0.18 \times 0.12$ 
       & 860  & 45 & $6.4 \times 10^4$ 	& 4.44 & {\tiny IRAS~20305+4010}\\
       & & & & & & & & {\tiny in IRDC G79.34+0.33}\\ 
S38 & J203222.4+401919  & 90 & $0.23 \times 0.12$ 
       & 410  & 22 & $2.0 \times 10^4$ 	& -- & in IRDC G79.34+0.33\\
S39 & J203229.0+401600  & 170 & $0.14 \times 0.09$ 
       & 440  & $<$15-23 & $<$4.2-6.6$\times 10^4$ 	& 37.71 & {\tiny MSX 79.2963+0.2835,}\\ 
       & & & & & & & & {\tiny cm$^1$ free-free}\\
S40 & J203232.2+401625  & 80 & $0.16 \times 0.10$ 
       & 240  & $<$8-13 & $<$1.6-2.6$\times 10^4$ 	& 73.03 & {\tiny MSX 79.3070+0.2768, cm$^9$ free-free}\\ 
\hline
\end{tabular}
\end{table*}
\setcounter{table}{0}
\begin{table*}[htbp]
\caption[]{Continued}
{\centering
\begin{tabular}{rcrcrrccl}
\hline
& Fragment & \speak$^{\rm a}$ & {\it FWHM}~$^{\rm a}$ & \sint$^{\rm a}$ & \msmm$^{\rm b}$ & $<n\htwo>$$^{\rm c}$ & MSX$^{\rm d}$ & Name, comments, and coincidence\\
& name & (mJy/ & (pc~$\times$~pc) & (mJy) & (\msun) & (\cmc) & 21~$\mu$m & with stellar activity signatures\\
&		& beam)	&		&		&		&	& (Jy) &\\
\hline\hline
S41 &J203233.9+401653  & 220 & $0.19 \times 0.11$ 
       & 800  & 27-42 & 3.7-5.9$\times 10^4$ 	& 49.4 & part of IRAS 20306+4005\\ 
S42 & J203239.9+384603  & 90 & $0.18 \times 0.13$ 
       & 340  & 18 & $2.0 \times 10^4$ 	& -- &\\ 
S43 & J203240.8+384631  & 140 & $0.23 \times 0.13$ 
       & 650  & 35 & $2.9 \times 10^4$ 	& Abs &\\ 
\hline\hline
& CygX-NW region & & & & & & &\\  
\hline
NW1 & J201938.8+405638  & 280 & $0.17 \times 0.15$ 
       & 1110  & $<$38-59 & $<$4.2-6.6$\times 10^4$ & 400.66 &  IRAS 20178+4046, cm$^1$ free-free\\ 
NW2 & J201940.5+405704  & 200 & $0.23 \times 0.07$ 
       & 680  & 36 & $7.7 \times 10^4$ & -- &\\ 
NW3 & J202029.9+412144  & 80 & $0.15 \times 0.07$ 
       & 190  & 10 & $4.2 \times 10^4$ & -- &\\ 
NW4 & J202030.0+412207  & 130 & $0.07 \times 0.05$ 
       & 180  & 10 & $2.2 \times 10^5$ & -- &\\ 
NW5 & J202030.7+412125  & 420 & $0.07 \times 0.06$ 
       & 640  & 22-34 & 3.2-5.0$\times 10^5$ & 85.85 &  EM* LkHA 225S, maser$^{14}$ (H$_2$O)\\
NW6 & J202032.1+412354  & 130 & $0.15 \times 0.03$ 
       & 250  & 13 & $2.0 \times 10^5$ & -- &\\ 
NW7 & J202032.2+412123  & 80 & $0.07 \times 0.03$ 
       & 100  & 6 & $2.5 \times 10^5$ & -- &\\ 
NW8 & J202323.0+411749  & 90 & $0.10 \times 0.04$ 
       & 150  & 8 & $1.2 \times 10^5$ & -- &\\ 
NW9 & J202324.0+411739  & 230 & $0.10 \times 0.04$ 
       & 370  & 19 & $3.0 \times 10^5$ & 27.47 & IRAS 20216+4107\\ 
NW10 & J202343.6+411657  & 160 & $0.20 \times 0.08$ 
       & 520  & 27 & $5.7 \times 10^4$ & -- &\\ 
NW11 & J202413.0+411700  & 110 & $0.13 \times 0.09$ 
       & 290  & 15 & $4.7 \times 10^4$ & -- &\\ 
NW12 & J202414.3+421143  & 140 & $0.06 \times 0.06$ 
       & 200  & 11 & $2.1 \times 10^5$ & Abs &\\ 
NW13 & J202419.5+421545  & 100 & $0.17 \times 0.09$ 
       & 290  & 15 & $3.5 \times 10^4$ & -- &\\ 
NW14 & J202431.7+420423  & 420 & $0.12 \times 0.09$ 
       & 940  & 32-50 & 1.2-2.0$\times 10^5$ & 15.58 & {\tiny Mol124 = IRAS 20227+4154,}\\
       & & & & & & & & {\tiny maser$^{11}$ (H$_2$O)}\\ 
\hline
\end{tabular}}
\normalsize
Notes:
\begin{list}{}{}
\item[ (a) ] {Peak flux, deconvolved {\it FWHM} size, and integrated flux derived from a 2D-Gaussian fit to the 1.2~mm map after background subtraction (cf. Sect.~\ref{s:census1} and Appendix~B).  An upper limit {\it FWHM} of 0.03~pc was assumed for unresolved fragments.}
\item[ (b) ] {Mass derived from Col.~5 using Eq.~(\ref{eq:mass}) and assuming $\kmm=0.01~\cmg$ and $\tdust=20~\K$. A second estimate with $\tdust=40~\K$ is made for \hii regions and the brightest infrared protostars of Cygnus~X.}
\item[ (c) ]  {Volume-averaged density derived from Col.~4 and Col.~6 using Eq.~(\ref{eq:density}).}
\item[ (d) ]  {The 21~$\mu$m fluxes are those given in the \emph{MSX} point source catalog (MSX~C6), except for CygX-N30, N43, N46, N47, N65 and S41, for which an aperture flux is measured on the MSX map. The MSX flux of two extended \hii regions is equally distributed between CygX-N46 and N47, and between CygX-S18, S19, S20. All fragments detected at 21~$\mu$m are also detected at 8~$\mu$m with the exception of CygX-N30, N46, N47, and N65. ``Abs'' means that sources are seen in absorption at $8~\mu$m.}
\item[ (e) ]  {Observations of high-density tracers toward CygX-N4 suggest that this 1.2~mm source does not correspond to a dense cloud fragment. Its nature is yet unclear but it is probably not a site for future or ongoing star formation.}
\end{list}
References: (1) \cite{KCW94}; (2) \cite{PMB05}; (3) \cite{hasch81}; (4) \cite{MCB01}; (5) \cite{ARM00}; (6) \cite{harr73}; (8) \cite{BE83}; (9) \cite{zoot90}; (10) \cite{mol98b}; (11) \cite{pall91}; (12) \cite{piph76}; (13) \cite{trin03}; (14) \cite{pall95}.
\end{table*}
\normalsize

\begin{table*}[htbp]
\caption[]{Properties of clumps (large-scale cloud structures) detected in the Cygnus~X complex}
\label{t:clumps}
{\centering
\begin{tabular}{rccrrcl}
\hline
& Structure  & {\it FWHM}~$^{\rm a}$ & \sint~$^{\rm a}$ & \msmm$^{\rm b}$ & $<n\htwo>$$^{\rm c}$ & 
Comments \\
&  name & (pc~$\times$~pc) & (mJy) & (\msun) & (\cmc) & \\
\hline\hline
& CygX-North region &  & & & &\\ 
\hline
Cl-N1 & J203534.1+422004  & $0.48 \times 0.31$ 
       & 8470  & 665 & $5.0 \times 10^4$ &  contains CygX-N1, N2, N3\\ 
Cl-N2 & J203605.0+420851  & $1.46 \times 0.94$ 
       & 15150  & 1189 & $3.1 \times 10^3$ &  Starless?\\ 
Cl-N3 & J203607.3+414001  & $0.64 \times 0.38$ 
       & 7010  & 550 & $1.9 \times 10^4$ &  contains CygX-N5, N6\\ 
Cl-N4 & J203651.5+413625  & $0.66 \times 0.65$ 
       & 14780  & 1161 & $1.7 \times 10^4$ &  contains CygX-N10\\ 
Cl-N5 & J203657.9+421132  & $0.58 \times 0.49$ 
       & 11690  & 917 & $2.5 \times 10^4$ &  contains CygX-N12, N13\\ 
Cl-N6 &  J203701.9+413453  & $0.81 \times 0.64$ 
       & 13580  & 1066 & $1.2 \times 10^4$ &  contains CygX-N14\\ 
Cl-N7 &  J203711.9+413338  & $1.45 \times 0.90$ 
       & 16370  & 1285 & $3.6 \times 10^3$ &  contains CygX-N15\\ 
Cl-N8 &  J203728.4+421630  & $1.80 \times 1.01$ 
       & 17060  & 1339 & $2.3 \times 10^3$ &  \\ 
Cl-N9 &  J203729.8+421409  & $1.32 \times 0.73$ 
       & 14350  & 1127 & $4.9 \times 10^3$ &  contains CygX-N18\\ 
Cl-N10 &  J203805.3+423937  & $0.87 \times 0.51$ 
       & 10000  & 785 & $1.1 \times 10^4$ &  Starless? contains CygX-N21, N22, N24\\ 
Cl-N11 &  J203833.2+420619  & $1.25 \times 0.74$ 
       & 14880  & 1168 & $5.5 \times 10^3$ &  Starless? contains CygX-N29\\ 
Cl-N12 &  J203836.3+423937 & $1.30 \times 0.56$ 
       & 10240  & 804 & $5.5 \times 10^3$ &  contains CygX-N28\\ 
Cl-N13 &  J203836.8+423734  & $0.48 \times 0.40$ 
       & 46930  & 3684 & $1.9 \times 10^5$ &  contains CygX-N30, N31, N32\\ 
Cl-N14 &  J203900.5+422238  & $0.88 \times 0.44$ 
       & 92460  & 7259 & $1.3 \times 10^5$ &  contains CygX-N36, N38, N41, N44, N48\\ 
Cl-N15 &  J203901.3+421935  & $0.56 \times 0.39$ 
       & 70210  & 5513 & $2.3 \times 10^5$ &  contains CygX-N42, N46, N47\\ 
Cl-N16 &  J203901.7+422507  & $0.91 \times 0.60$ 
       & 47360  & 3718 & $3.8 \times 10^4$ &  contains CygX-N37, N43, N51, N53, N54\\ 
Cl-N17 &  J203927.2+412009 & $0.78 \times 0.59$ 
       & 15980  & 1254 & $1.7 \times 10^4$ &  contains CygX-N58, N59\\ 
Cl-N18 &  J203936.4+412001 & $0.91 \times 0.36$ 
       & 5710  & 449 & $1.0 \times 10^4$ &  Starless? contains CygX-N60\\ 
Cl-N19 &  J203955.4+420020  & $1.79 \times 1.10$ 
       & 18110  & 1422 & $2.2 \times 10^3$ &  contains CygX-N61\\ 
Cl-N20 &  J204003.6+412814  & $1.03 \times 0.77$ 
       & 14080  & 1105 & $6.6 \times 10^3$ &  contains CygX-N62\\ 
Cl-N21 &  J204005.1+413210  & $0.31 \times 0.20$ 
       & 3350  & 263 & $7.0 \times 10^4$ &  contains CygX-N63\\ 
Cl-N22 &  J204027.8+415707  & $0.68 \times 0.37$ 
       & 5200  & 408 & $1.4 \times 10^4$ &  contains CygX-N64, N65\\ 
Cl-N23 &  J204033.3+415045  & $0.80 \times 0.57$ 
       & 8700  & 683 & $9.4 \times 10^3$ &  Starless? contains CygX-N67, N69\\ 
\hline\hline
& CygX-South region &  & & & &\\ 
\hline
Cl-S1 &  J202038.5+393753  & $0.51 \times 0.34$ 
       & 13920  & 1093 & $6.3 \times 10^4$ & contains CygX-S6, S7, S8, S9\\ 
Cl-S2 &  J202044.8+393525  & $0.67 \times 0.45$ 
       & 4870  & 382 & $9.4 \times 10^3$ & contains CygX-S10\\	
Cl-S3 &  J202716.1+372231  & $1.23 \times 0.69$ 
       & 26640  & 2092 & $1.1 \times 10^4$ &  contains CygX-S15, S16, S17\\ 
Cl-S4 &  J202726.9+372248  & $0.73 \times 0.53$ 
       & 33420  & 2624 & $4.6 \times 10^4$ &  contains CygX-S18, S19, S20, S21\\ 
Cl-S5 &  J202924.9+401116  & $0.42 \times 0.38$ 
       & 10090  & 792 & $5.2 \times 10^4$ &  contains CygX-S25, S26 \\ 
Cl-S6 &  J202959.4+401555  & $1.25 \times 0.60$ 
       & 11560  & 908 & $5.9 \times 10^3$ &  Starless? contains CygX-S29\\ 
Cl-S7 &  J203112.1+400314  & $0.62 \times 0.49$ 
       & 10710  & 841 & $2.1 \times 10^4$ &  contains CygX-S30, S31\\ 
Cl-S8 &  J203119.2+385731  & $0.86 \times 0.56$ 
       & 8740  & 686 & $8.6 \times 10^3$ &  contains CygX-S32\\ 
Cl-S9 &  J203156.7+401833  & $1.39 \times 0.60$ 
       & 12140  & 953 & $5.2 \times 10^3$ &  Starless? IRDC G79.33P3, contains CygX-S34\\ 
Cl-S10 &  J203221.9+401955  & $0.92 \times 0.79$ 
       & 17490  & 1373 & $9.3 \times 10^3$ &  contains CygX-S36, S37, S38\\ 
Cl-S11 &  J203232.3+401635  & $0.99 \times 0.67$ 
       & 14010  & 1100 & $8.5 \times 10^3$ &  contains CygX-S39, S40, S41\\ 
Cl-S12 &  J203240.7+384614  & $0.67 \times 0.52$ 
       & 6480  & 509 & $1.0 \times 10^4$ &  Starless? contains CygX-S42, S43\\ 
\hline\hline
& CygX-NW region &  & & & &\\ 
\hline
Cl-NW1 &  J201939.0+405651  & $0.73 \times 0.64$ 
       & 14190  & 1114 & $1.5 \times 10^4$ &  contains CygX-NW1, NW2\\ 
Cl-NW2 &  J202030.5+412132  & $0.64 \times 0.41$ 
       & 6890  & 541 & $1.7 \times 10^4$ &  contains CygX-NW3, NW4, NW5, NW7\\ 
 Cl-NW3 &  J202344.5+411701  & $0.97 \times 0.71$ 
       & 11480  & 901 & $6.5 \times 10^3$ &  Starless?, contains CygX-NW10\\ 
Cl-NW4 &  J202420.3+421527  & $0.96 \times 0.66$ 
       & 8160  & 640 & $5.3 \times 10^3$ &  Starless?, contains CygX-NW13\\ 
Cl-NW5 &  J202431.7+420419  & $0.73 \times 0.60$ 
       & 8670  & 681 & $9.8 \times 10^3$ &  contains CygX-NW14\\ 
\hline
\end{tabular}}
\normalsize
Notes:
\begin{list}{}{}
\item[ (a) ] {Deconvolved {\it FWHM} size and integrated flux derived from a 2D-Gaussian fit to the 1.2~mm map smoothed to a $55\arcsec$ beam.}
\item[ (b) ] {Mass derived from Col.~4 using Eq.~(\ref{eq:mass}) and assuming $\kmm=0.01~\cmg$ and $\tdust=15~\K$.}
\item[ (c) ]   {Volume-averaged density derived from Col.~3 and Col.~5 using Eq.~(\ref{eq:density}).}
\end{list}
\end{table*}
\normalsize

\newpage
\setcounter{figure}{11}
\begin{figure*}
\centerline{\includegraphics[angle=0,width=6.8cm]{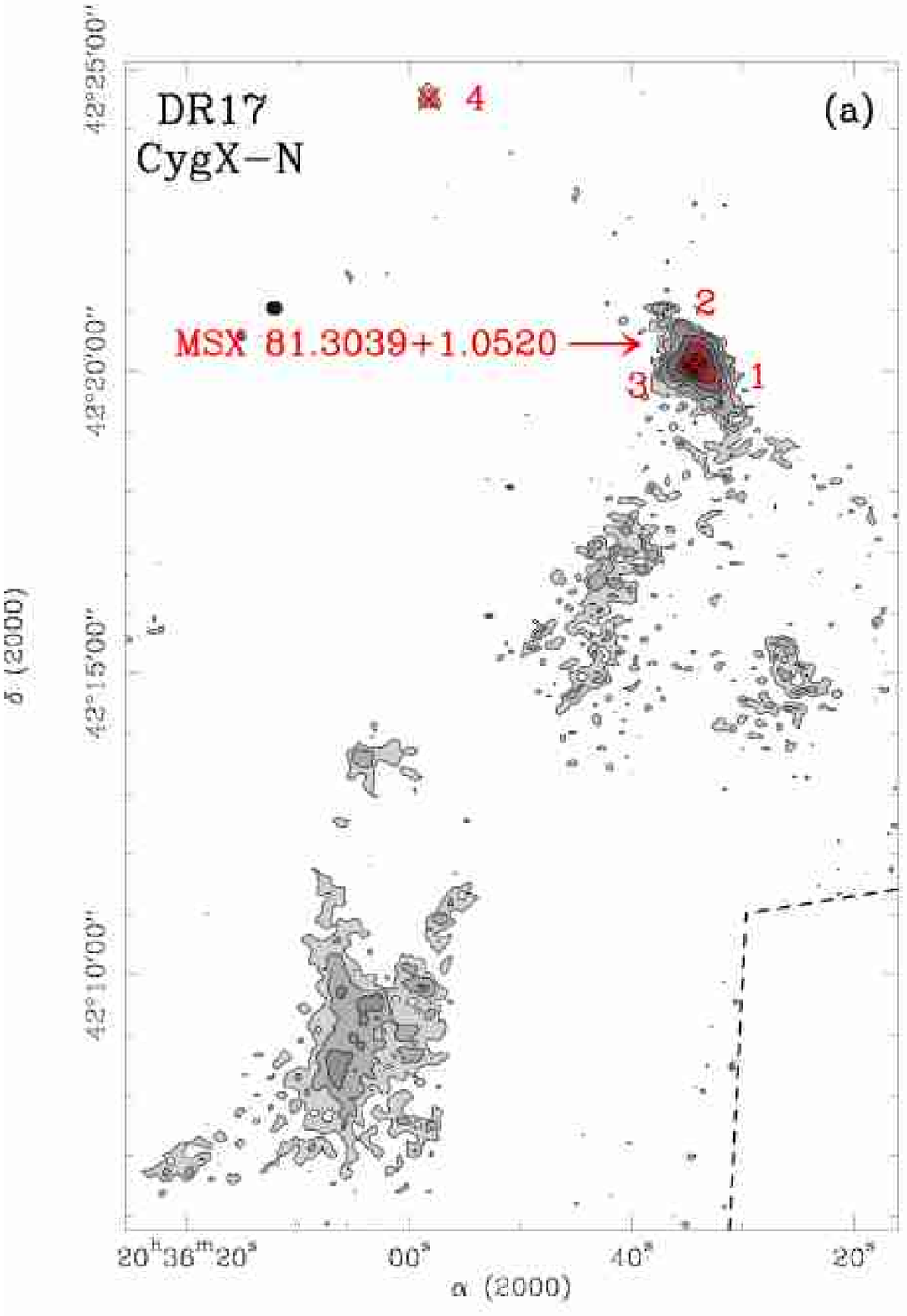} 
\hskip -1.2cm \includegraphics[angle=0,width=6.8cm]{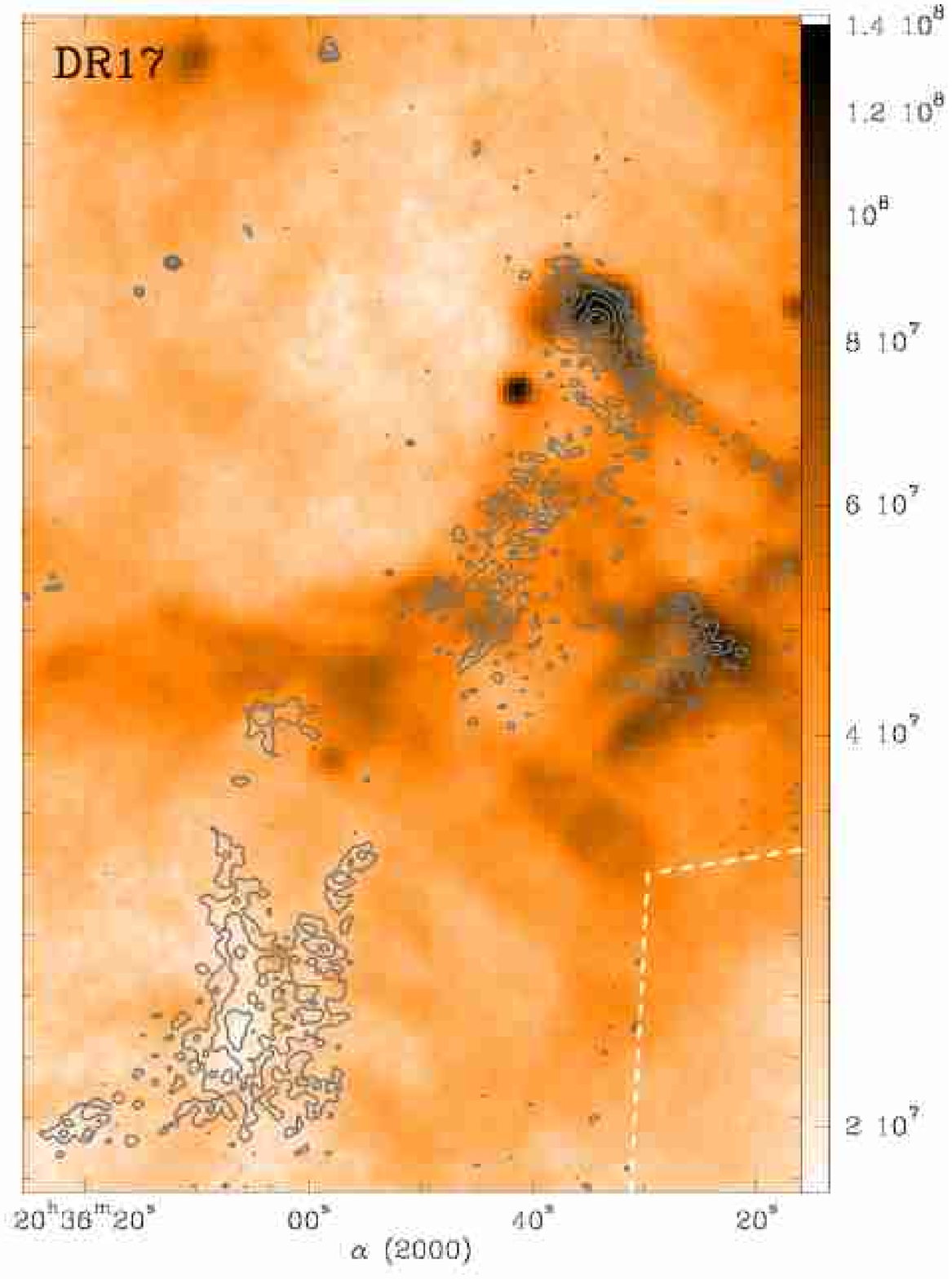}}
\vskip -0.8cm
\centerline{\includegraphics[angle=270,width=6.38cm]{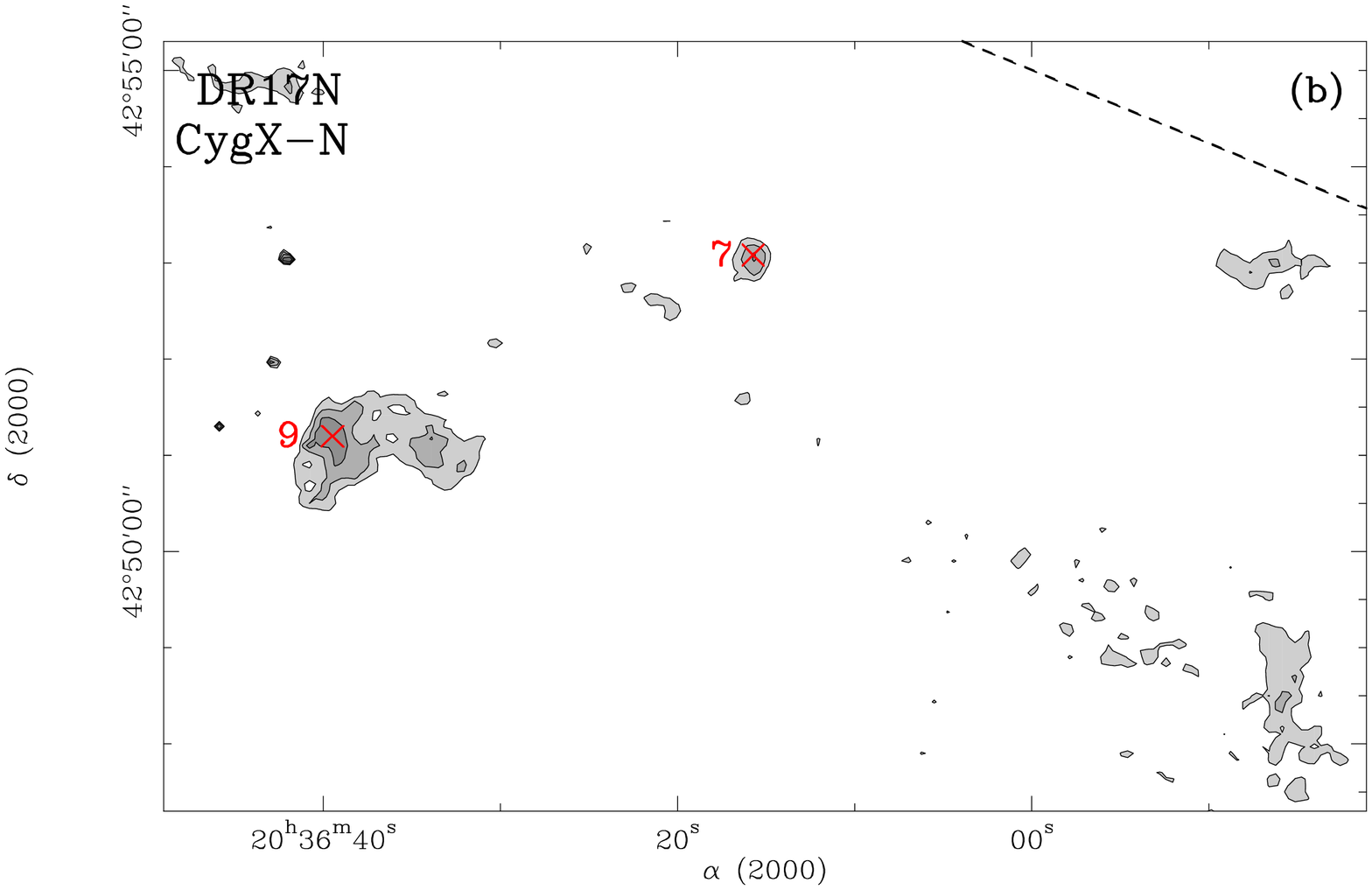}
\hskip -0.5cm \includegraphics[angle=270,width=6.38cm]{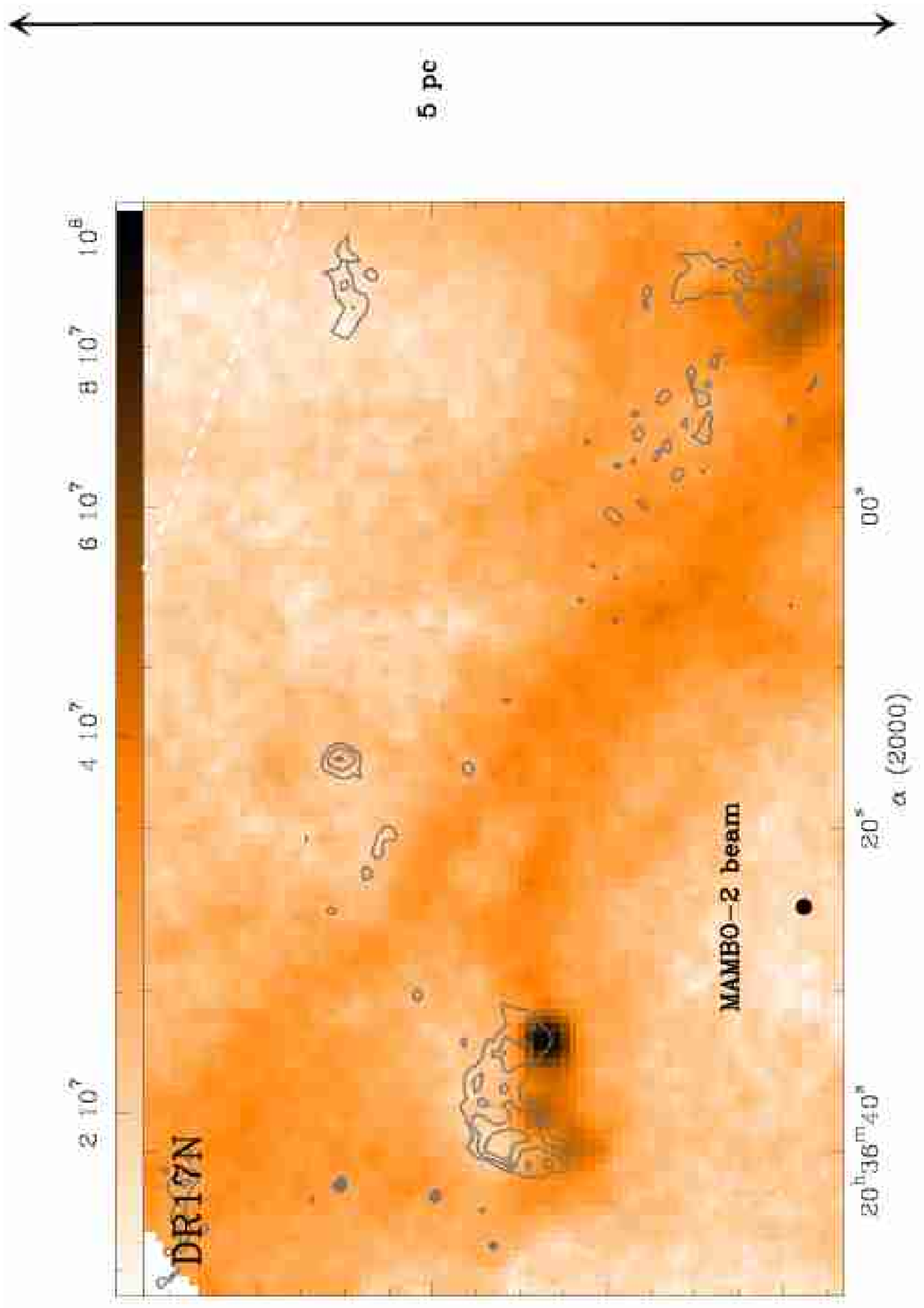}}
\vskip -0.4cm
\centerline{\includegraphics[angle=270,width=8.7cm]{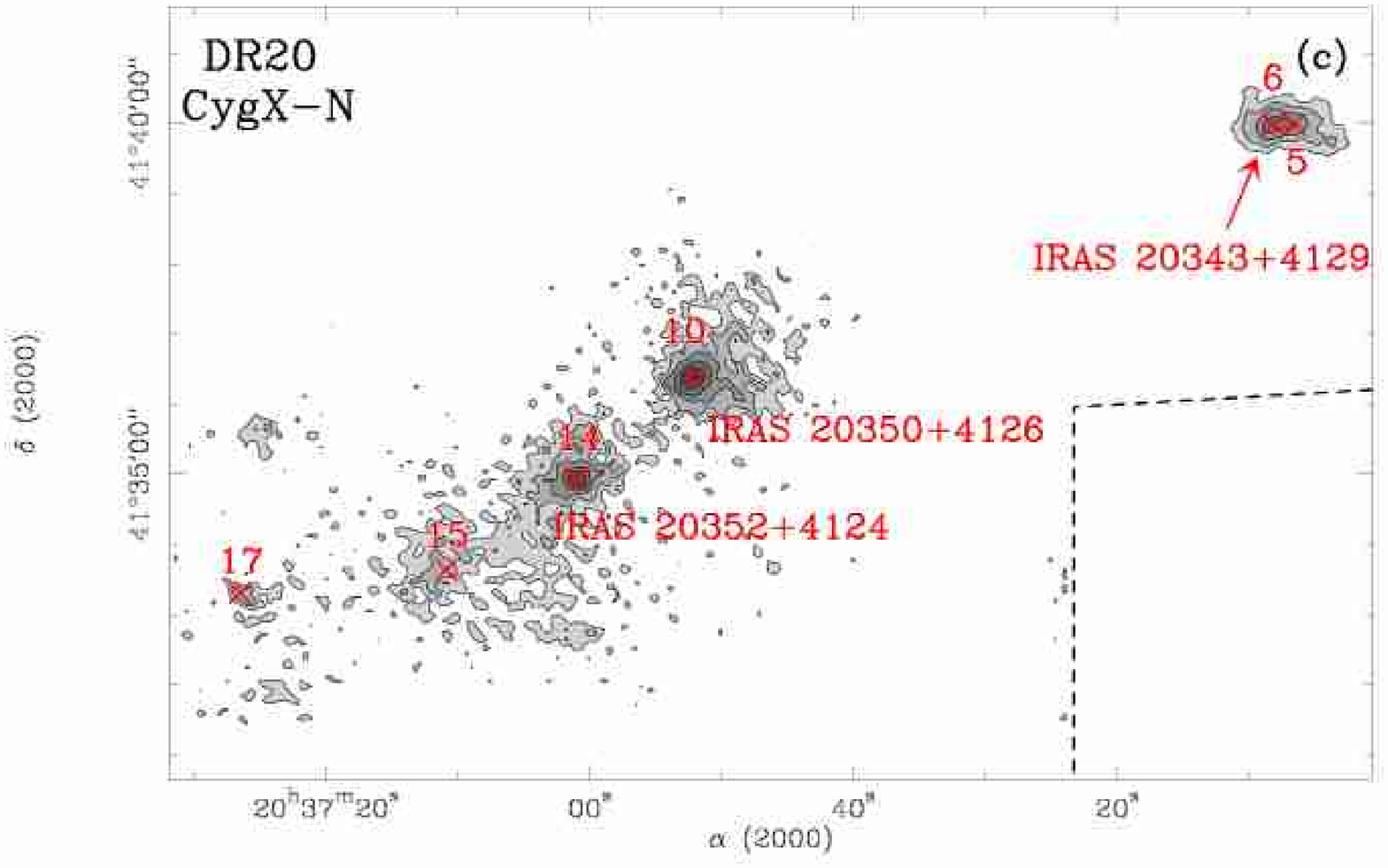}
\hskip -1.2cm \includegraphics[angle=270,width=8.7cm]{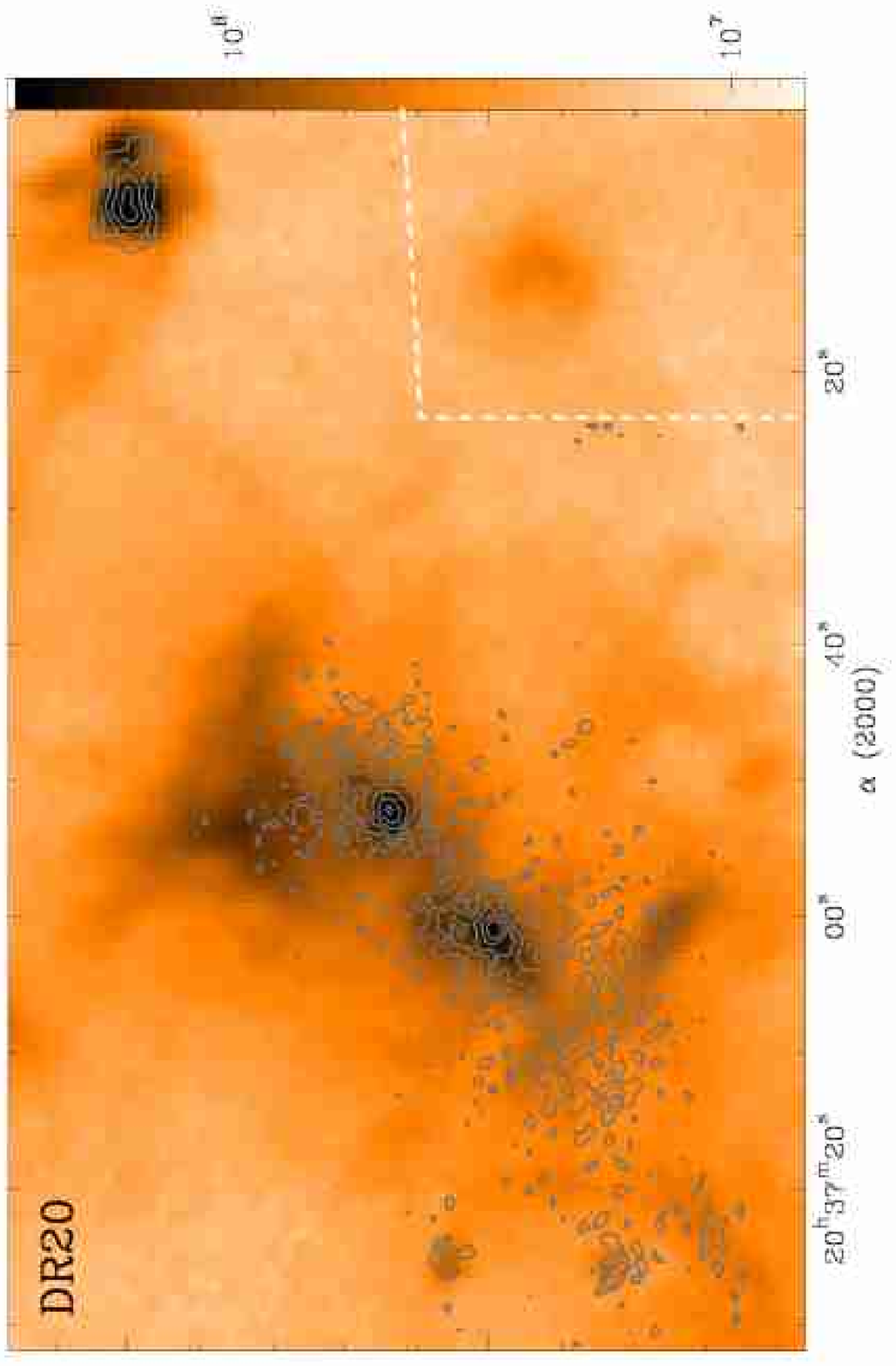}}
\caption[]
{MAMBO maps of CygX-North ({\bf left}: gray-scale and contours, {\bf right}: contours overlaid on 8~$\mu$m images obtained by \emph{MSX} and converted to Jy sr$^{-1}$) extracted from Fig.~\ref{f:mambo}a. Regions shown are south and north of DR17  ({\bf a--b}) and around DR20 ({\bf c}). The 1.2~mm and 8~$\mu$m images have $11\arcsec$ and $20\arcsec$ angular resolutions, respectively. The compact cloud fragments discovered in MAMBO images (see Table~\ref{t:densecores}) are labeled and marked by crosses in the gray-scale plot. The infrared sources which coincide with a MAMBO cloud fragment are also indicated. Contour levels are logarithmic and go from 40 to $800~\mjb$ in {\bf a} and {\bf b}, and from 75 to $800~\mjb$ in {\bf c}.}
\label{f:msxnorth_app}
\end{figure*} 
\setcounter{figure}{11}
\begin{figure*}
\centerline{\includegraphics[angle=270,width=8.4cm]{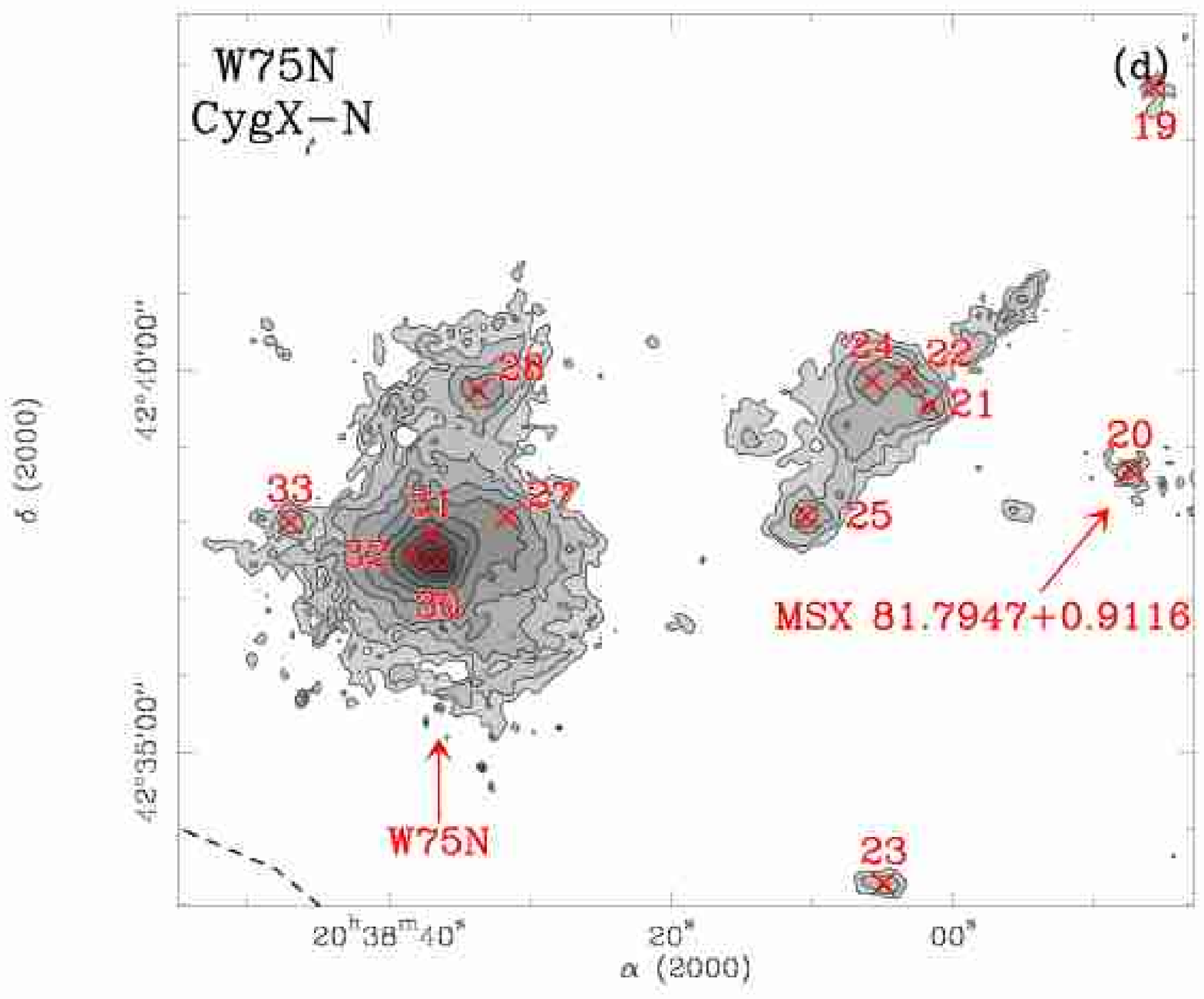}
\hskip -2.6cm \includegraphics[angle=270,width=8.4cm]{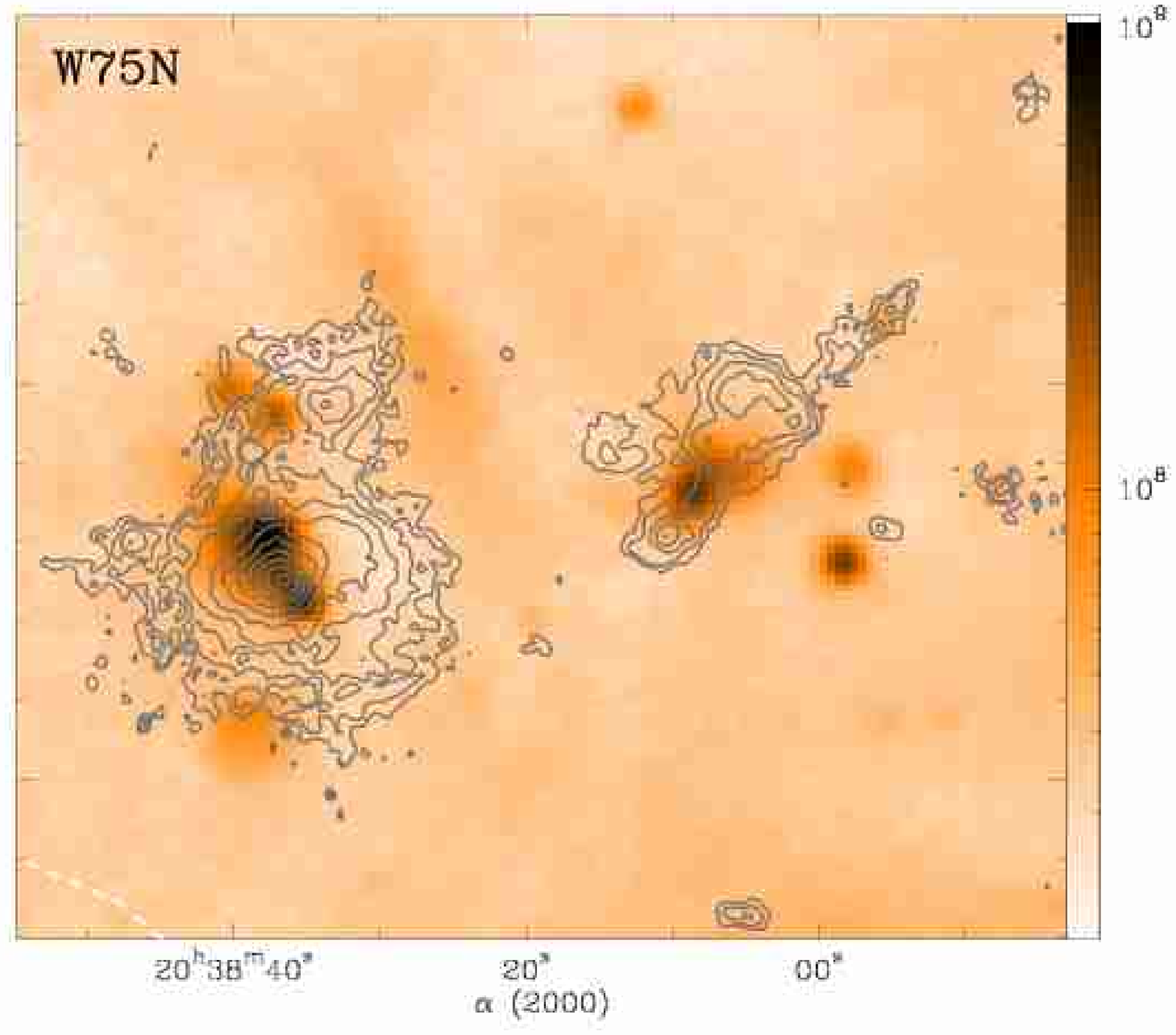}}
\vskip -0.2cm
\centerline{\includegraphics[angle=270,width=10.8cm]{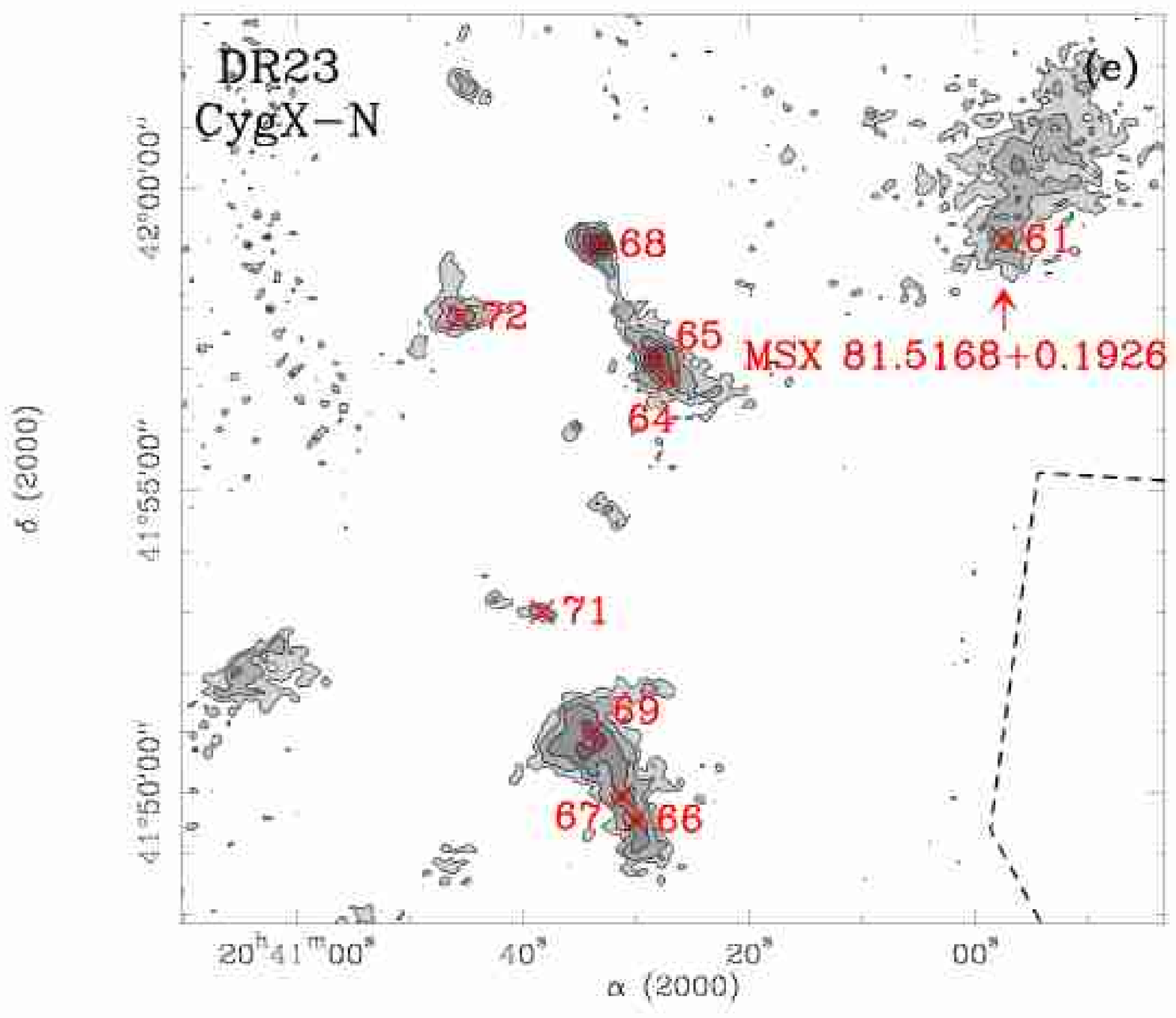}
\hskip -3.8cm \includegraphics[angle=270,width=10.8cm]{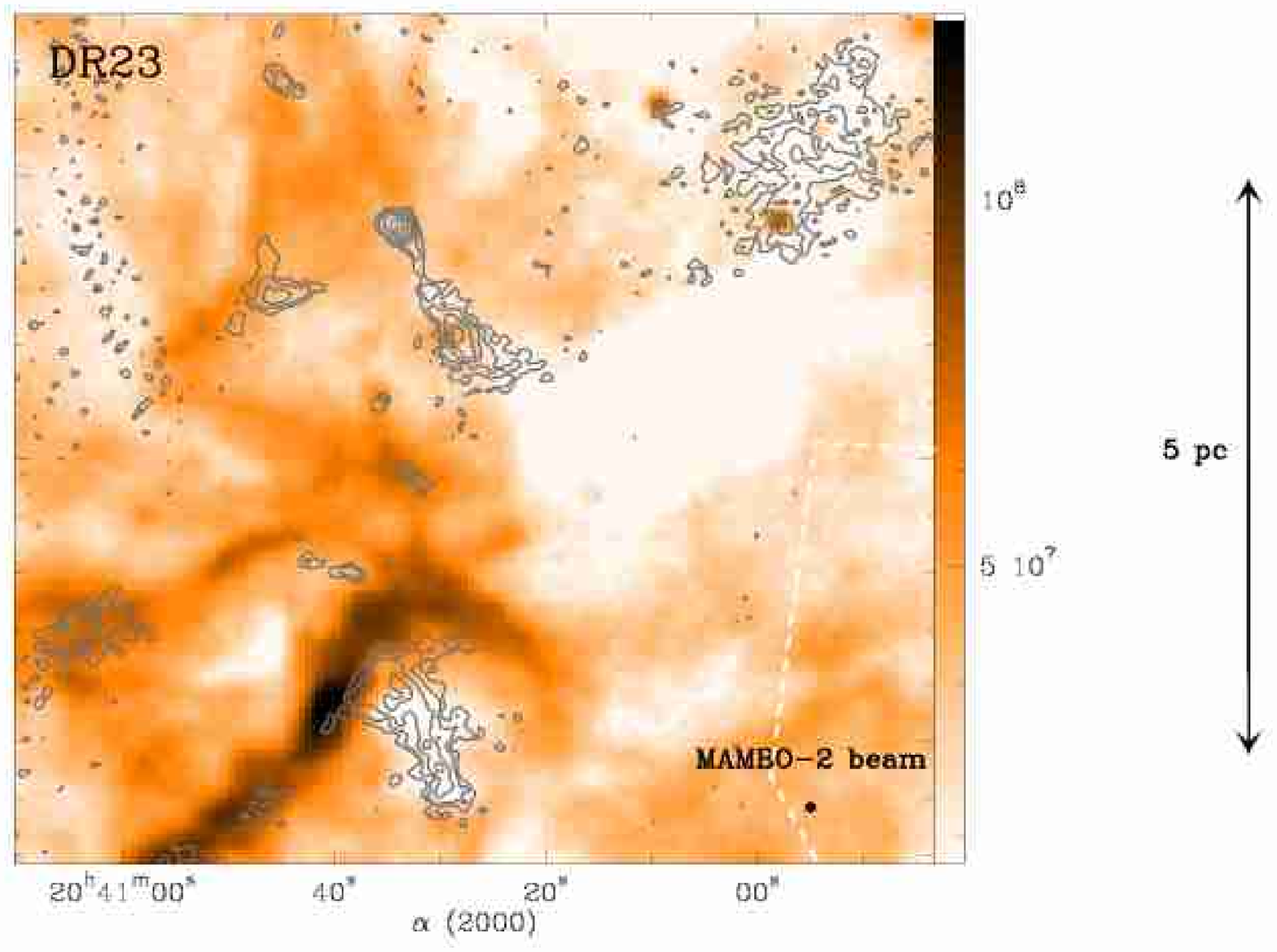}}
\vskip -0.4cm
\centerline{\includegraphics[angle=0,width=6.4cm]{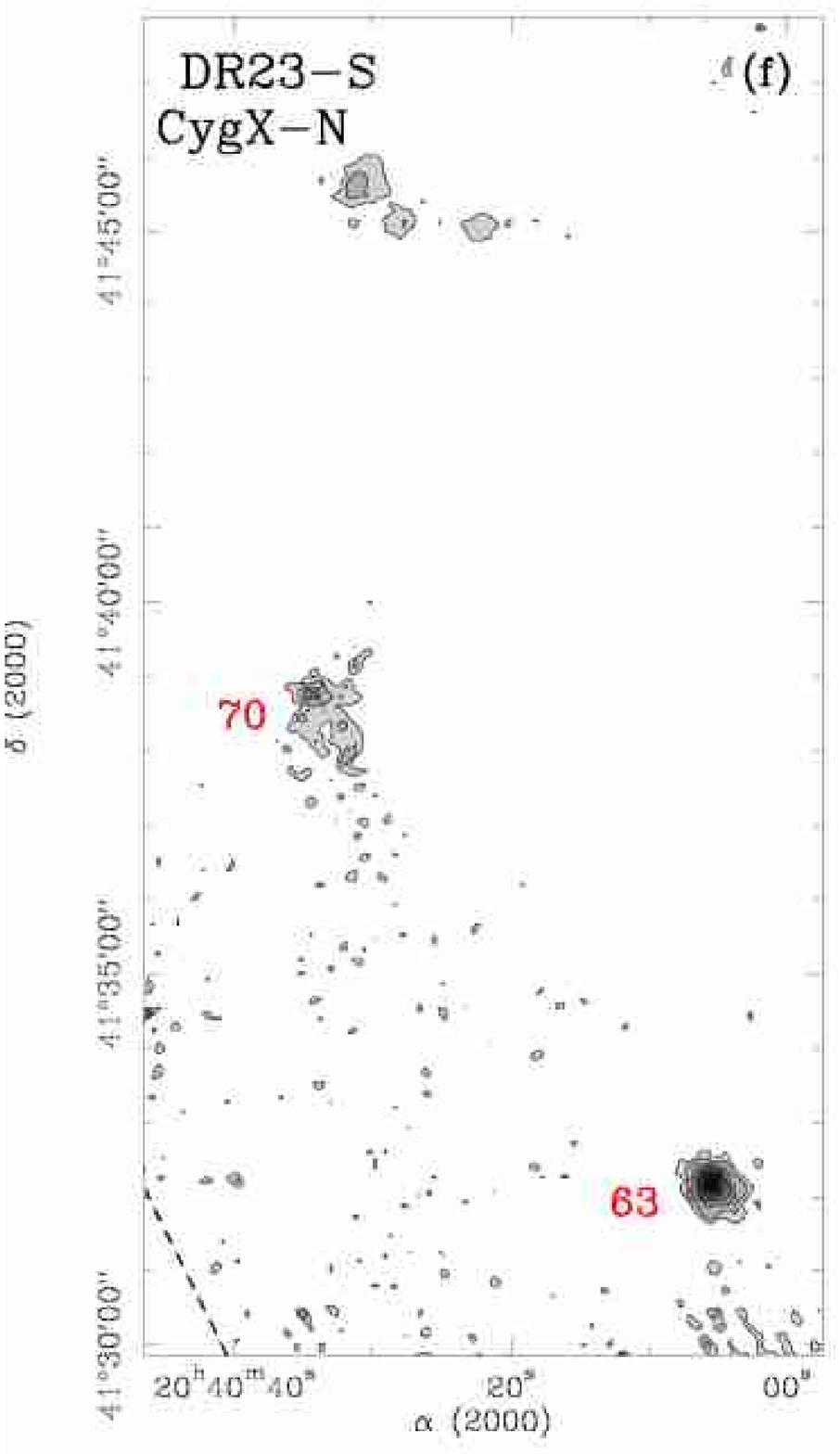}
\hskip -2.1cm \includegraphics[angle=0,width=6.4cm]{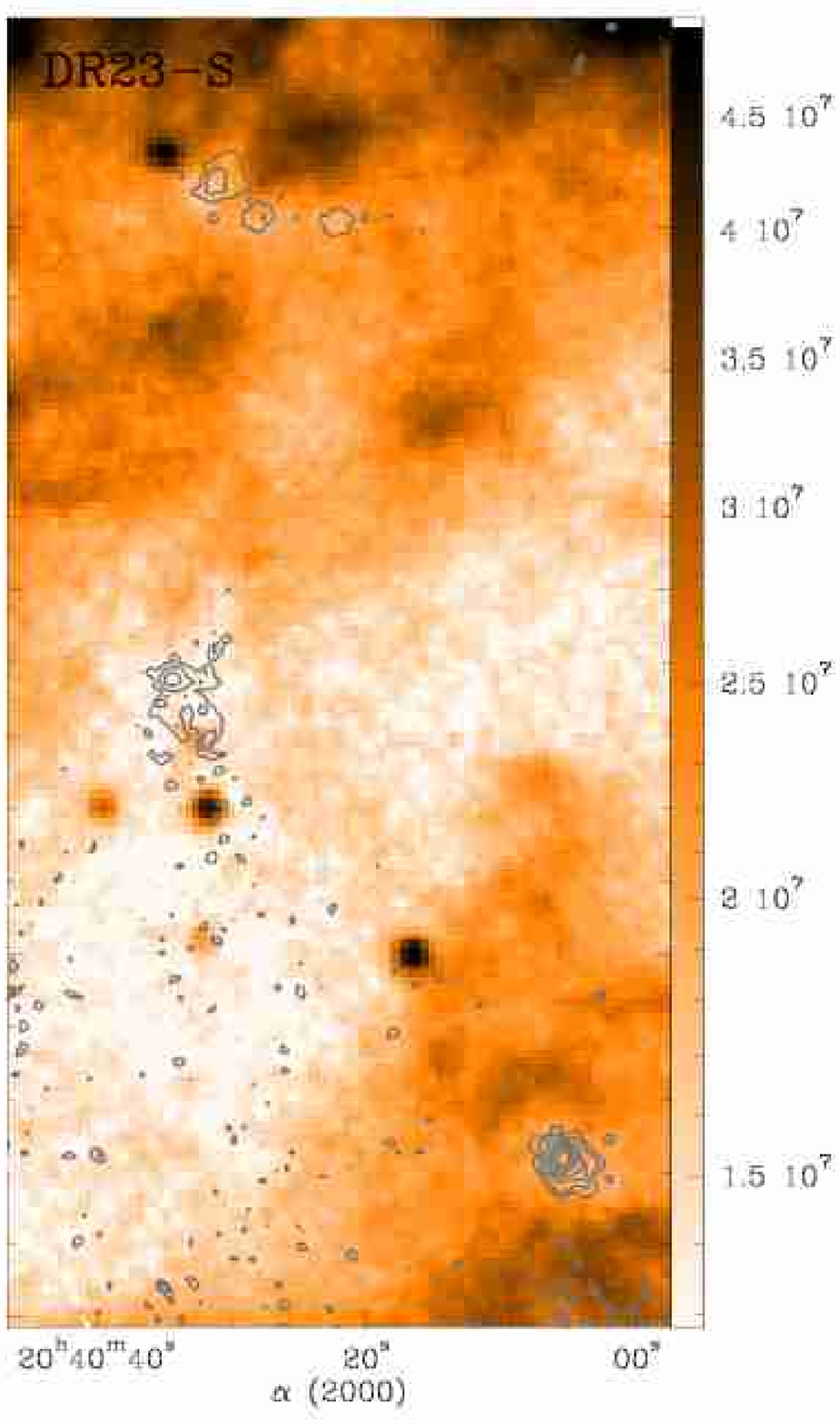}}
\vskip -0.4cm
\caption[]{(continued) Same caption for the region around W75N ({\bf d}) and DR23 ({\bf e}), and south of DR23 ({\bf f}). Contour levels are logarithmic and go from 40 to $4\,800~\mjb$ in {\bf d}, and from 40 to $800~\mjb$ in {\bf e} and {\bf f}.}
\end{figure*}

\setcounter{figure}{12}
\begin{figure*}
\centerline{\hskip -0.4cm \includegraphics[angle=270,width=8.7cm]{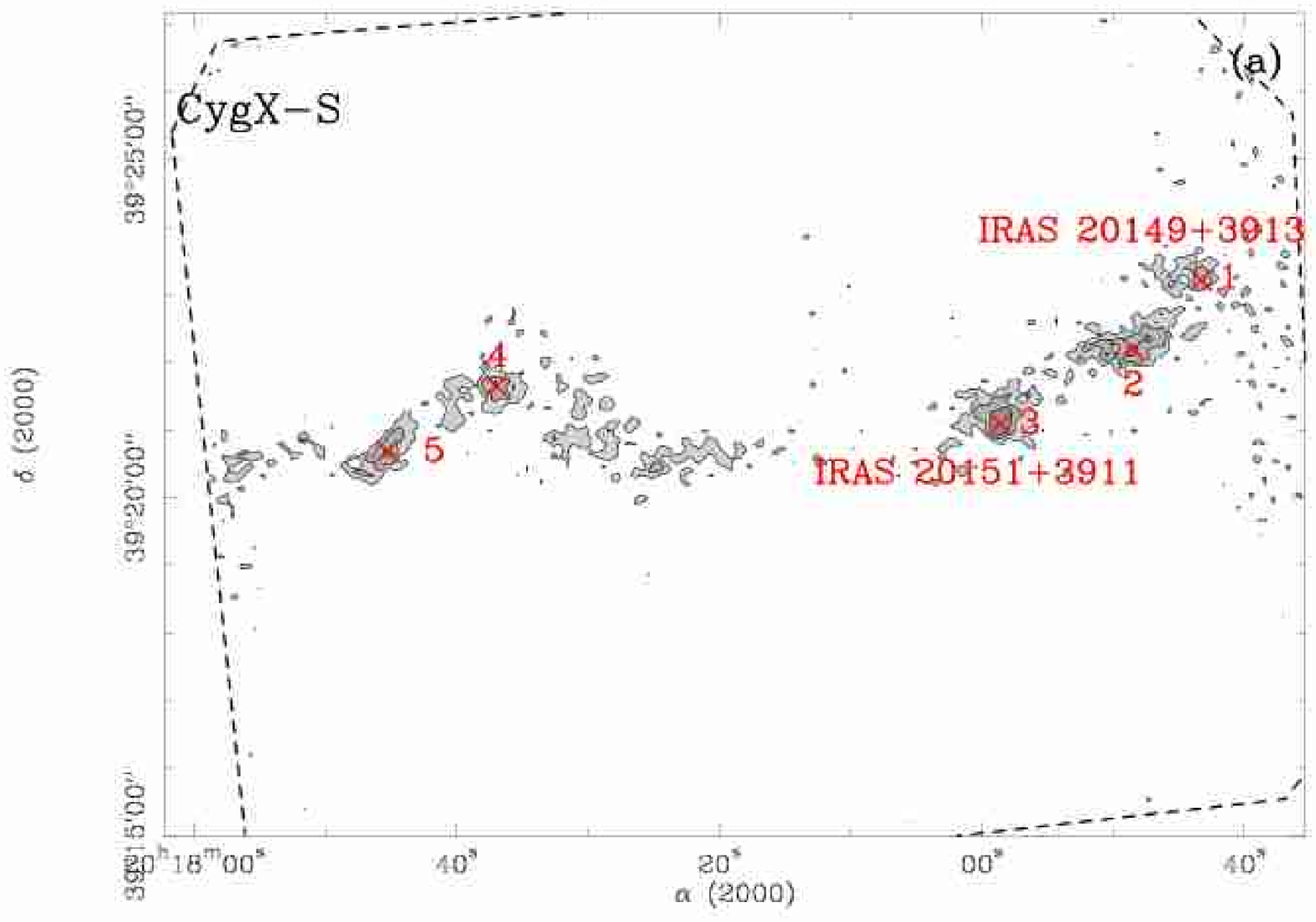} 
\hskip -1.5cm \includegraphics[angle=270,width=8.7cm]{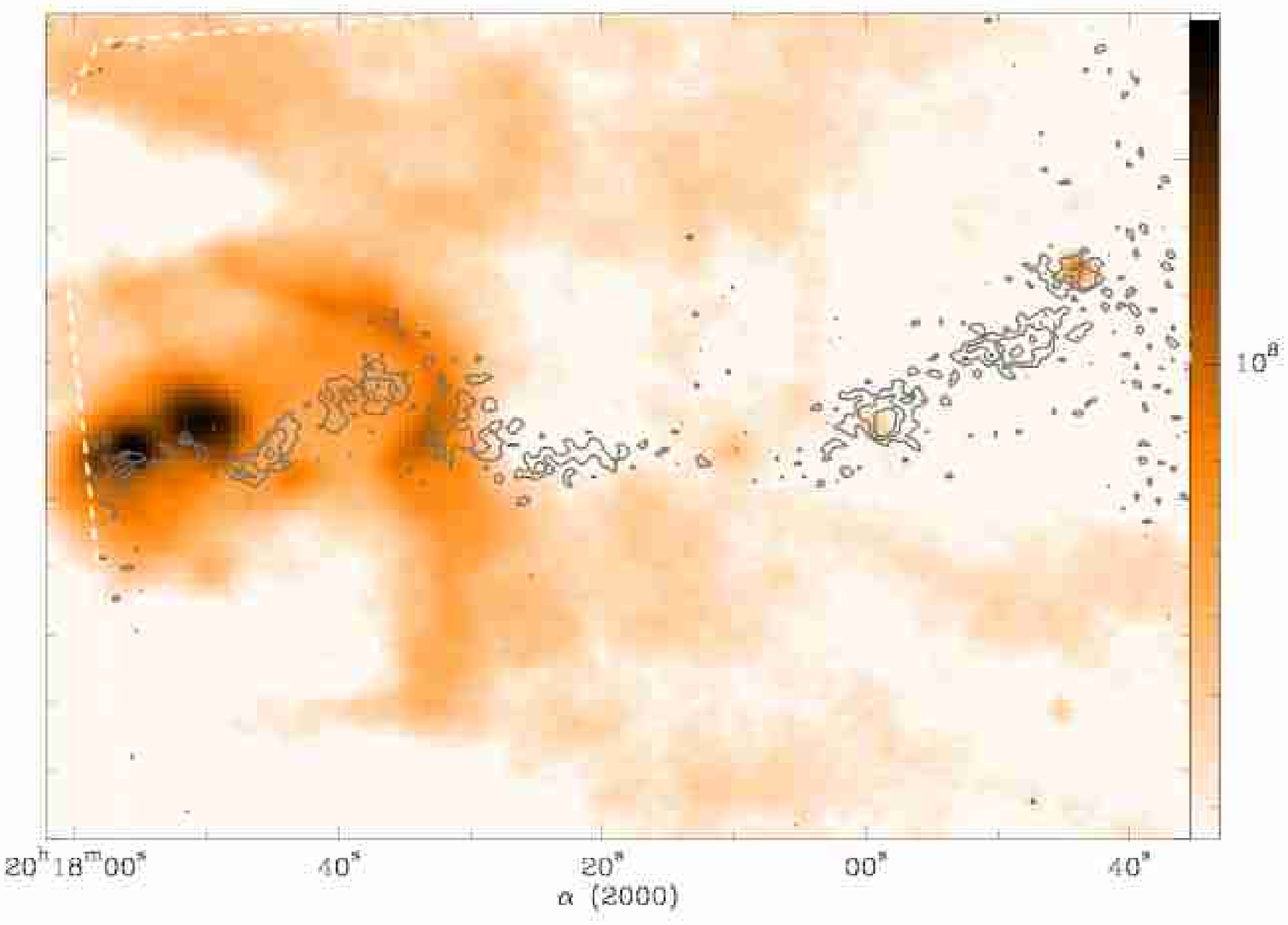}}
\vskip -0.1cm
\centerline{\includegraphics[angle=270,width=8.9cm]{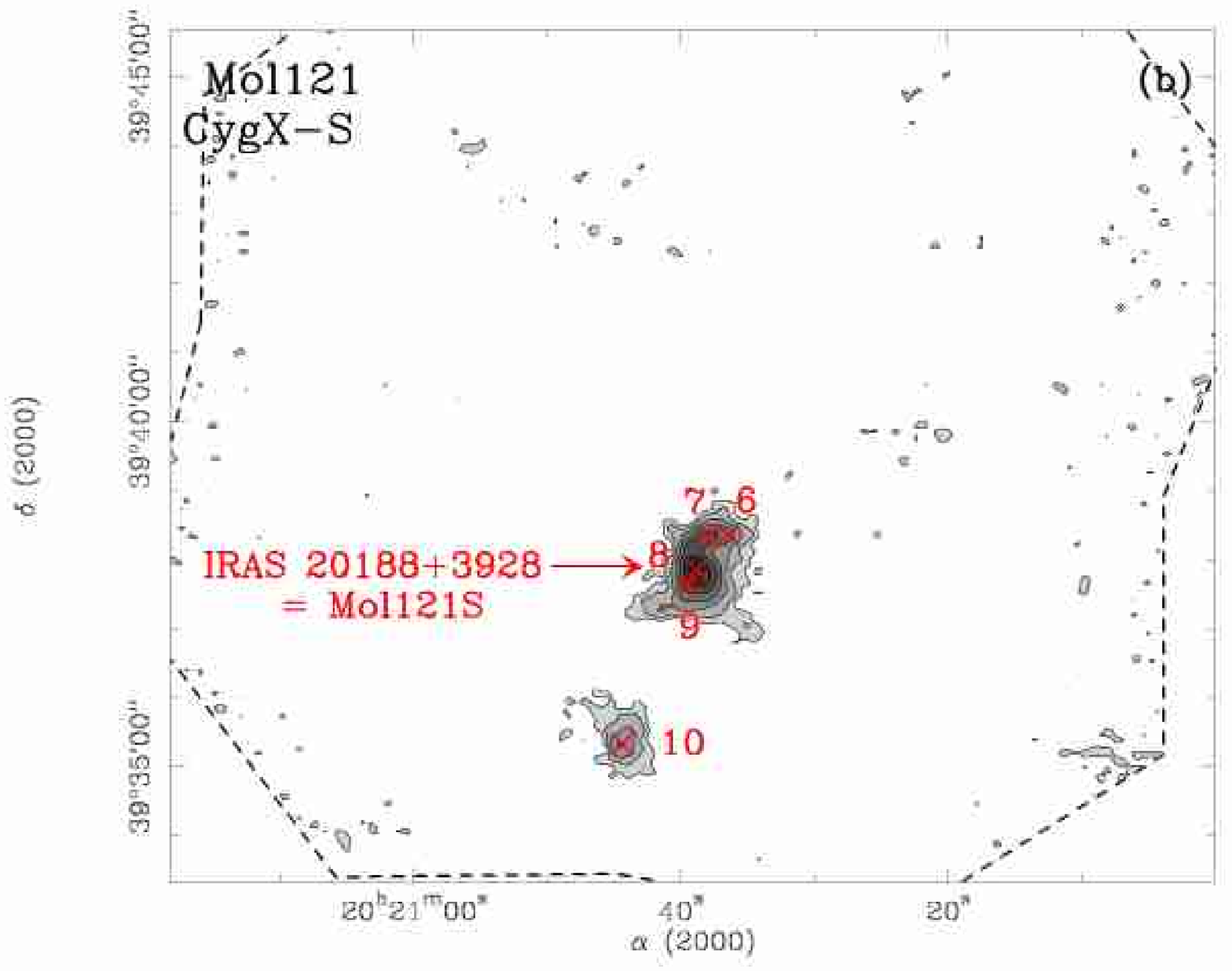}
\hskip -2.4cm \includegraphics[angle=270,width=8.9cm]{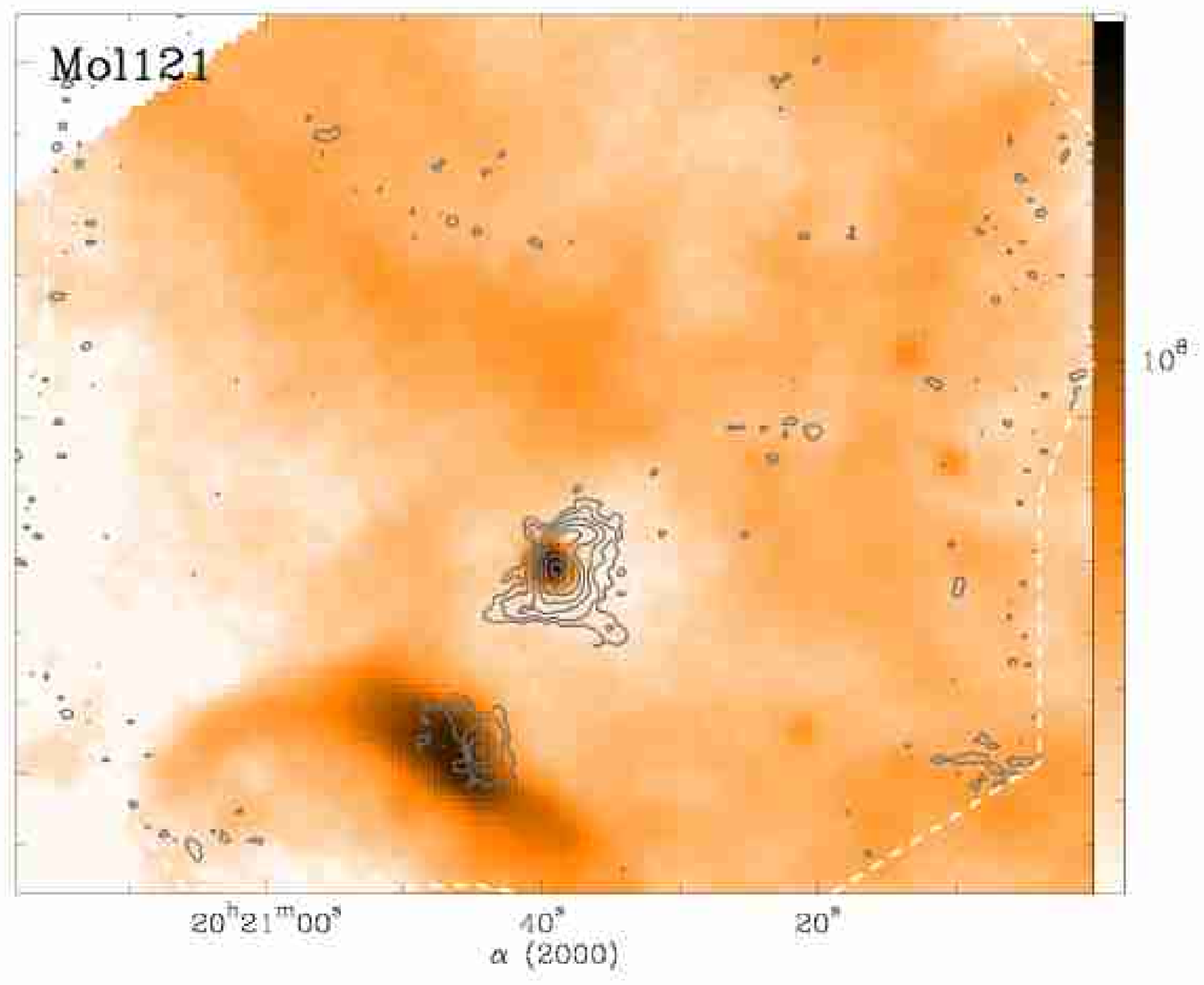}}
\vskip 0.cm
\centerline{\includegraphics[angle=270,width=7.7cm]{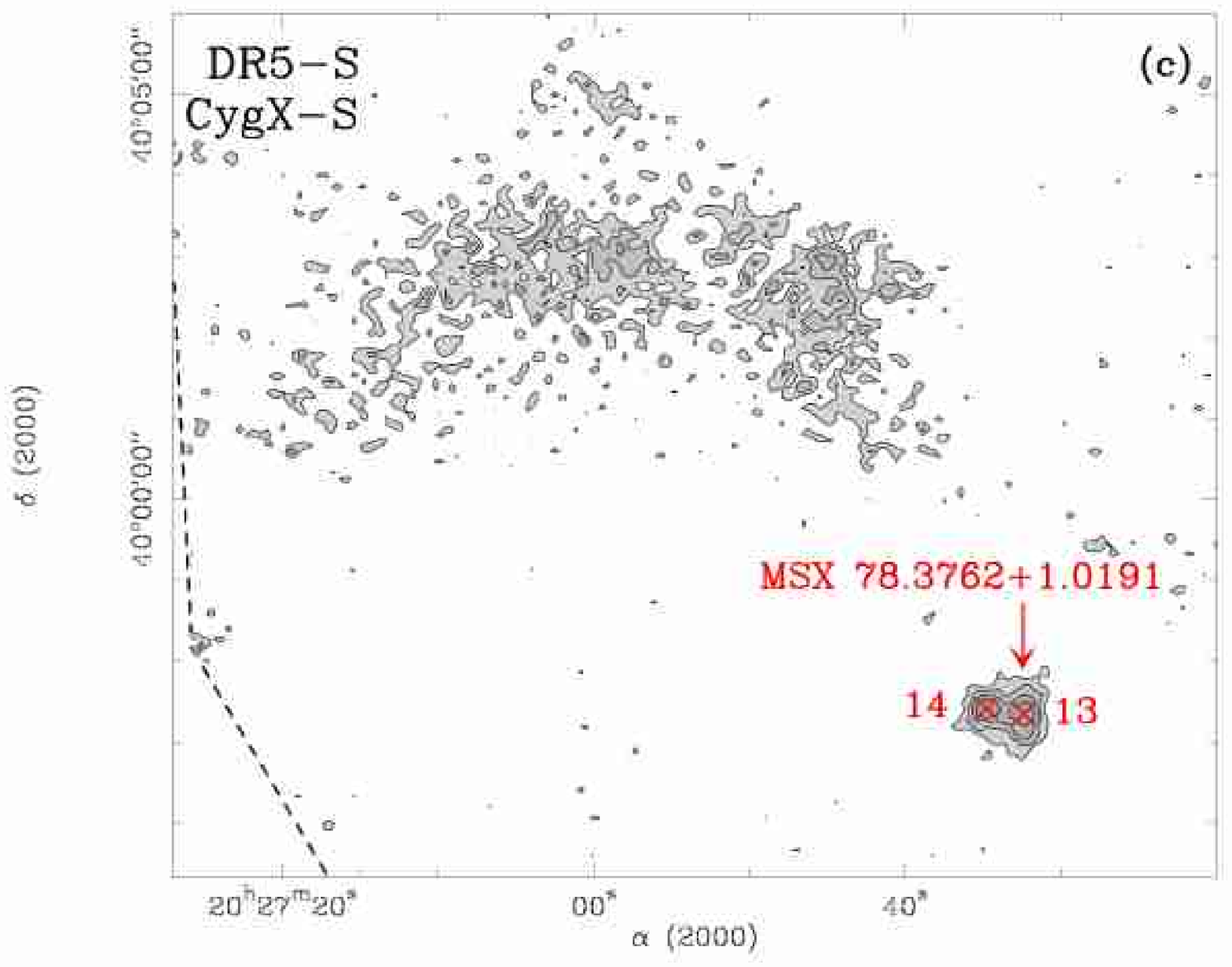}
\hskip -2.0cm \includegraphics[angle=270,width=7.7cm]{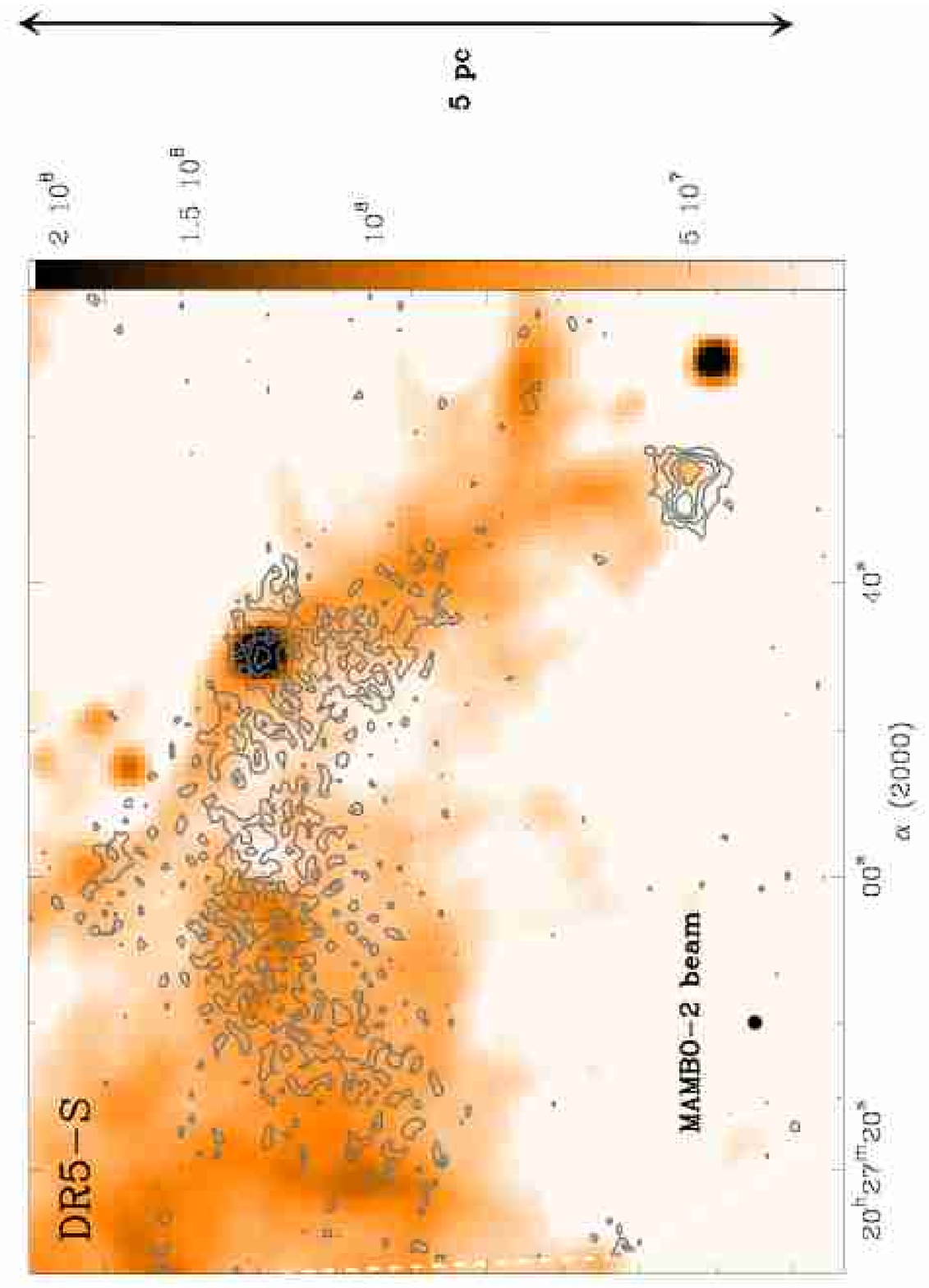}}
\vskip 0.1cm
\centerline{\includegraphics[angle=270,width=6.9cm]{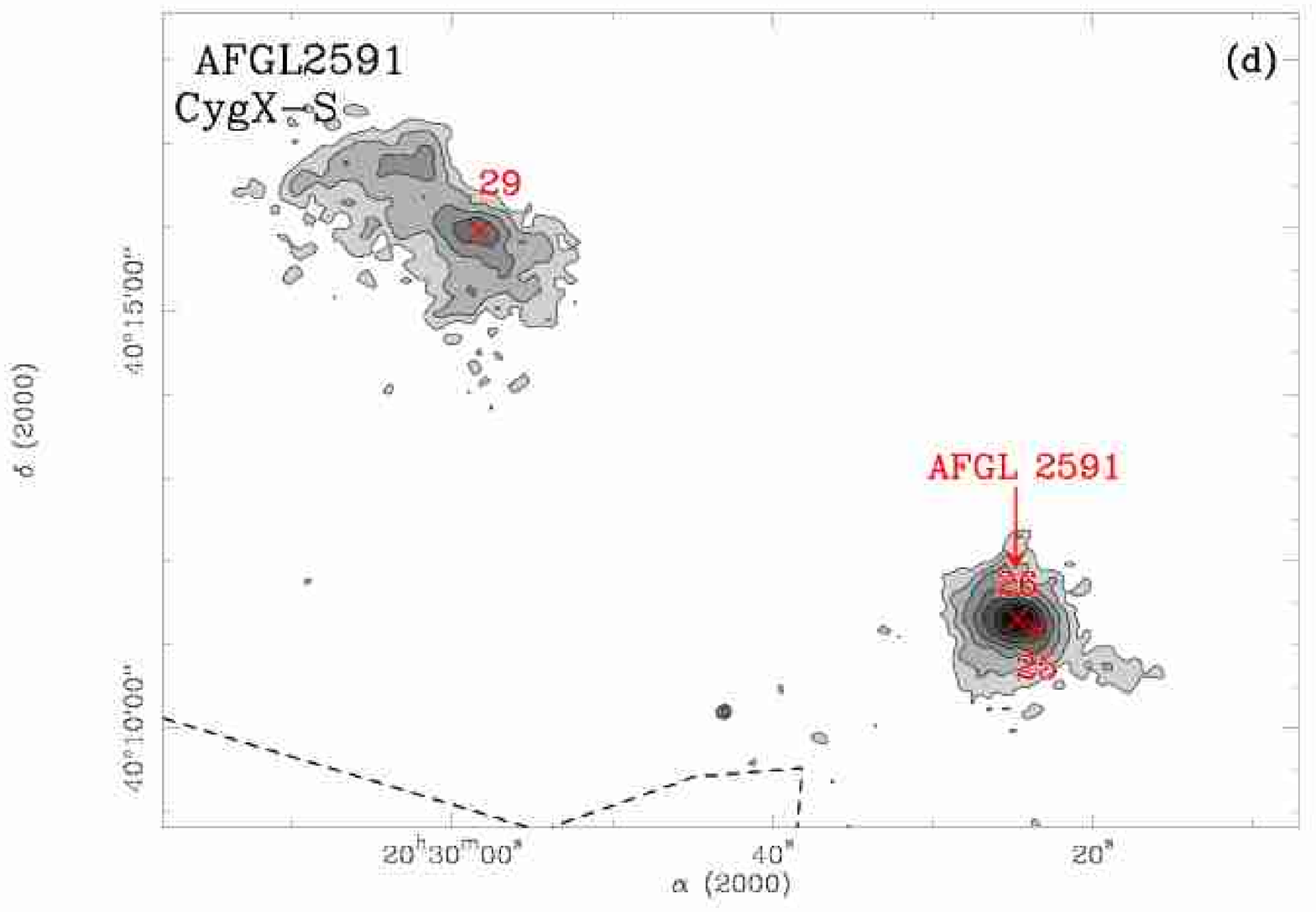}
\hskip -1.1cm \includegraphics[angle=270,width=6.9cm]{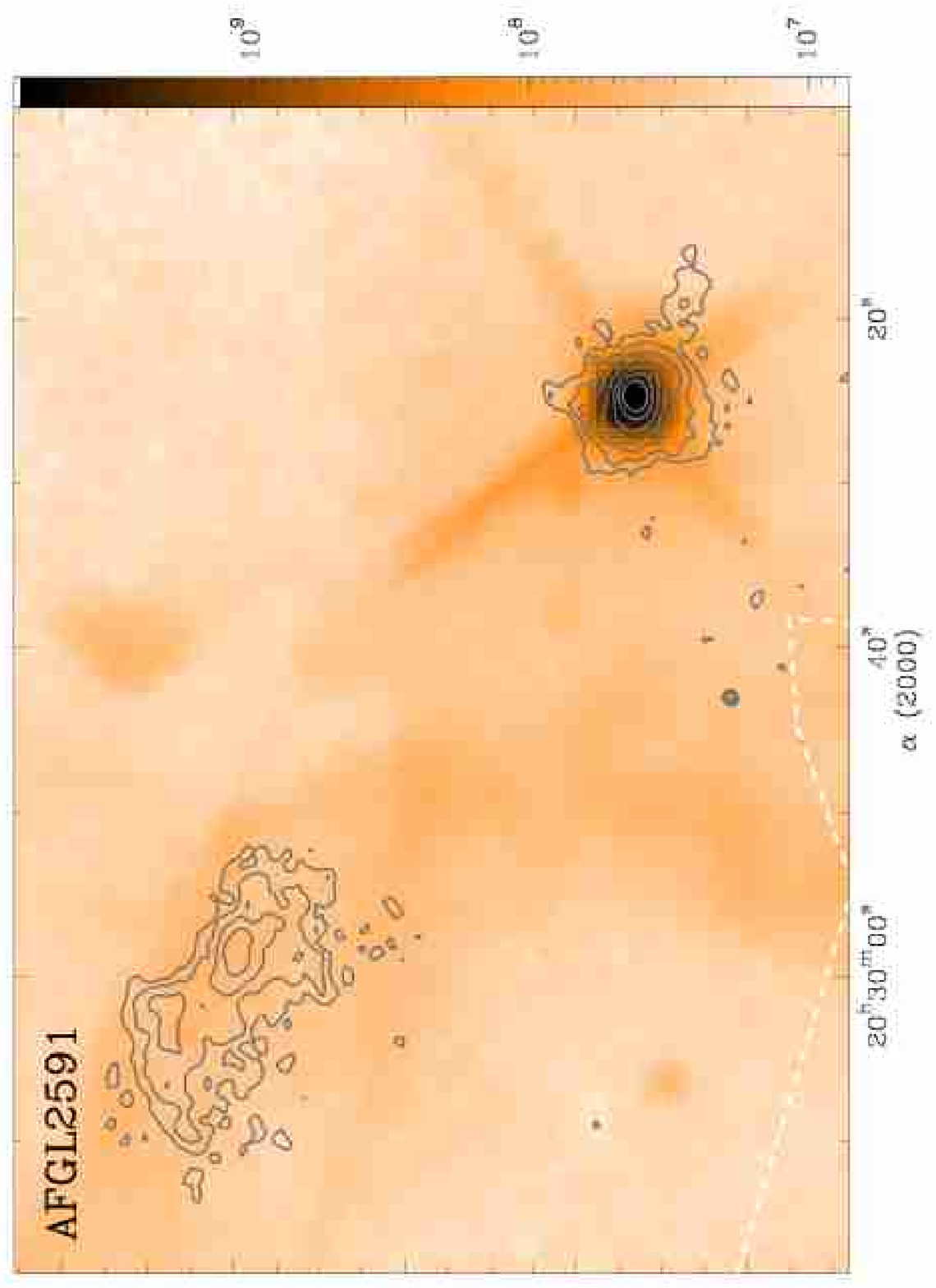}}
\caption[]{Same convention as Fig.~\ref{f:msxnorth_app} for MAMBO maps of CygX-South extracted from Fig.~\ref{f:mambo}b. Regions shown are the most western part of CygX-South ({\bf a}), around Mol121 ({\bf b}), around AFGL~2591 ({\bf c}), and south of DR5  ({\bf d}). Contour levels are logarithmic and go from 60 to $180~\mjb$ in {\bf a}, from 60 to $2\,000~\mjb$ in {\bf b}, from 40 to $200~\mjb$ in {\bf c}, and from 40 to $1\,000~\mjb$ in {\bf d}.}
\label{f:msxsouth_app}
\end{figure*}
\setcounter{figure}{12}
\begin{figure*}
\centerline{\includegraphics[angle=270,width=7.7cm]{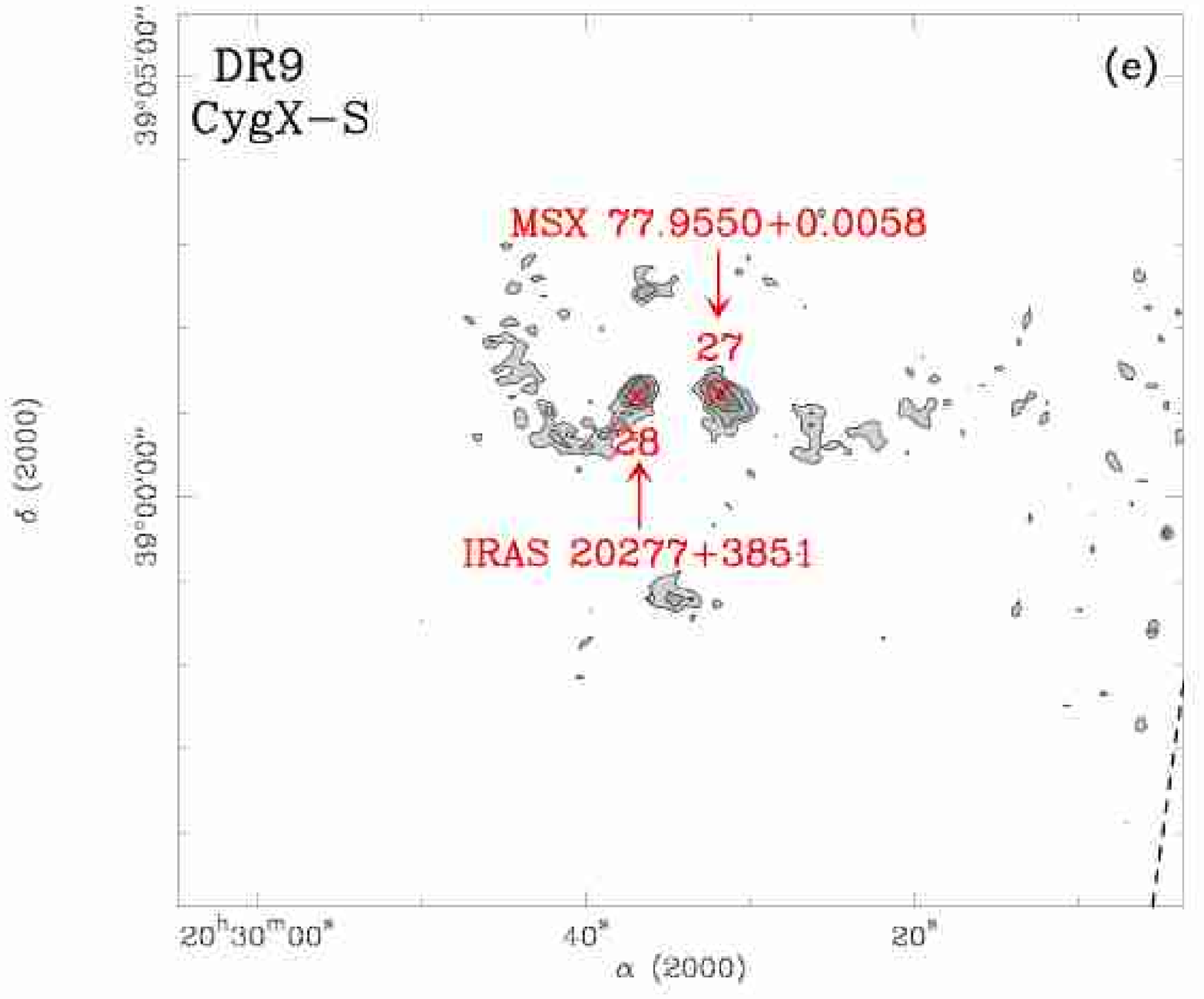}
\hskip -2.4cm \includegraphics[angle=270,width=7.7cm]{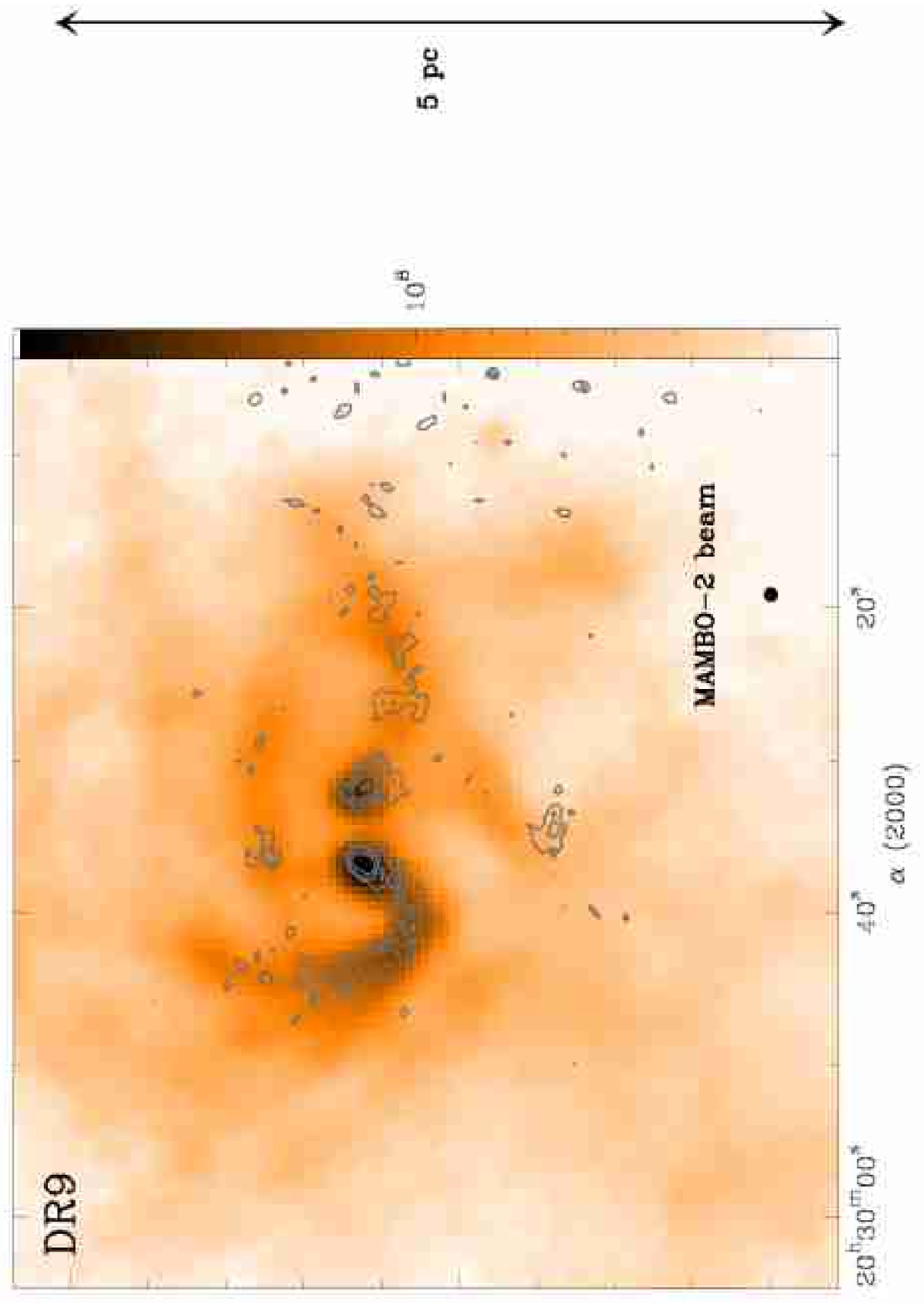}}
\vskip -2.8cm
\centerline{\includegraphics[angle=0,width=11.35cm]{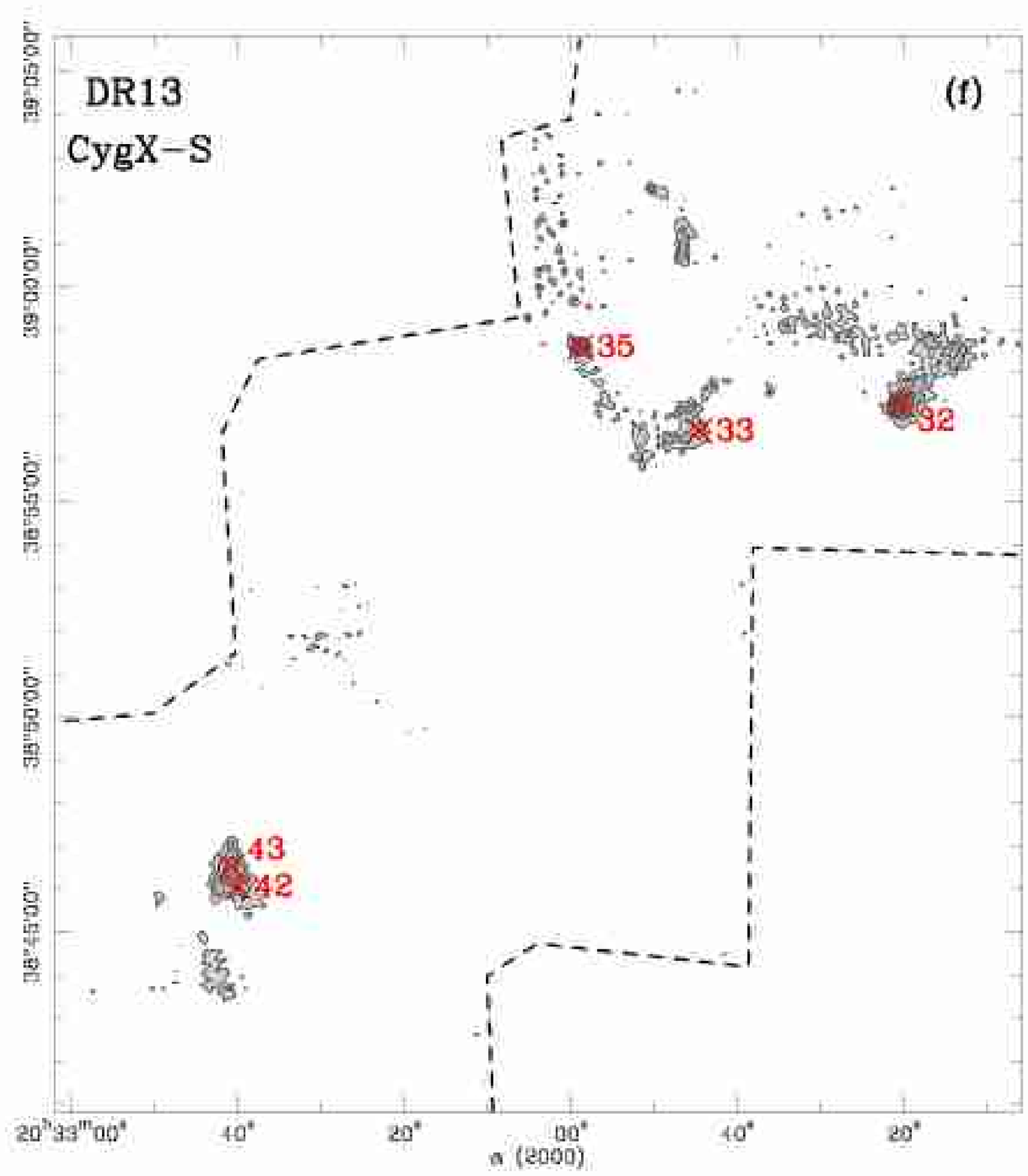}
\hskip -1.8cm \includegraphics[angle=0,width=11.35cm]{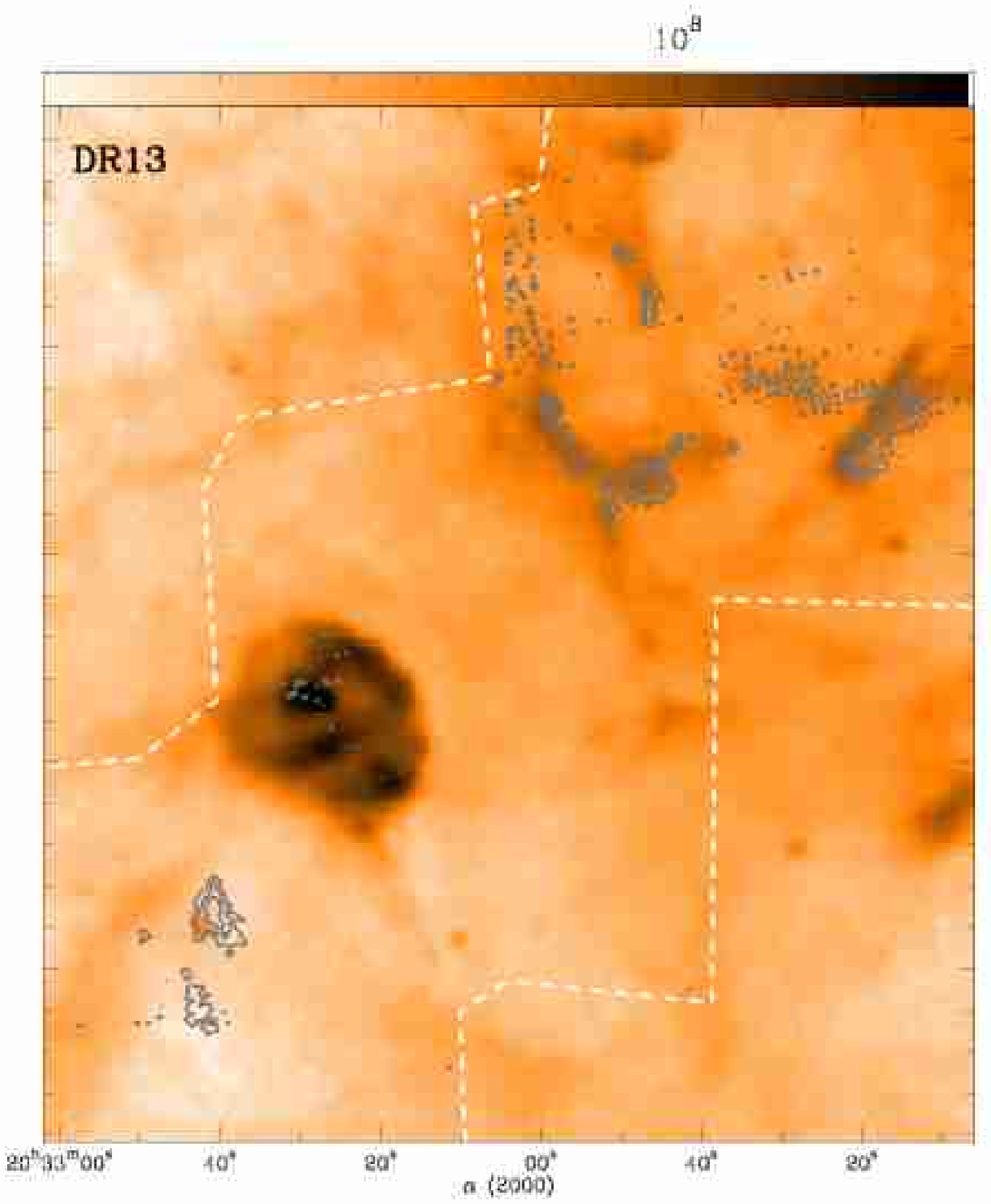}}
\vskip -1.cm
\caption[]{(continued) Same caption for the region around DR9 ({\bf e}) and DR13  ({\bf f}). Contour levels are logarithmic and go from 40 to $200~\mjb$ in {\bf e}, and from 60 to $300~\mjb$ in {\bf f}.}
\end{figure*}

\setcounter{figure}{13}
\begin{figure*}
\centerline{\includegraphics[angle=270,width=7.9cm]{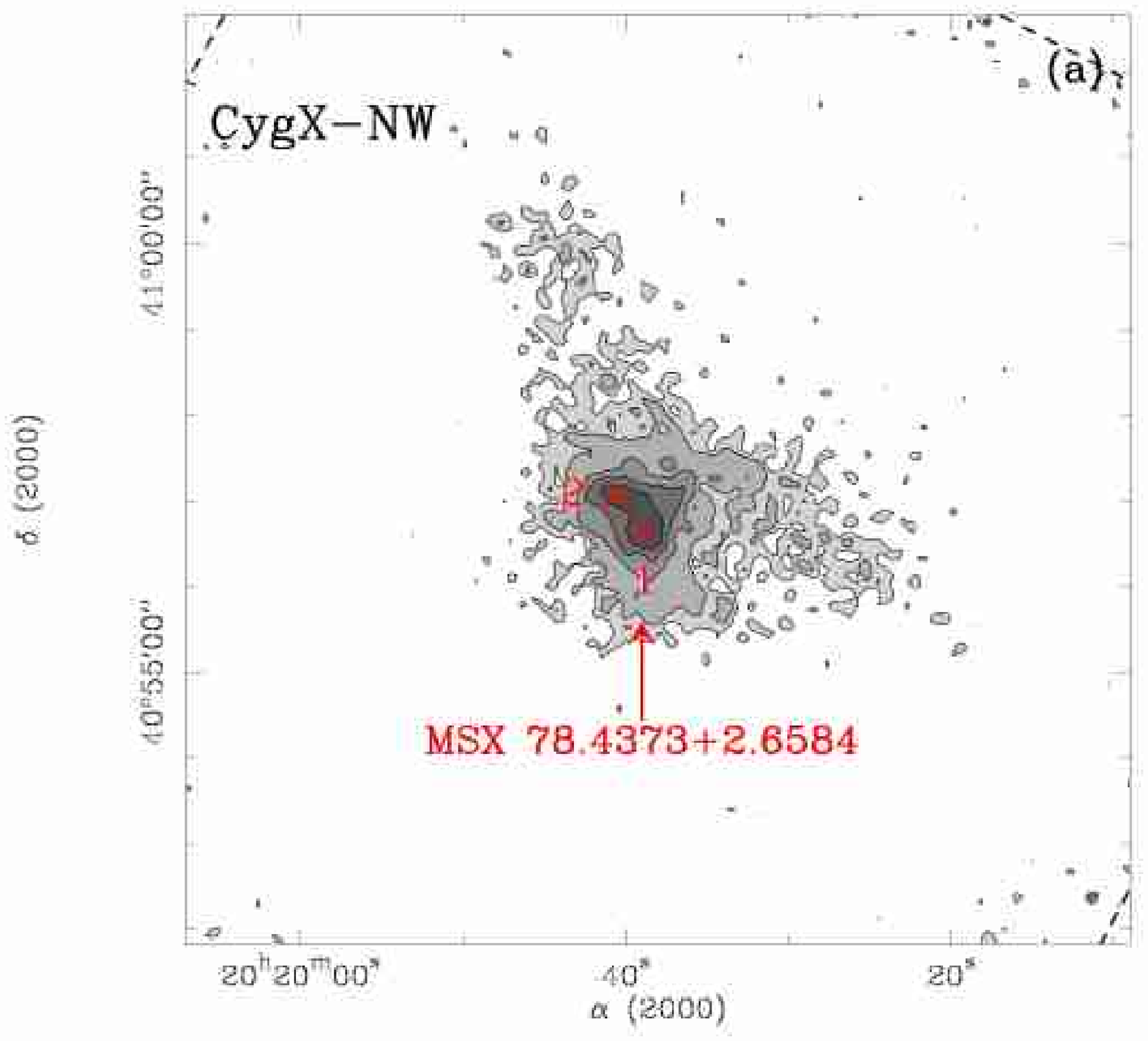} 
\hskip -3.0cm\includegraphics[angle=270,width=7.9cm]{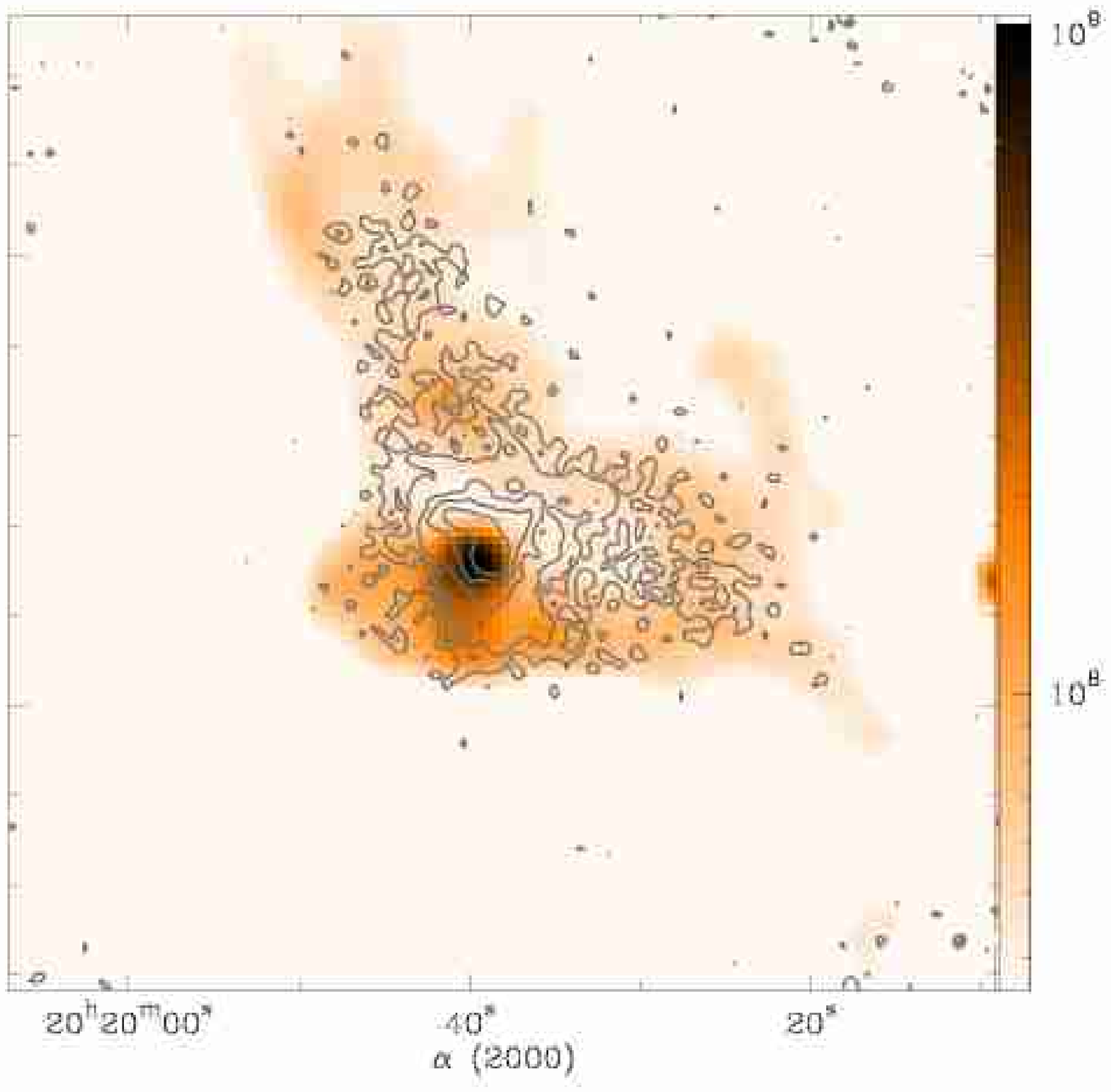}}
\vskip 0.3cm
\centerline{\includegraphics[angle=270,width=6.5cm]{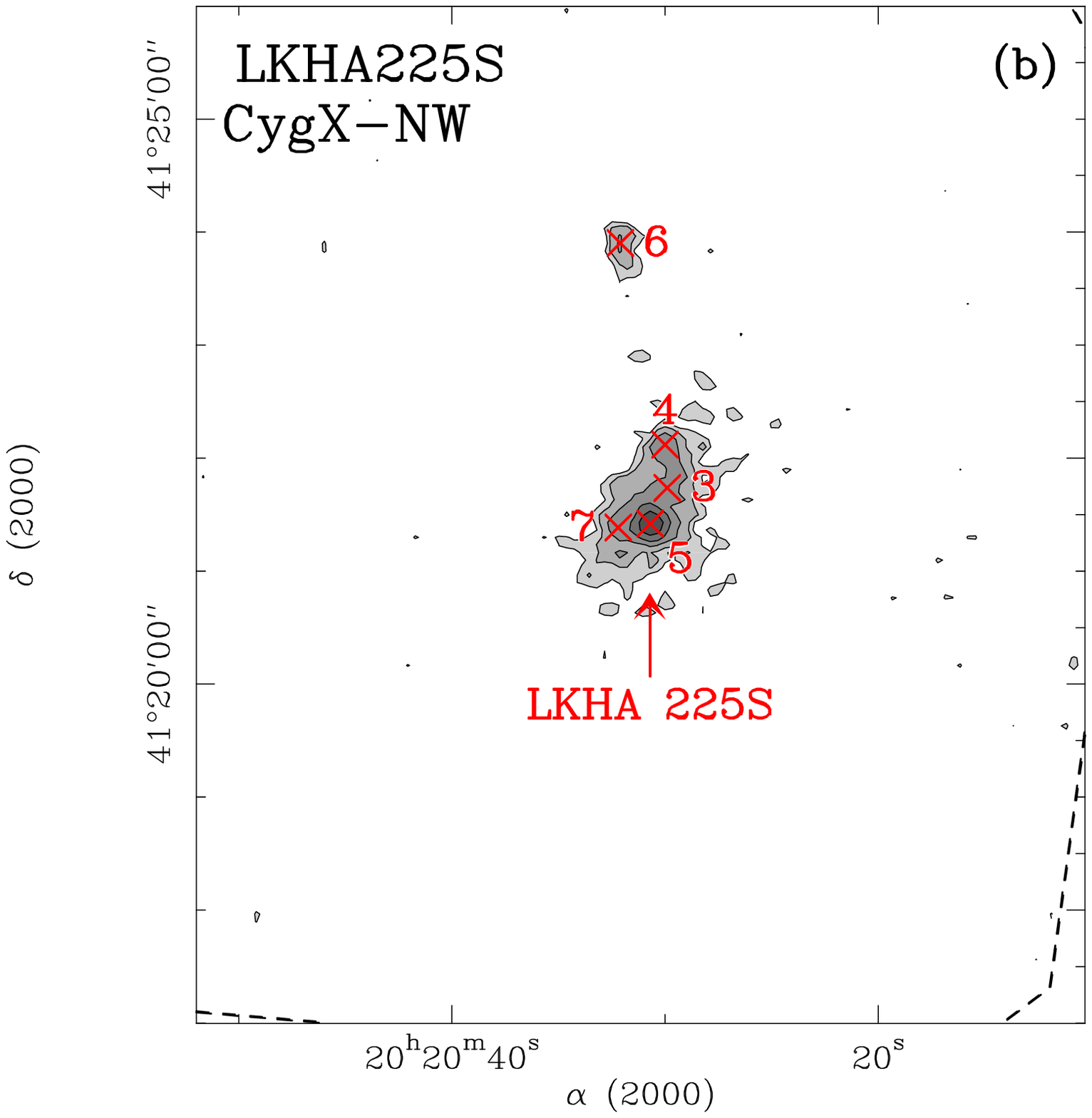}
\hskip -3.0cm\includegraphics[angle=270,width=6.5cm]{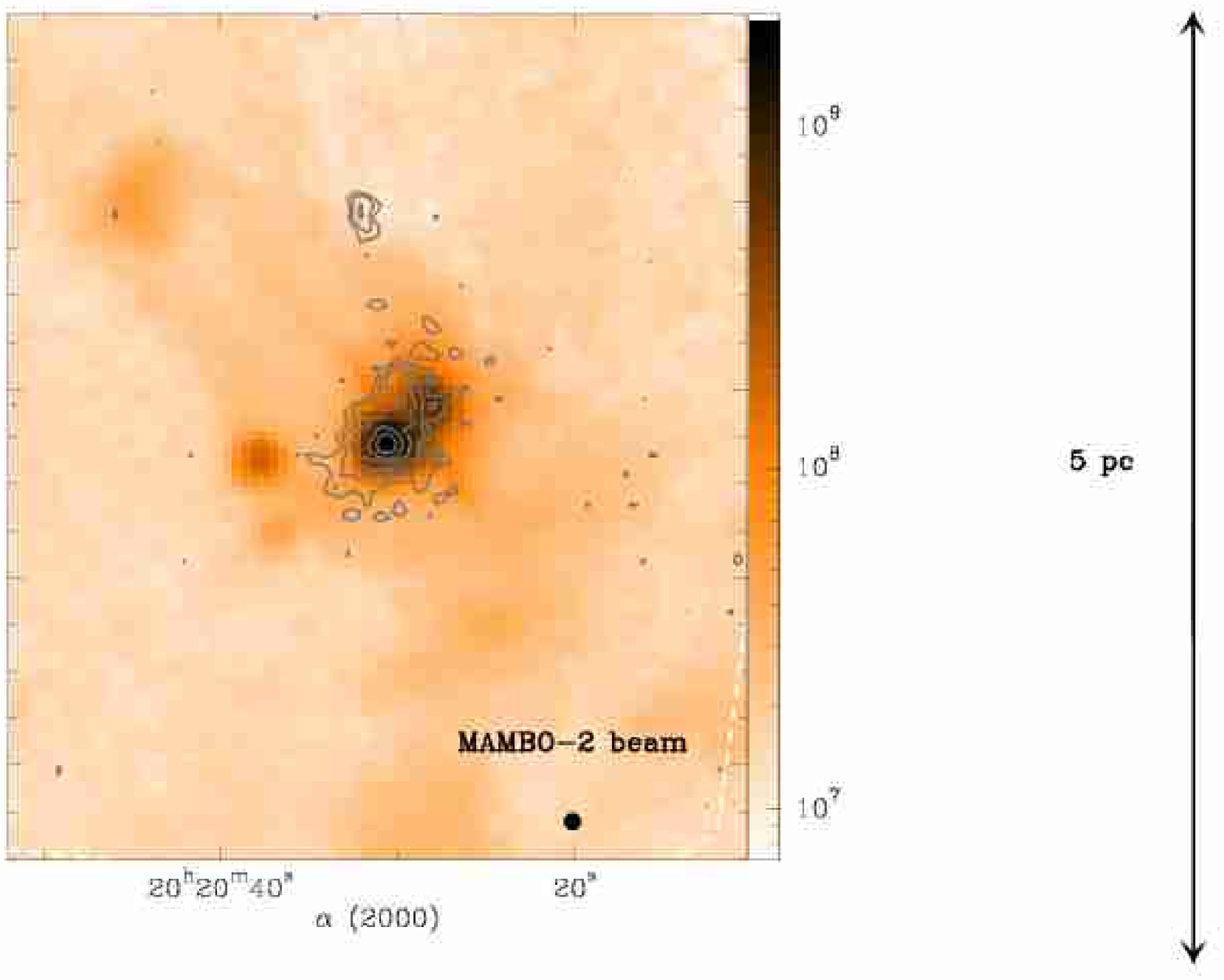}}
\vskip 0.cm
\centerline{\hskip -0.4cm \includegraphics[angle=270,width=6.6cm]{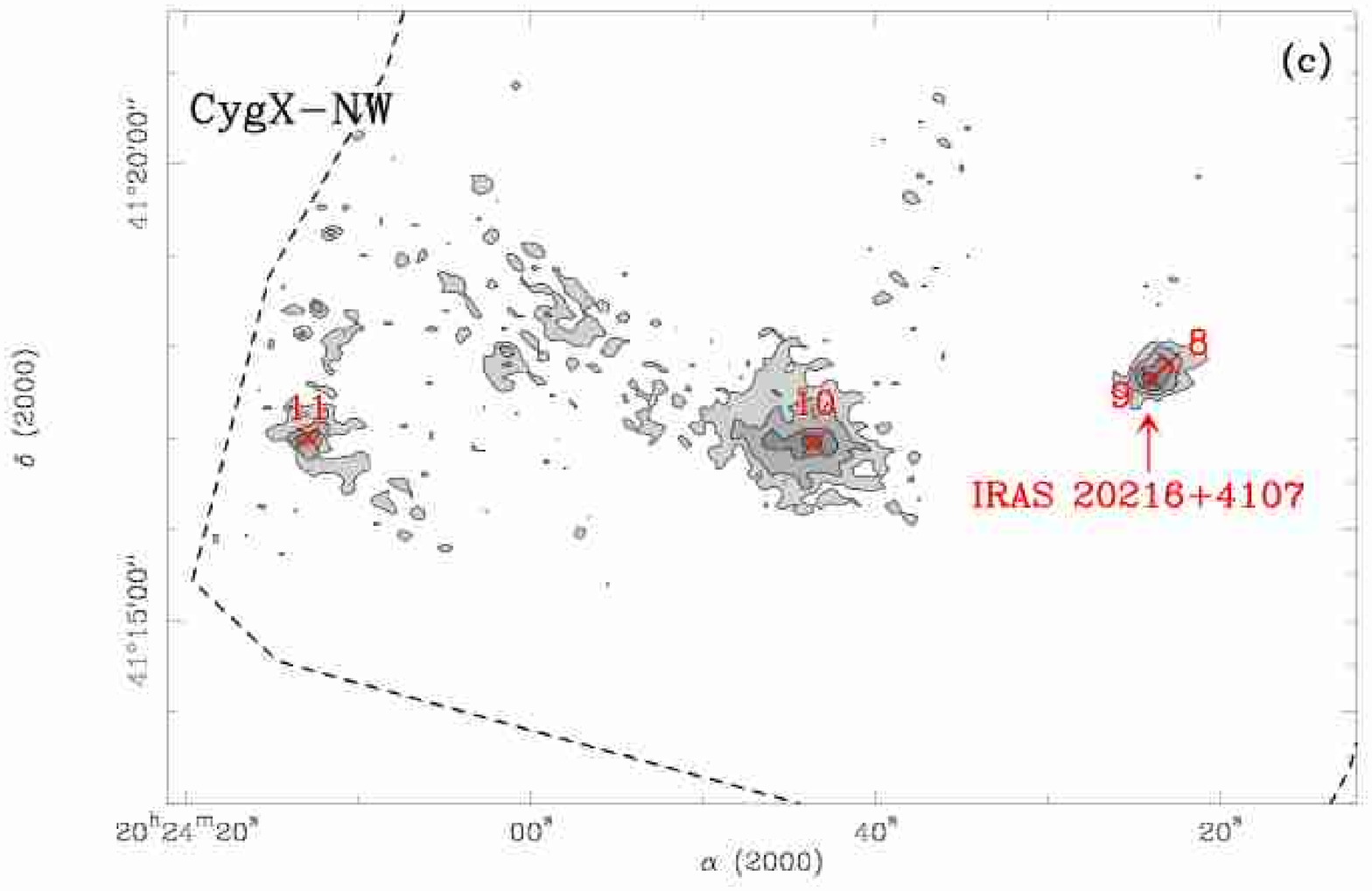}
\hskip -1.cm\includegraphics[angle=270,width=6.6cm]{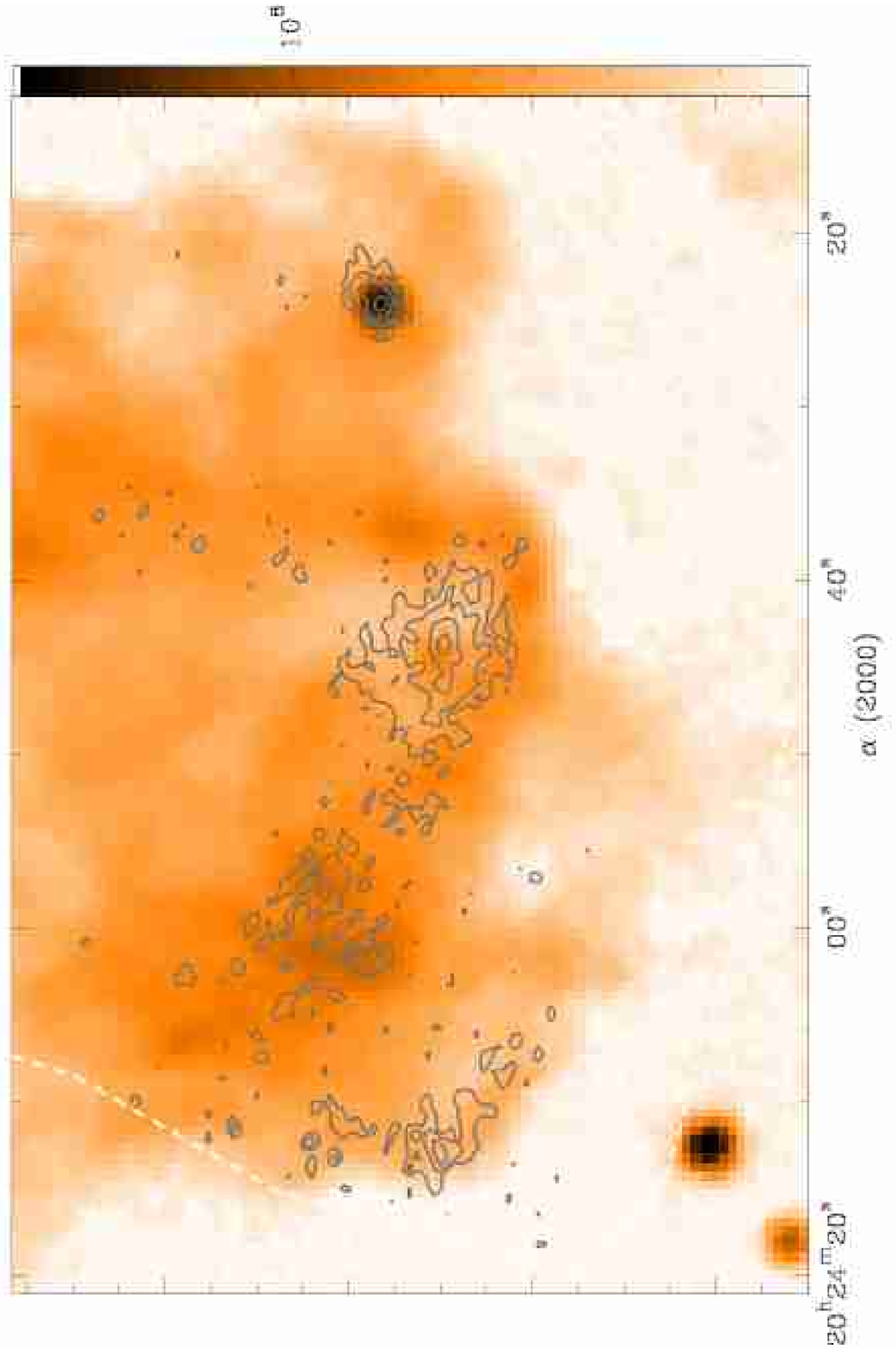}}
\caption[]{Same convention as Fig.~\ref{f:msxnorth_app} for MAMBO maps of CygX-NW extracted from Fig.~\ref{f:mambo}c. Regions shown are those of Fig.~\ref{f:mambo}c4  ({\bf a}), of Fig.~\ref{f:mambo}c3  ({\bf b}), and of Fig.~\ref{f:mambo}c2 ({\bf c}). Contour levels are logarithmic and go from 40 to $300~\mjb$ in {\bf a}, from 60 to $400~\mjb$ in {\bf b}, and from 60 to $250~\mjb$ in {\bf c}.}
\label{f:msxnw_app}
\end{figure*}

\setcounter{figure}{13}
\begin{figure*}
\vskip -0.5cm
\centerline{\includegraphics[angle=0,width=9.5cm]{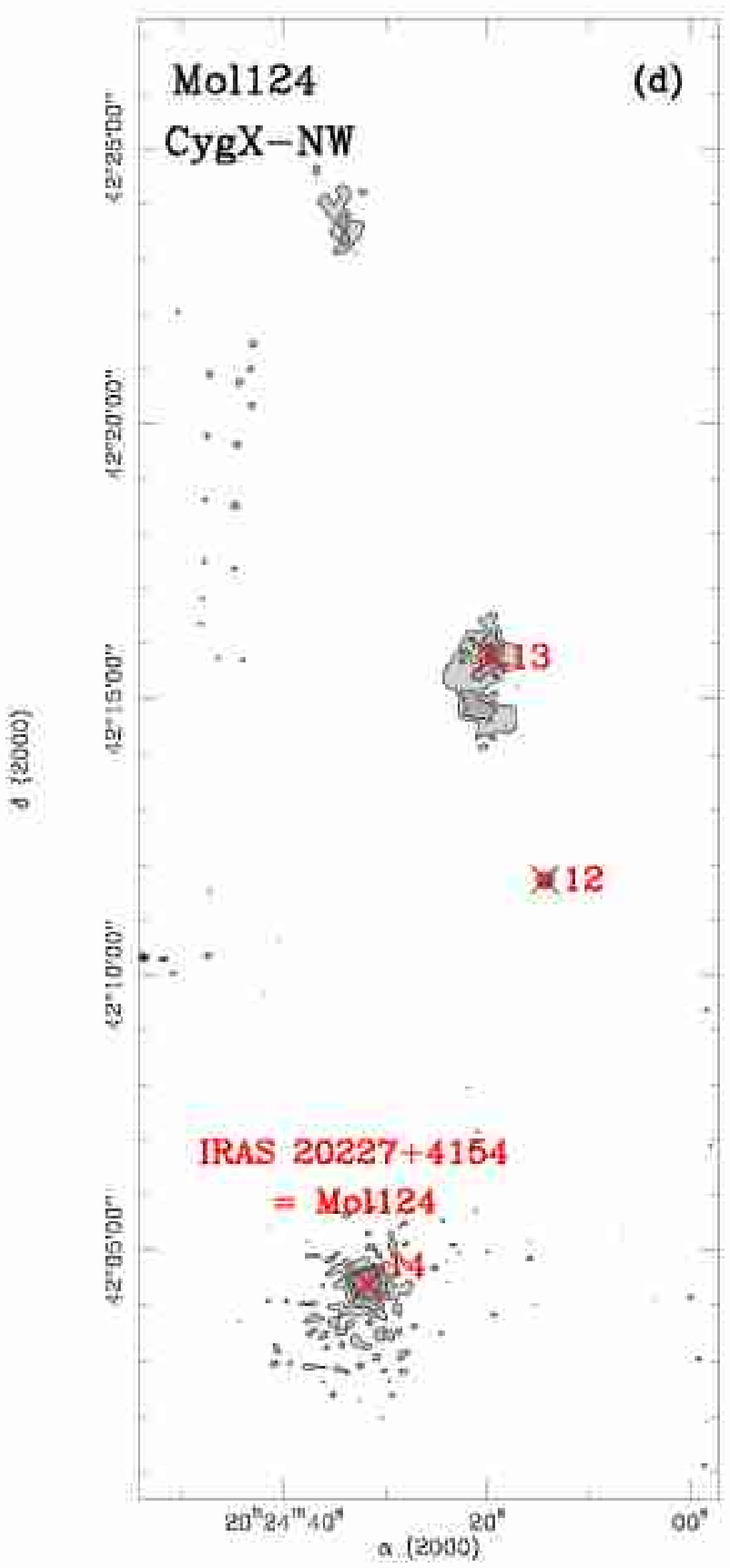}
\hskip -5.cm\includegraphics[angle=0,width=9.5cm]{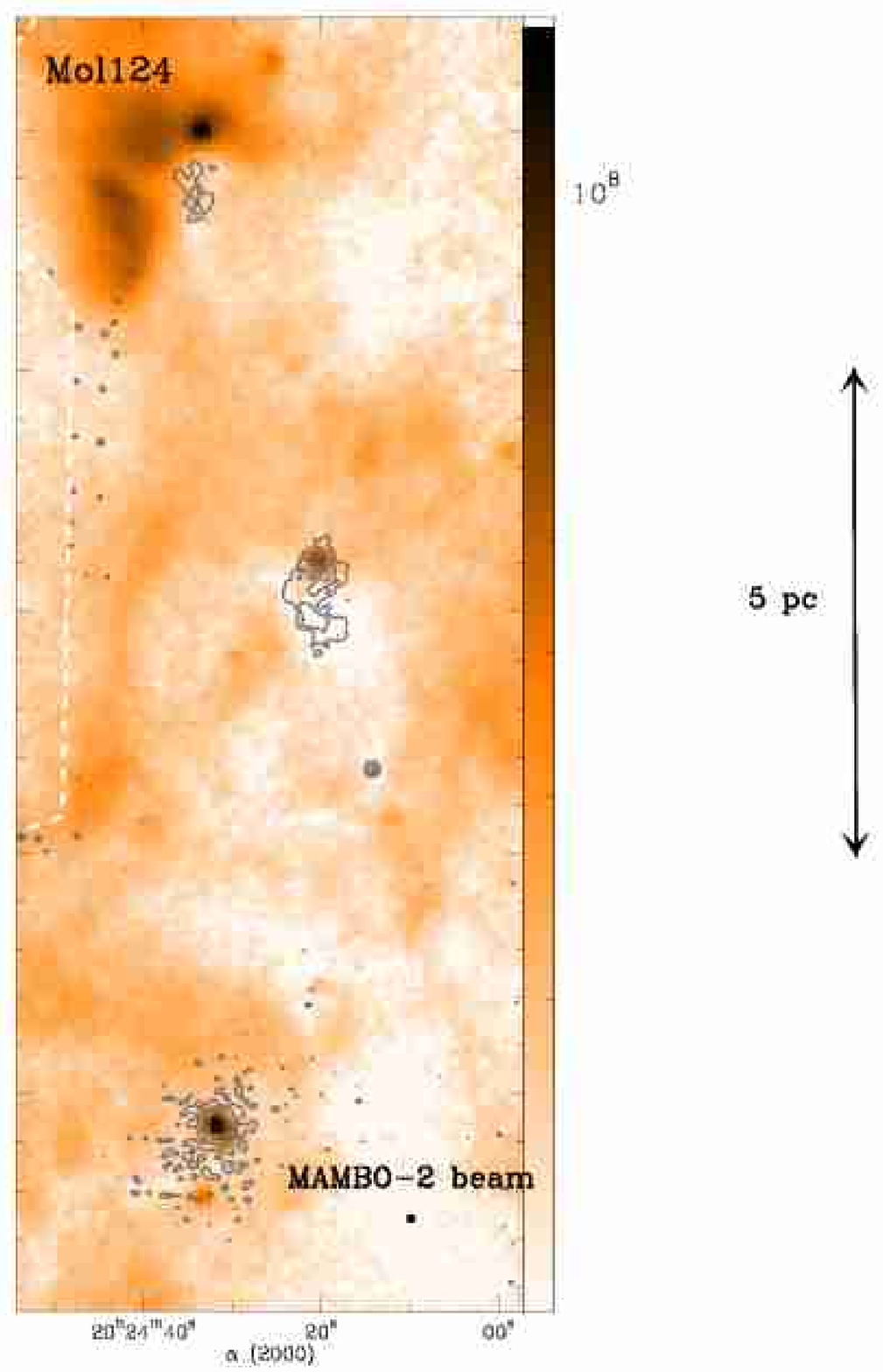}}
\vskip -1.cm
\caption[]{(continued) Same caption for the region around Mol124 ({\bf d}). Contour levels are logarithmic and go from 60 to $180~\mjb$ in {\bf d}.}
\end{figure*}

\begin{figure*}
\vskip 1.5cm
\centerline{\includegraphics[angle=0,width=17cm]{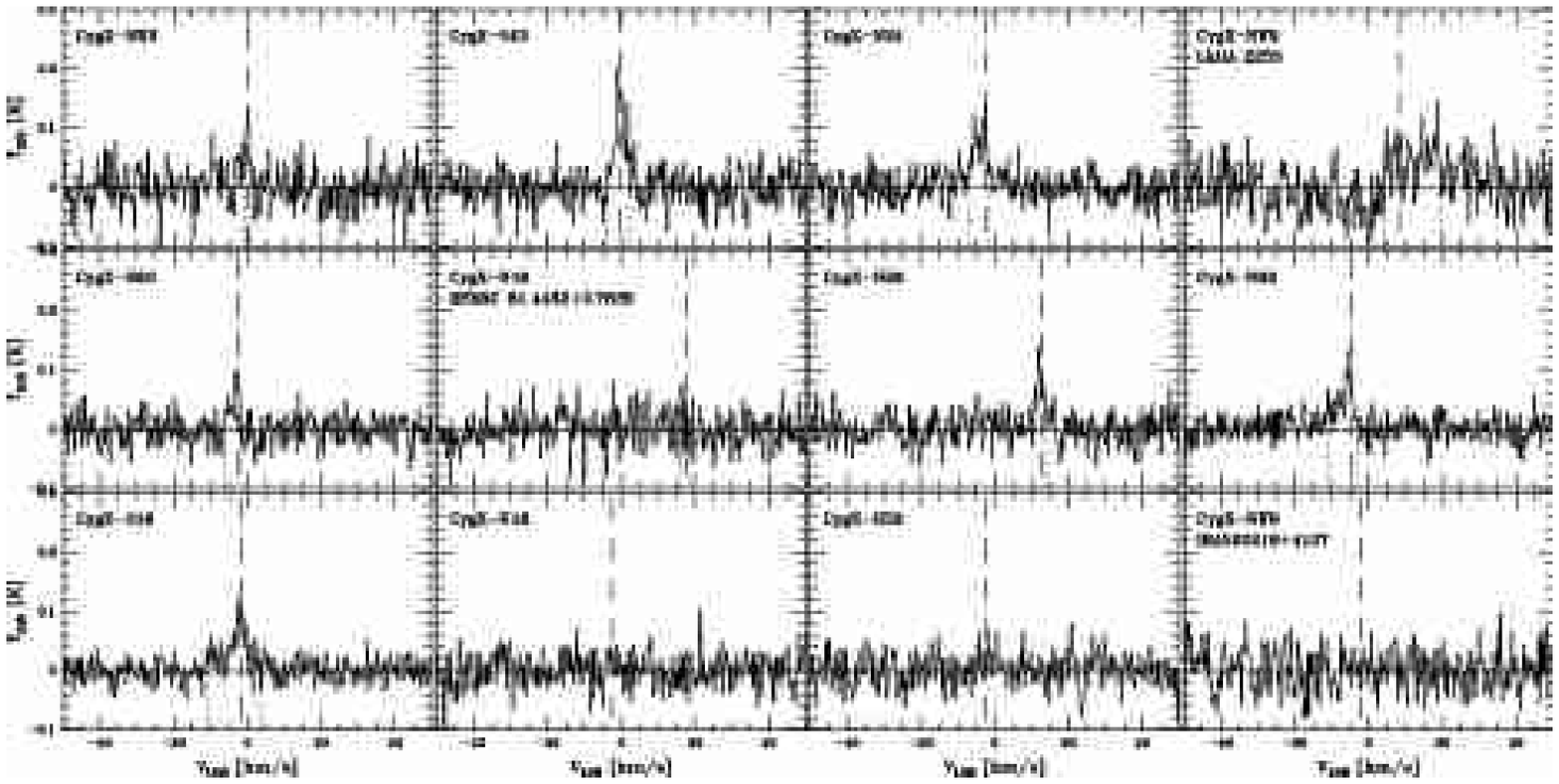}}
\caption{SiO($2-1$) emission toward 12 additional fragments, ordered by decreasing $M_{\rm 1.2mm}$.}
\label{f:sio_app}
\end{figure*}

\end{document}